\documentclass[fleqn,usenatbib]{mnras}
\usepackage{mathtools}

\usepackage{breqn}
\usepackage[T1]{fontenc}
\usepackage{ae,aecompl,ulem}
\usepackage{ragged2e}

\usepackage{soul}
\usepackage{textcomp}

\usepackage{graphicx}	
\usepackage{amsmath}	
\usepackage{amssymb}	
\usepackage{float}
\usepackage{multirow}
\usepackage{units}
\usepackage{placeins}

\usepackage{pdflscape}
\usepackage{rotating} %
\usepackage{epstopdf}
\usepackage{comment}

\usepackage[table]{xcolor}
\definecolor{Lightblue}{rgb}{0.68,0.85,0.9}
\definecolor{magenta2}{rgb}{0.8, 0.6, 0.8}
\definecolor{orange}{rgb}{0.93, 0.57, 0.13}

\title[Accretion Geometry of BH-XRBs: A Multi-Mission Study]{Probing the accretion geometry of black hole X-ray binaries: A multi-mission spectro-polarimetric and timing study}

\author[Majumder et al]{Seshadri Majumder$^{1}$\thanks{E-mail: smajumder@iitg.ac.in},
	Ankur Kushwaha$^{2}$, 
        Swapnil Singh$^{2}$,
	Kiran M. Jayasurya$^{2}$, 
	Santabrata Das$^{1}$\thanks{E-mail: sbdas@iitg.ac.in}, 
	\newauthor Anuj Nandi$^{2}$\thanks{E-mail: anuj@ursc.gov.in} \\
	$^{1}$Department of Physics, Indian Institute of Technology Guwahati, Guwahati, 781039, India.\\
	$^{2}$Space Astronomy Group, ISITE Campus, U. R. Rao Satellite Centre, Outer
	Ring Road, Marathahalli, Bangalore, 560037, India. \\
}

\begin{document}
	\label{firstpage}
	\maketitle

\begin{abstract}

We present a comprehensive spectro-polarimetric and timing analysis of twelve black hole X-ray binaries, namely Cyg X$-$1, 4U 1630$-$47, Cyg X$-$3, LMC X$-$1, 4U 1957$+$115, LMC X$-$3, Swift J1727.8$-$1613, GX 339$-$4, Swift J151857.0$-$572147, IGR J17091$-$3624, MAXI J1744$-$294 and GRS 1915$+$105, using quasi-simultaneous observations from {\it{IXPE}}, {\it{NICER}}, {\it{NuSTAR}}, and {\it{AstroSat}}. Timing analyses reveal type-B and type-C Quasi-periodic Oscillations across different spectral states, often associated with episodic radio ejections. Broadband ($0.7-60$ keV) spectral modeling, employing disc, Comptonization, and reflection components, reveals degeneracies in constraining disc-corona geometries. Polarimetric measurements in $2-8$ keV band detect significant polarization degrees (PDs) ranging from $3-20.6\%$ ($1.2-21.4\%$) in harder (softer) states, with moderate to strong energy dependence, except for LMC X$-1$, Swift J$151857.0-572147$, and MAXI J1744$-$294, where no significant polarization is detected. We report polarization detections of Cyg X$-3$ (PD $\sim 21.4\%$, SIMS), LMC X$-3$ (PD $\sim 2.4\%$, HSS) and IGR J$17091-3624$ (PD $\sim 9\%$, LHS) using recent {\it IXPE} observations. A positive correlation is found between PD and Comptonized photon fraction ($cov_{\rm frac}$), while an anti-correlation is observed with disc-to-Comptonized flux ratio ($F_{\rm ratio}$) across spectral states. Combined timing, spectral, and polarimetric results, together with constraints from radio jet observations, suggest a radially extended corona within a truncated disc for Cyg X$-1$, Swift J$1727.8-1613$, IGR J$17091-3624$, and GX $339-4$, whereas the disc-corona geometry remains poorly constrained for 4U $1957+115$, LMC X$-3$, and 4U $1630-47$. We discuss the implications of these findings for understanding accretion geometries and highlight prospects for future X-ray polarimetric studies.

\end{abstract}


\section{Introduction}	

Black hole X-ray binaries (BH-XRBs) are believed to be the ideal candidates for understanding the physical processes that govern the radiation mechanism around the compact objects. BH-XRBs often exhibit distinct spectral states over timescales ranging from days to months, strongly associated with the underlying accretion dynamics of the system \cite[and references therein]{Morgan-etal1997, Done-etal1999, Chakrabarti-etal2000, Homan-etal2001, Belloni-etal2005, Remillard-etal2006, Nandi-etal2012, Iyer-etal2015, Sreehari-etal2019, Baby-etal2020, Majumder-etal2022, Athulya-etal2022, Prabhakar-etal2023, Nandi-etal2024, Majumder-etal2025b}. Empirically, the high-soft state (HSS) is characterized by the multi-temperature black-body emission that is likely to emerge from an optically thick and geometrically thin accretion disc \cite[]{Shakura-etal1973}. In contrast, the canonical low-hard state (LHS), represented by a power-law profile with high energy cut-off, is found to be dominated by the hard emission produced from the Compton upscattering of disc photons into the `hot' electron cloud (equivalently X-ray corona). Moreover, depending on the relative contributions from the disc and coronal components, hard/soft intermediate states (HIMS/SIMS) are observed.

Needless to mention that the possible geometry of the Comptonizing corona in BH-XRBs remains an unsettled issue to date. Several alternative scenarios featuring the models of the sandwich corona \cite[]{Haardt-etal1993, Stern-etal1995}, radially elongated corona at the truncated inner accretion disc \cite[]{Eardley-etal1975, Chakrabarti-etal1995, Poutanen-etal1997, Iyer-etal2015} and vertically extended corona as the base of the jet \cite[]{Miyamoto-etal1991, Markoff-etal2005, Mendez-etal2022, Zhang-etal2023} have been proposed over the years. However, the overall geometry of the disc-corona-jet remains elusive mostly due to the model degeneracies that complicate the interpretation of the observational data.

Notably, the temporal properties of BH-XRBs show rapid X-ray variability over different timescales. This variability is usually observed in the power density spectrum (PDS) and is closely correlated to the spectral states. In particular, transient phenomena like strong and stable Low-Frequency Quasi-periodic Oscillations (LFQPOs) on a wide range of frequencies distinguish the spectral states and act as the precursor of the state transitions in BH-XRBs \citep{Remillard-etal2006, Done-etal2007, Nandi-etal2012,Iyer-etal2015}. For example, the LHS and HIMS are characterized by the appearance of strong, coherent, variable peaked type-C LFQPO of frequency $\sim 0.1-15$ Hz superposed on a flat-top noise (FTN) component in the PDS \cite[and references therein]{Remillard-etal2006, Nandi-etal2012}. The origin of these type-C QPOs is often explained through various mechanisms, including oscillations of radiative shock waves within the accretion disc \citep{Molteni-etal1996, Chakrabarti-etal2008, Das-etal2014}, relativistic Lense-Thirring precession of the inner hot flow or the truncated disc \citep{Stella-etal1998, Ingram-etal2009}, precession of small-scale jets \citep{Ma-etal2021} and outward drift of the truncated inner disc radius enveloped by the corona \citep{Karpouzas-etal2020, Bellavita-etal2022, Bellavita-etal2025}. The wide array of interpretations introduces a degeneracy among different disc-corona-jet configurations used in explaining QPO phenomena (see \citealt{Ingram-etal2019} for a review). 

On the other hand, relatively weak type-B/type-A QPOs appear at a narrow frequency range of around $\sim 6-8$ Hz \citep{Casella-etal2005} during the SIMS. In this state, FTN is absent, and the PDS continuum shows weak red noise characterized by a simple power-law in PDS continuum. These type-B/type-A QPOs are often found to be closely connected with the radio ejections generally observed in the SIMS \citep{Soleri-etal2008, Fender-etal2009, Kylafis-etal2020, Homan-etal2020, Garcia-etal2021, Liu-etal2022, Zhang-etal2023}. Usually, soft states are characterized by less variability in the PDS without the detection of QPO like features \citep{Belloni-etal1999, Belloni-etal2005, Nandi-etal2012, Radhika-etal2014,Radhika-etal2016}.

Furthermore, X-ray polarimetric study is also considered a powerful diagnostic tool to infer the accretion geometry of the BH-XRBs. The recent launch of {\it IXPE} \cite[]{Weisskopf-etal2022}, a polarimetric mission sensitive to low-energy ($2-8$ keV) X-rays, enables the opportunity to investigate in-depth polarimetric properties of BH-XRBs. So far, {\it IXPE} has observed twelve BH-XRBs, namely Cyg X$-$1, 4U 1630$-$47, Cyg X$-$3, LMC X$-$3, LMC X$-$1, 4U 1957$+$115, Swift J1727.8$-$1613, Swift J151857.0$-$572147, GX 339$-$4, IGR J17091$-$3624, MAXI J1744$-$294 and GRS 1915$+$105 with significant polarized emission detected in eight sources \cite[see also][]{Dovciak-etal2024}.

Despite the significant advancements in X-ray polarimetry, interpreting the observed polarization degree (PD) and polarization angle (PA) within the framework of theoretical models remains a formidable challenge. The classical work by Chandrasekhar \citep{Chandrasekhar-1960} on semi-infinite electron scattering predicts a low PD of $\sim 2\%$ from the accretion disc of highly inclined systems. However, 4U 1630$-$47 shows a remarkably higher PD of around $8.3\%$ in its disc-dominated thermal state, deviating substantially from these theoretical expectations \citep{Kushwaha-etal2023a, Ratheesh-etal2024}. Similarly, soft state observations of Cyg X$-1$ show polarization perpendicular to the disc plane, whereas theoretical models predict it to be parallel. This possibly indicates a rapidly rotating black hole in Cyg X$-1$, responsible for the returning radiation effects seen in the spectro-polarimetric data \citep{Steiner-etal2024, Krawczynski-etal2024}. Furthermore, PDs of about $4\%$ have been observed in the LHS of Cyg X$-$1 and Swift J1727.8$-$1613 \citep{Krawczynski-etal2022b, Ingram-etal2024}, whereas existing models predict only $\sim 1\%$ PD from a wedge-shaped corona in such low-inclined systems \citep{Krawczynski-etal2022b}. To reconcile this discrepancy, \cite{Poutanen-etal2023} invoked the model of \cite{Beloborodov-etal1998} of an outflowing corona with mildly relativistic motion to explain the high PD levels. It is worth mentioning that the recent detection of $\sim 9\%$ PD in the LHS of IGR J17091$-$3624 \citep{Ewing-etal2025} further indicates the limitations of current theoretical frameworks and highlights the need for more comprehensive models of X-ray polarization.

Moreover, in certain cases, the spectro-temporal findings of BH-XRBs appear to be contradictory with interpretations derived from X-ray polarimetry. For instance, in Swift J1727.8$-$1613, the evolution of type-C QPOs suggests the presence of a jet-like corona aligned perpendicular to the disc plane \citep{Liao-etal2024}. In contrast, simultaneous polarimetric observations of the same source indicate a radially extended corona resided at the equatorial plane of the disc \citep{Veledina-etal2023, Ingram-etal2024}. Therefore, it is evident that the polarimetric findings offer challenges to the existing X-ray spectro-temporal models, indicating that a deeper understanding of the accretion dynamics and disc-corona-jet geometry in BH-XRBs is yet to be unveiled.

Being motivated by this, we carry out in-depth timing and spectro-polarimetric analyses of eleven BH-XRBs using quasi-simultaneous archival observations from {\it IXPE}, {\it NICER}, {\it NuSTAR} and {\it AstroSat}, covering a broad energy range of $0.5-100$ keV. We conduct a comprehensive spectro-temporal study of these sources using {\it NICER}, {\it NuSTAR} and {\it AstroSat} data and confirm the presence of distinct spectral states of the BH-XRBs during the observational campaigns. Further, we deduce the polarization properties of the sources using {\it IXPE} data in the $2-8$ keV energy range, followed by a detailed spectro-polarimetric correlation study. Notably, the timing and spectro-polarimetric results serve as powerful diagnostics offering deeper insights into the complex accretion dynamics and geometry of BH-XRBs under consideration.

The paper is organized as follows. In \S2 and \S3, we present the source selection with observation details and the data reduction procedures, respectively. The analysis and results of the timing and spectro-polarimetric study are presented in \S4. Finally, we discuss the results and present conclusions in \S5 and \S6, respectively.

\begin{figure*}
    \begin{center}
	\includegraphics[width=0.8\textwidth]{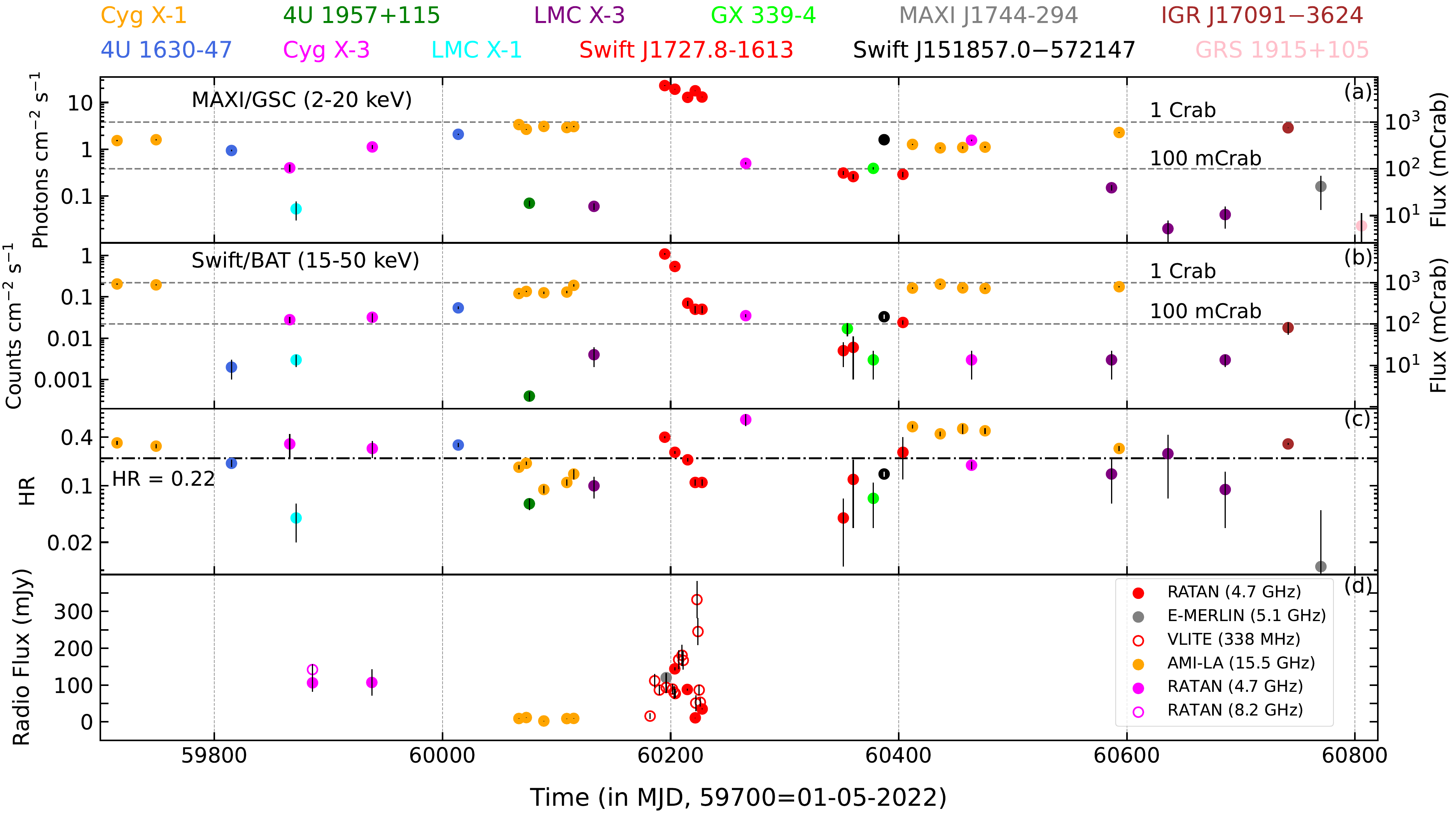}
    \end{center}
    \vskip -0.4cm
    \caption{Long-term X-ray monitoring of the sources with {\it MAXI/GSC} ($2-20$ keV), {\it Swift/BAT} ($15-50$ keV) and the corresponding hardness ratio obtained with {\it MAXI/GSC} are presented in panel (a), (b) and (c), respectively. Each data point for a given source represents averaged out respective quantity over the exposure of corresponding {\it IXPE} observation. In panel (d), the variation of radio flux is shown from the quasi-simultaneous radio observations available from various facilities. The $1\sigma$ errorbars of the respective quantities are small and remain within the markers for most of the cases. For a given source, the data points plotted with MJD, correspond to the respective epochs as mentioned in Table \ref{tab:log}.  
    }
    \label{fig:lcurve}
\end{figure*}

\vspace{-0.5cm}

\section{Source Selection and observations}

In this work, we analyze all the {\it IXPE} \citep{Weisskopf-etal2022} observations of BH-XRBs conducted so far along with simultaneous/quasi-simultaneous {\it NICER}, {\it NuSTAR} and {\it AstroSat} observations depending on data availability. As of now, {\it IXPE} observed twelve BH-XRBs during its first three and a half years of operational period. The {\it IXPE}, {\it NICER} and {\it NuSTAR} observations used in this work are publicly available in the {HEASARC}\footnote{\url{https://heasarc.gsfc.nasa.gov/db-perl/W3Browse/w3browse.pl}} database and the {\it AstroSat} data is archived at the {ISSDC}\footnote{\url{https://webapps.issdc.gov.in/astro_archive/archive/Home.jsp}} website. The quasi-simultaneous multi-mission observations of the sources are separated by at most three days from the {\it IXPE} epoch except LMC X$-1$ for which the {\it NuSTAR} observation lies four days after the {\it IXPE} observation. The sources exhibiting marginal spectro-temporal variability between observations from different missions offer an opportunity for a combined multi-mission study. We mention that {\it IXPE} monitoring of GRS $1915+105$ was carried out on $10^{\rm th}$ May 2025 for about $\sim 142$ ks. However, the source remained below the detection level with {\it IXPE}, showing a {\it MAXI/GSC} flux of $\sim6$ mCrab (see Fig. \ref{fig:lcurve}). Therefore, GRS $1915+105$ is excluded from the detailed analyses in the following sections. All the observations analyzed in this study are tabulated in Table \ref{tab:log}. A brief overview of each of the sources under consideration are presented below.

\noindent $\bullet$ {\bf Cyg X$-$1:} This is the first galactic BH-XRB discovered in 1971 \cite[]{Webster-etal1972}, continues to be one of the most extensively studied celestial objects. Recent measurements reveal that it hosts a black hole with a mass of $21.2\pm2.2$ $M_\odot$, located at a distance of $2.2\pm0.2$ kpc, with a supergiant O-type star as its binary companion \citep{Miller-Jones-etal2021}. Meanwhile, numerous studies confirm the presence of a maximally rotating black hole having spin greater than $0.99$ at the center core of the binary system \citep{Tomsick-etal2013, Gou-etal2014, Zhao-etal2021, Kushwaha-etal2021}. Cyg X$-$1 has remained persistently bright, mostly in the LHS, though it has transitioned to the HSS through short-lived intermediate states over the past few decades \citep{Kushwaha-etal2021}. Interestingly, significant polarized emission is detected in both LHS and HSS of the source \citep{Krawczynski-etal2022b, Jana-etal2024, Steiner-etal2024}.

\noindent $\bullet$ {\bf 4U 1630$-$47:} This is a recurrent X-ray transient discovered by {\it Uhuru} \cite[]{Jones-etal1976}. Since discovery, the source has exhibited more than $20$ quasi-period outbursts \cite[]{Baby-etal2020, Chatterjee-etal2022} typically in every $600-700$ days \cite[]{Parmar-etal1995}. Several efforts relying on indirect measurements constrained its mass as $10 \pm 0.1$ $M_\odot$ \cite[]{Seifina-etal2014}, distance as $10-11$ kpc \cite[]{Seifina-etal2014, Kalemci-etal2018} and high inclination $i\sim 60-70^{\circ}$ \cite[]{Tomsick-etal1998, Seifina-etal2014}. Spectral modeling \cite[]{Pahari-etal2018} and spectro-polarimetric fitting in the high soft state confirm its spin as $\sim 0.92$ \cite[]{Kushwaha-etal2023a}. Notably, 4U 1630$-$47 is the second source observed by {\it IXPE} for which a remarkably high polarization degree ($\sim 6-8\%$) has been detected in both thermal and steep powerlaw (SPL) states \citep{Ratheesh-etal2024, Kushwaha-etal2023a, Rawat-etal2023a, Rodriguez-etal2023}.

\noindent $\bullet$ {\bf Cyg X$-$3:} It was discovered over five decades ago \cite[]{Giacconi-etal1967} and is a high-mass X-ray binary system hosting a compact object accreting from a Wolf-Rayet donor star \cite[]{van-Kerkwijk-etal1992}. Despite extensive studies over the years, the nature of the compact object remains uncertain. Recent polarization measurements suggest that it could be a concealed Galactic ultra-luminous X-ray source (ULX) surpassing the Eddington limit due to anisotropic emissions \cite[]{Veledina-etal2024a}. Recent measurements from {\it VLBA} observations have precisely constrained the distance of the source as ${9.67}_{-0.48}^{+0.53}$ kpc \cite[]{Reid-etal2023}.

\noindent $\bullet$ {\bf LMC X$-$1:} This is the first discovered extra-galactic persistent BH-XRB located in the Large Magellanic Cloud (LMC) \cite[]{Mark-etal1969} at a well-constrained distance of $\sim 48.1$ kpc \cite[]{Pietrzynski-etal2013}. Dynamical measurements have determined the black hole's mass to be $10.9\pm1.4$ $M_\odot$ with a moderate inclination of $36.4^{^\circ}\pm1.9^{^\circ}$ \cite[]{Orosz-etal2009}. The measurements using the continuum fitting method in the thermally dominated state constrain LMC X$-1$ spin as $\sim 0.92$ \cite[]{Gou-etal2009, Bhuvana-etal2021}, indicating the presence of a rapidly rotating black hole. LMC X$-$1 is the first source for which a null-detection of X-ray polarization has been reported from {\it IXPE} data \citep{Podgorny-etal2023}.

\noindent $\bullet$ {\bf 4U 1957$+$115:} This is a bright and persistent low-mass X-ray binary, first discovered by the {\it Uhuru} mission in $1973$ \cite[]{Giacconi-etal1974}. The source distance is found to be in the range of $5-40$ kpc, with inclination within $\sim 13^{\circ}-78^{\circ}$ and mass in $\sim 6-10 M_\odot$ \citep{Maitra-etal2014, Gomez-etal2015}. Nevertheless, several studies suggest the presence of a maximally spinning black hole in 4U 1957$+$115 \cite[]{Barillier-etal2023}. Polarimetric study of the source reveals several relativistic effects close to the black hole, leading to low PD of $\sim 1.9\%$ \citep{Marra-etal2024}.

\noindent $\bullet$ {\bf LMC X$-$3:} This is a persistent extra-galactic BH-XRB in Large Magellanic Cloud (LMC) at a distance of $48.1$ kpc \cite[]{Orosz-etal2009}, discovered by {\it UHURU} in $1971$. A slowly rotating black hole of spin $0.25_{-0.29}^{+0.20}$ \cite[]{Steiner-etal2010, Bhuvana-etal2021} and mass $6.98\pm0.56 ~ {\rm M}_\odot$ \cite[]{Orosz-etal2014, Bhuvana-etal2021} is found to be present at the central core of the system. LMC X$-3$ is a relatively high inclined system ($i\sim69.24^{\circ}\pm0.72^{\circ}$) \cite[]{Orosz-etal2014}. Being a persistent source, it mostly remains in the HSS \cite[]{Bhuvana-etal2021} except for a few occasions during which the LHS is also observed \cite[]{Smale-etal2012}. An exceptionally anomalous low state of peak luminosity $\sim 10^{35}$ erg $\rm s^{-1}$ is also observed for a few instances \cite[]{Torpin-etal2017}. Interestingly, LMC X$-$3 is another BH-XRB for which a relatively high polarization degree ($\sim 3\%$) has been reported from soft state observations \citep{Majumder-etal2024a, Svoboda-etal2024a}.

\noindent $\bullet$ {\bf Swift J1727.8$-$1613:} This is a recently discovered BH-XRB transient, first detected by {\it Swift/BAT} on August 24, 2023. Immediate monitoring of the source with {\it MAXI/GSC} in the $2-20$ keV energy range revealed an `unusual' peak in X-ray flux, increasing from $150$ mCrab to $7$ Crab within a few days of detection \cite[]{Nakajima-etal2023, Nandi-etal2024}. Recent measurements estimate the source distance as $\sim 2.7$ kpc, with a spin of $\sim 0.98$ and an inclination of $\sim 40^{\circ}$ \cite[]{Mata-etal2024, Peng-etal2024a}. X-ray polarimetric constraints with {\it IXPE} in various spectral states of Swift J$1727.8-1613$ also aim to unveil the disc-corona geometry of the source \citep{Veledina-etal2023, Ingram-etal2024, Svoboda-etal2024b, Podgorny-etal2024}.

\noindent $\bullet$ {\bf GX 339$-$4:} It was discovered by the OSO$-7$ mission in $1973$ \cite[]{Markert-etal1973} and typically undergoes outbursts every $2-3$ years. The black hole mass is constrained to be within $8.28-11.89$ $\rm M_\odot$ \cite[]{Sreehari-etal2019a}, while the source distance is estimated as $8.4\pm0.9$ kpc \cite[]{Parker-etal2016}. Recent studies suggest the source inclination angle between $30^{\circ}-34^{\circ}$, though spin measurements remain highly model-dependent, yielding values ranging from negative to moderate positive spins \cite[]{Zdziarski-etal2024}. Further, \cite{Mastroserio-etal2024} reported the alignment of X-ray polarization direction with that of radio jet orientation in GX $339-4$.

\noindent $\bullet$ {\bf Swift J151857.0$-$572147:} This is a newly discovered Galactic BH-XRB transient, first detected by {\it Swift/BAT} during its $2024$ outburst \cite[]{Kennea-etal2024}. Subsequent observations with {\it NICER} and {\it Insight-HXMT} constrained the source distance, inclination, and spin parameter as $\sim 5.8$ kpc, $\sim 2.1^{\circ}$, and $\sim 0.84$, respectively \cite[]{Peng-etal2024b}. Additionally, coherent type-C QPOs were detected with {\it Insight-HXMT} in the intermediate state of the outburst, which have been linked to the shock instability model of transonic accretion flow around black holes \cite[]{Chatterjee-etal2024}. Notably, \cite{Mondal-etal2024} reported low polarization for this source, whereas \cite{Ling-etal2024} found a null-detection.

\noindent $\bullet$ {\bf IGR J17091$-$3624:} This is a galactic BH-XRB transient system, discovered by {\it INTEGRAL/IBIS} in $2003$ \citep{Kuulkers-etal2003}. Interestingly, IGR J$17091-3624$ exhibits several exotic X-ray variability signatures, including the well known ``heartbeat'' class and therefore, often termed as the twin of GRS $1915+105$ \citep{Capitanio-etal2012, Radhika-etal2018, Katoch-etal2021, Wang-etal2024, Shui-etal2024}. Using reflection spectroscopy, the inclination of the source is estimated as $\sim 30^{\circ}-45^{\circ}$ \citep{Xu-etal2017, Wang-etal2018}. In addition, spectral modeling suggests a black hole of mass $10-12.3$ $M_\odot$ likely to be present at the center of IGR J$17091-3624$ \citep{Iyer-etal2015, Radhika-etal2018}. Based on the luminosity estimates, the distance of the source is predicted to lie within $11-17$ kpc \citep{Rodriguez-etal2011}. Further, exceptionally high polarization degree has been reported for this source in LHS \citep{Ewing-etal2025, Debnath-etal2025}.

\noindent $\bullet$ {\bf MAXI J1744$-$294:} The sources is located near the Galactic center, is the most recently discovered X-ray transient by {\it MAXI/GSC} on January 2, 2025 at a flux level of $\sim 133$ mCrab \citep{Kudo-etal2025}. The energy spectra of the source, obtained from follow-up {\it NuSTAR} observations, are found to be well described with the model comprising a thermal disc component of temperature $\sim 0.7$ keV, a power-law component of photon index $\sim 2.3$ and iron line profile at $\sim 6.6$ keV \citep{Mandel-etal2025}. Based on these initial findings, the source is inferred to be a BH-XRB transient observed in the HSS \citep{Mandel-etal2025}. Assuming a distance of $8$ kpc, the luminosity of the source is estimated to be $\sim 1.5 \times 10^{37}$ erg $\rm s^{-1}$ in $2-10$ keV energy range \citep{Mandel-etal2025}. 

\section{Data reduction}

{\it IXPE}, NASA's first space-based low-energy X-ray polarimetry mission, was launched on December 9, 2021 \citep{Weisskopf-etal2022}. It comprises three identical gas-pixel detector units (DUs) designed to measure the polarimetric properties of astrophysical sources using the principle of photo-electron tracking in the $2-8$ keV energy band. For each detected photon, the Stokes parameters (I, Q, U) are determined based on the azimuthal angle of the electric field vector, which is reconstructed by analyzing the photo-electron track within the detector plane. The additive nature of the Stokes parameters enables the summation of individual photon contributions to obtain the resultant I, Q, and U, which are used to calculate the polarization degree (PD) and polarization angle (PA).

For the polarimetric analysis, we utilize the cleaned and calibrated level-2 event files from the three DUs of {\it IXPE} in the $2-8$ keV energy range. Data reduction is performed using the latest \texttt{IXPEOBSSIMv31.0.3} software \citep{Baldini-etal2022}, following standard procedures outlined in \citealt[]{Kislat-etal2015, Strohmayer-etal2017, Kushwaha-etal2023a, Majumder-etal2024a, Majumder-etal2024b}. The source and background regions are defined as a $60$ arcsec circular region centered at the source coordinates and an annular region with radii between $180$ and $240$ arcsec, respectively \citep{Kushwaha-etal2023a, Majumder-etal2024a}. The \texttt{XPSELECT} task is used to extract source and background events from these regions. We use \texttt{XPBIN} task with various algorithms, such as \texttt{PCUBE}, \texttt{PHA1}, \texttt{PHA1Q} and \texttt{PHA1U}, to generate the necessary data products for model-dependent polarization as well as spectro-polarimetric studies, respectively. The latest response files provided by the software team are used in the fitting of Stokes spectra. The {\it IXPE} light curves of $2000$ s time bin are extracted from the event files using \texttt{XSELECT}.

{\it NICER} \citep{Gendreau-etal2016} is the X-ray mission onboard the International Space Station (ISS), sensitive in the soft X-ray energy band of $0.2-12$ keV and capable of excellent observations for in-depth spectro-temporal studies.
The {\it NICER/XTI} data is reduced using the standard data extraction software (\texttt{NICERDASv13}) integrated within \texttt{HEASOFT V6.34} with the appropriate calibration database. We use the tool \texttt{nicerl2} with standard filtering criteria to generate the clean event files. The spectral products are generated using the task \texttt{nicerl3-spect} with background model \texttt{3c50} \citep{Remillard-etal2022}. The {\it NICER} light curves of desired time resolutions are generated using \texttt{nicerl3-lc} task.

We analyze the {\it NuSTAR} \citep{Harrison-etal2013} data using the dedicated software \texttt{NUSTARDAS} available in \texttt{HEASOFT V6.34}. We use the task {\it nupipeline} to generate the cleaned event files from both {\it FPMA} and {\it FPMB} instrument onboard {\it NuSTAR}. A circular region of $60$ arcsec radii at the source position and away from it is considered to extract the source and background spectra, instrument response and ancillary files, respectively, using the task {\it nuproduct}. All the {\it NuSTAR} spectra are grouped with $25$ counts per bin to obtain better statistics during the spectral fitting.

{\it AstroSat} \citep{Agrawal-etal2006} data are extracted using the latest reduction software \texttt{LaxpcSoftv3.4.4}.The standard procedures of {\it AstroSat/LAXPC} data extraction are performed following \citealt[]{Antia-etal2017, Antia-etal2021, Antia-etal2022, Sreehari-etal2019, Sreehari-etal2020, Majumder-etal2022}. We use {\it LAXPC20} data for generating the light curves in different energy ranges for the respective sources except Swift J$1727.8-1613$, for which {\it LAXPC10} data in low gain mode was available for analysis \cite[see][for details]{Nandi-etal2024}. Note that, level-1 data corresponding to single events from all layers of the detector are considered in each observation.

We utilize on-demand\footnote{\url{http://maxi.riken.jp/mxondem/}} processed {\it MAXI/GSC} \citep{Matsuoka-etal2009} data and scaled map transient analysis data\footnote{\url{https://swift.gsfc.nasa.gov/results/transients/}} of {\it Swift/BAT} \citep{Krimm-etal2013} for the individual sources in different energy ranges. Note that, {\it MAXI/GSC} count rates in $2-20$ keV energy range are converted into the flux unit (Crab) following the relation $1$ Crab $=3.8$ photons $\rm cm^{-2}$ $\rm s^{-1}$ as suggested by the instrument team\footnote{\url{http://maxi.riken.jp/top/readme.html}}. Similarly, for {\it Swift/BAT} data in the $15-50$ keV energy range, we obtain the corresponding photon flux\footnote{\url{https://swift.gsfc.nasa.gov/results/transients/}} in units of mCrab, where $1$ mCrab $= 0.000220$ counts $\rm cm^{-2}$ $\rm s^{-1}$.

\section{Analysis and Results}

\subsection{Temporal Analysis}

\begin{figure*}
    \begin{center}
	\includegraphics[width=0.7\textwidth]{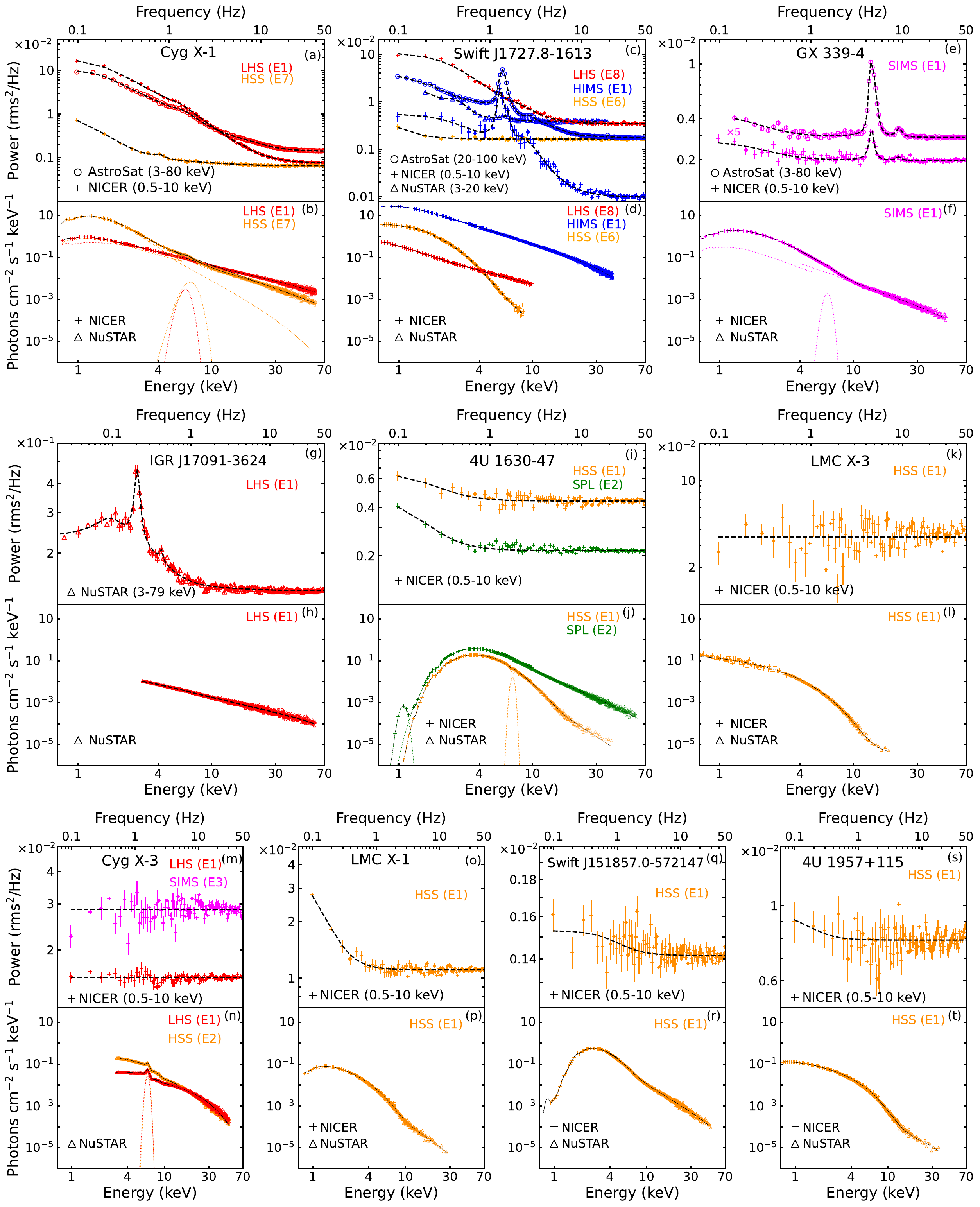}
    \end{center}
    \vskip -0.4cm
    \caption{Results of temporal (top panel) and spectral (bottom panel) analyses of ten BH-XRBs in different spectral states (color coded) obtained using quasi-simultaneous {\it IXPE}, {\it NICER}, {\it NuSTAR} and {\it AstroSat} observations. {\it Top panels:} PDS obtained from {\it NICER} ($0.5-10$ keV) and/or {\it NuSTAR}/{\it AstroSat} observations in different energy ranges are depicted for the respective sources in various epochs. The power spectrum of GX $339-4$ from {\it NICER} observation is rescaled by a factor of $5$. {\it Bottom panels:} Energy spectra of the sources from {\it NICER} and/or {\it NuSTAR} data are shown during the respective epochs. PDS and energy spectra in LHS, HIMS, SIMS, HSS and SPL states are presented using red, blue, magenta, orange and green colors, respectively. PDS from \textit{NuSTAR} and \textit{AstroSat} data are plotted using circle and triangle, irrespective of the spectral states. See the text for details.
    }
    \label{fig:spec_temp}
\end{figure*}

\subsubsection{Multi-mission Monitoring}

We study the variability properties of each source using multi-mission data during the epochs of {\it IXPE} observations. In Fig. \ref{fig:lcurve}, we present the count rate obtained with {\it MAXI/GSC} ($2-20$ keV) and {\it Swift/BAT} ($15-50$ keV) monitoring in panels (a) and (b) along with the variation of hardness ratio ($HR$) from {\it MAXI/GSC} in panel (c), respectively, for all sources under consideration. Here, $HR$ is defined as the flux ration in $6-20$ KeV and $2-6$ keV. Note that, flux values associated with the count rates in soft (hard) energy bands with {\it MAXI/GSC (Swift/BAT)} are also mentioned in units of mCrab in the $y$-axis (right side) of the respective panels. Further, radio flux densities obtained from different observational campaigns, such as {\it RATAN} \citep{Veledina-etal2024a, Ingram-etal2024}, {\it E-MERLIN} \citep{Williams-Baldwin-etal2023}, {\it VLITE} \citep{Peters-etal2023} and {\it AMI-LA} \citep{Steiner-etal2024} near the {\it IXPE} epochs, are shown in panel (d) for the respective sources based on their availability. We observe that all sources show marginal variation in both {\it MAXI/GSC} and {\it Swift/BAT} count rates over the entire {\it IXPE} exposure for a given epoch. Hence, in Fig. \ref{fig:lcurve}a-c, we present the average 
values of the respective quantities over the entire {\it IXPE} exposure of each epoch.

Interestingly, for Swift J$1727.8-1613$, we observe that both {\it MAXI/GSC} and {\it Swift/BAT} flux decrease as its outburst progresses. The $HR$ also sharply decreases and anti-correlates with the accompanied radio flux density (see Fig. \ref{fig:lcurve}c-d). The variation of both low and high energy fluxes along with $HR$ in presence of radio ejection suggest that the source evolves through spectral state transitions. In particular, we notice exceptionally bright ($\gtrsim 3$ Crab) intermediate states ($HR \sim 0.1-0.5$, E1-E5) followed by relatively faint ($\sim 30$ mCrab) softer states ($HR \sim 0.04-0.1$, E6-E7) before the appearance of dimmed ($\sim 100$ mCrab) hard state ($HR \sim 0.3$, E8) (see Table \ref{tab:log2}). Moreover, these variability features resemble the `canonical' state transition as HIMS $\rightarrow$ SIMS $\rightarrow$ HSS $\rightarrow$ LHS (decay), commonly exhibited by BH-XRB transients \cite[see][for details]{Nandi-etal2012}. Similarly, for Cyg X$-1$, we observe a marginal increase in the {\it MAXI/GSC} count rate and a corresponding decrease in the {\it Swift/BAT} count rate during epochs E3 to E7. In the subsequent observations (E8 to E11), the {\it MAXI/GSC} count rate decreases while the {\it Swift/BAT} count rate rises. In addition, $HR$ shows a significant drop from $0.4$ to $0.1$ followed by an increase in values exceeding $0.4$. This trend possibly indicates the spectral state transitions between LHS (E1-E2, E8-E11) and HSS (E3-E7, E12), respectively (see Table \ref{tab:log}).

However, only marginal variations are seen in both low- and high-energy count rates for several sources, such as  LMC X$-1$, LMC X$-3$, 4U $1957+115$, GX $339-4$, Swift J$151857.0-572147$ and MAXI J$1744-294$. For all of them, the hardness ratio remains low ($HR \lesssim 0.1$) and these sources are also detected at flux levels below approximately $100$ mCrab. These possibly suggest that the sources are likely in softer spectral states. In contrast, Cyg X$-3$ exhibits a clear sequence of state transitions, evolving from the LHS to the HSS, then to the SIMS, and back to HSS. This evolution is accompanied by moderate changes in count rates and hardness ratio, and is likely associated with strong radio flares ($\sim 100$ mJy) (see Table \ref{tab:log}, Fig. \ref{fig:lcurve}, and \citealt[]{Veledina-etal2024a}). For 4U $1630-47$, HSS \citep{Kushwaha-etal2023a} and SPL \citep{Rodriguez-etal2023} states are observed with reasonable variations in the count rates and $HR$. It is worth noting that IGR $17091-3624$ is observed in the LHS at $\sim 1$ Crab {\it MAXI/GSC} flux with $HR\sim 0.4$, very similar to the LHS observations of Cyg X$-1$. Furthermore, GRS $1915+105$ is observed with {\it IXPE} in its present faint and obscured state with a flux level of $\sim 6$ mCrab only (see Fig. \ref{fig:lcurve}). We find that observations in harder spectral states (LHS and HIMS) and relatively softer spectral states (HSS, SIMS and SPL) of the sources are separated in two distinct regions of $HR$ variations (see Fig. \ref{fig:lcurve}). In particular, $HR \sim 0.22$ separates two regions, with $HR > 0.22$ indicating harder and $HR < 0.22$ corresponds to softer spectral states with few exceptions.

\vspace{-0.6cm}
\subsubsection{Power Density Spectra}

We investigate the power density spectra (PDS) of all sources in different spectral states using {\it NICER} observations (see Table \ref{tab:log}) in $0.5-10$ keV energy range. In addition, PDS properties in hard X-rays ($>10$ keV) are also examined using available {\it NuSTAR} and {\it AstroSat} observations of the sources. We generate 0.01 s time binned light curves and compute the PDS up to Nyquist frequency of $50$ Hz with $1024$ newbins per interval using \texttt{powspec} inside \texttt{HEASOFT V6.34}. The individual power spectra are averaged out to get the resultant PDS in units of $\rm rms^2/Hz$ which is geometrically rebinned in the frequency space with a factor of $1.03$ \cite[see][]{Belloni-etal2002, Belloni-etal2005, Sreehari-etal2019, Majumder-etal2022}. In the upper panels of Fig. \ref{fig:spec_temp}, we present the PDS of the respective sources in different spectral states. We model each PDS with a combination of \texttt{constant} and multiple \texttt{Lorentzian} components. Further, we compute the total percentage rms amplitude ($rms_{\rm tot}\%$) as a measure of variability in $0.1-50$ Hz and tabulate it in Table \ref{table:spec_para}. 

Interestingly, we observe that the PDS in several spectral states exhibit distinct variable features. In particular, PDS of Cyg X$-1$ in the LHS show significant variability with $rms_{\rm tot}\% \sim 31.1-52.4$ ($0.5-10$ keV) and a clear power spectral break at $\sim 1.5$ Hz with {\it NICER} observations. Further, power spectral properties in hard X-rays ($3-80$ keV) with quasi-simultaneous {\it AstroSat} observations exhibit similar break frequencies ($\sim 1.5$ Hz) with $rms_{\rm tot}\% \sim 37$ in LHS (see Table \ref{tab:log} and Fig. \ref{fig:spec_temp}a). We also observe marginal (significant) PDS variability of $rms_{\rm tot}\% \sim 9.1$ ($33$) with low (high) energy {\it NICER} ({\it AstroSat}) observations of Cyg X$-1$ in the HSS. Note that similar variability properties including power spectral breaks were also reported by \cite{Kushwaha-etal2021} in LHS and HSS of Cyg X$-1$.

For Swift J$1727.8-1613$, the presence of type-C QPO feature at $\sim 1.4$ Hz is observed during HIMS (epoch 1, hereafter E1) in $0.5-100$ keV with {\it NICER}, {\it NuSTAR} and {\it AstroSat} quasi-simultaneous observations (see also \citealt[]{Nandi-etal2024, Liao-etal2024}). The evolution of the QPO frequency up to $\sim 6.7$ Hz is seen with {\it NICER} in the later epochs (E2-E5). However, a power distribution, similar to Cyg X$-1$ (LHS), is observed in the LHS of Swift J$1727.8-1613$ without the presence of power spectral break, whereas marginal variability with $rms_{\rm tot}\% \sim 1.4$ is noticed during HSS (E6-E7). 

The remaining sources (4U $1630-47$, LMC X$-1$, LMC X$-3$, 4U $1957+115$, Cyg X$-3$ and Swift J$151857.0-572147$) in SIMS and HSS display weak variability signatures mostly represented by constant noise distribution and occasionally marginal power variation at lower frequencies. Conversely, during the recurrent outburst of GX $339-4$, a sharp type-B QPO peak at $\sim 4.5$ Hz ($rms_{\rm QPO}\% \sim 10.2$) and harmonic feature at $\sim 9.5$ Hz along with weak broadband noise ($rms_{\rm tot}\% \sim 15.4$) are observed in SIMS (see Fig. \ref{fig:spec_temp} and \citealt[]{Aneesha-etal2024}). Note that, the QPO features with enhanced $rms_{\rm QPO}\%$ from $3.76$ ($0.5-10$ keV) to $12.6$ ($3-80$ keV) are observed with {\it NICER} and {\it AstroSat}, respectively. For IGR J$17091-3624$, a sharp type-C QPO at $\sim 0.2$ Hz ($rms_{\rm QPO}\% \sim 15$) along with a weak harmonic at $\sim 0.4$ Hz are observed during the recent outburst (see Fig. \ref{fig:spec_temp}).

Moreover, the overall power spectral properties of the sources, including break frequencies, noise distribution, and the detection of type-C and type-B QPO features confirm the presence of distinct spectral states during the respective {\it IXPE} campaigns. These findings further corroborate the results indicating distinct spectral states, obtained from the multi-mission monitoring of the sources (see \S 4.1.1).

\vspace{-0.6cm}
\subsection{Wide-band Spectral Analysis}

\subsubsection{Spectral Modeling}

\begin{figure}
    \begin{center}
        \includegraphics[width=0.9\columnwidth]{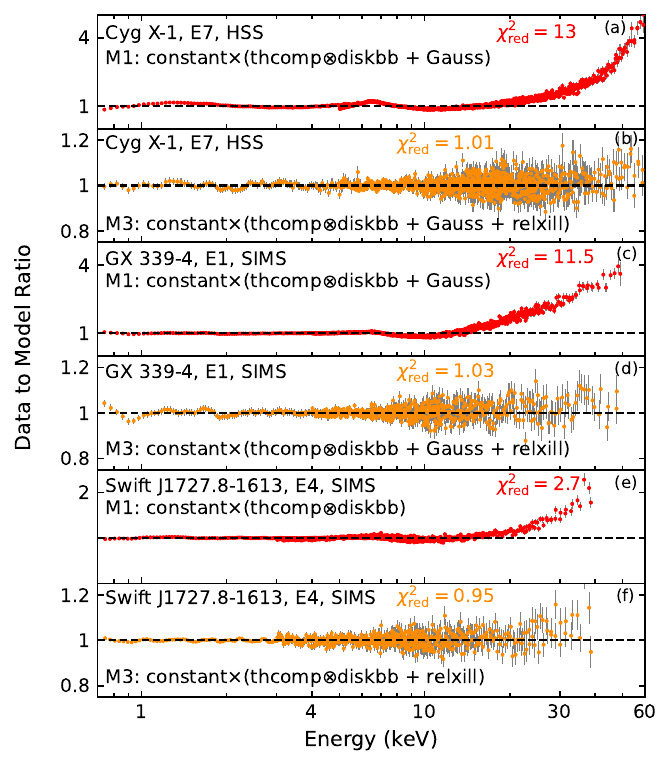}
    \end{center}
    \vskip -0.5cm
    \caption{Variation of the data to model ratio of the HSS/HIMS spectra of Cyg X$-1$, GX $339-4$ and Swift J$1727.8-1613$ fitted with model combination \texttt{M1} and \texttt{M3}. The improvement in the respective fits by incorporating the reflection model component over \texttt{M1} is depicted in panels (b), (d) and (f). See the text for details.}
    \label{fig:res}
\end{figure}

We investigate the wide-band ($0.7-60$ keV) spectral characteristics of each source using quasi-simultaneous {\it NICER} and {\it NuSTAR} observations. Depending on the availability of spectral coverage from different instruments, the spectra are modeled in different energy ranges (see Table \ref{table:spec_para}). For some observations, the spectra at higher energies are seen to be background dominated and/or of large uncertainties. Accordingly, we remove those parts of the spectra and proceed for modeling. To start with, we adopt the model combination \texttt{constant$\times$Tbabs$\times$(thcomp$\otimes$diskbb)} (hereafter \texttt{M1}) comprising of a convolution thermal Comptonization component \texttt{thcomp} \citep{Zdziarski-etal2020} along with the standard disc component \texttt{diskbb} \citep{Makishima-etal1986}. Here, \texttt{Tbabs} \citep{Wilms-etal2000} accounts for the inter-galactic absorption column density and the local absorption to the source. Note that, the component \texttt{thcomp} relies on the assumption of spherical geometry of the Comptonizing corona as the source of hot thermal electrons and up-scatters a fraction of the seed photon distribution of \texttt{diskbb} into higher energies. The model \texttt{thcomp} includes a parameter known as the covering fraction ($cov_{\rm frac}$), which represents the fraction of soft photons that are Comptonized in the corona. Consequently, $1 - cov_{\rm frac}$ represents the photon fraction that escape to the observer without undergoing Comptonization. The \texttt{constant} is used to adjust the cross-calibration between the spectra of different instruments wherever applicable. With this, the model \texttt{M1} is found to provide the best description of the spectra for all sources except a few observations of Cyg X$-1$ (E6 and E7), Swift J$1727.8-1613$ (E4 and E5) and GX $339-1$ (E1), which are discussed in the next section. Additionally, we note that a partial covering fraction absorption component \texttt{pcfabs} and a Gaussian absorption line \texttt{gabs} are required to model the low energy absorption features and the disc-wind regulated absorption line at $\sim 7$ keV, seen in the spectra of 4U 1630$-47$ during epoch-E1. Further, we use \texttt{Gaussian} components at $\sim 1.8$ keV and $\sim 2.2$ keV \citep{Kushwaha-etal2023a} to account for the instrumental features of {\it NICER} and at $\sim 1$ keV for the possible low energy emission line, depending on the strength of the lines in different observations. We also use \texttt{Gaussian} line at $\sim 6.4$ keV in several observations to account for the prominent iron K$\alpha$ line emission. It may be noted that the {\it NICER} spectra of Cyg X$-3$ exhibit multiple emission lines at various energies, which restricts us from constraining the parameters of model \texttt{M1}. Therefore, we use the {\it NuSTAR} data only in the spectral study of Cyg X$-3$. Further, the convolution model \texttt{cflux} is used without \texttt{Tbabs} component to estimate the bolometric flux associated with each model component and the entire spectral flux in $1-100$ keV energy range. Using the estimated distance of each source (see Table \ref{table:spec_para}), we compute the bolometric luminosity in units of Eddington luminosity. The best-fitted wide-band spectra (using \texttt{M1}) in different energy ranges and spectral states for the respective sources are shown in bottom panels of Fig. \ref{fig:spec_temp}.

\vspace{-0.7cm}
\subsubsection{Alternative Spectral Models}

We aim to explore the possible alternative disc-corona geometries in explaining the high-energy tail of the observed spectra. To achieve this, we use the \texttt{comptt} \citep{Titarchuk-etal1994} component as a proxy for \texttt{thcomp} in modeling thermal Comptonization. Note that, the `geometry switch' selected inside \texttt{comptt} model as `$\leq 1$' enables a disc-like geometry for the Comptonizing corona. Considering this, we proceed in the spectral fitting of all sources with the model combination \texttt{constant$\times$Tbabs$\times$(diskbb + comptt)} (hereafter \texttt{M2}). Interestingly, we find that the model \texttt{M2} provides a fit equally as good as \texttt{M1} for all observations under consideration except a few observations of Cyg X$-1$ (E6 and E7), Swift J$1727.8-1613$ (E4 and E5) and GX $339-4$ (E1). This leads to a possible degeneracy between the selected models in describing the observed spectra and the geometry of the Comptonizing region. 

It is worth noting that both \texttt{M1} and \texttt{M2} model combinations fail to provide a satisfactory fit to the observed spectra of Cyg X$-1$ (E6 and E7), Swift J$1727.8-1613$ (E4 and E5) and GX $339-4$ in the HSS/SIMS, resulting in a $\chi_{\rm red}^{2} \sim 2.7-13$ (for \texttt{M1}). This is mostly due to the presence of strong reflection features including iron K$\alpha$ line profile at $\sim 6.4$ keV and reflection hump at $\sim 30$ keV. To fit these reflection features, we include \texttt{relxill} \citep{Garcia-etal2014} component to model \texttt{M1}. Note that, \texttt{relxill} assumes an empirical power-law to model the emissivity profile of the corona of arbitrary geometry and calculates the primary spectrum considering a \texttt{cutoffpl} profile. With this, the model combination \texttt{constant$\times$Tbabs$\times$(thcomp$\otimes$diskbb + relxill)} (hereafter \texttt{M3}) delineates the spectral data in the HSS/SIMS of Cyg X$-1$, Swift J$1727.8-1613$ and GX $339-4$ with significant improvement in $\chi_{\rm red}^{2}$ in the range $\sim 0.95-1.11$. A \texttt{Gaussian} at $\sim 6.5$ keV is also required along with the components of \texttt{M3} to obtain the best fit for Cyg X$-1$ and GX $339-4$ observations. We fix the spin and inclination of the sources to the previously estimated values reported in \cite{Kushwaha-etal2021, Peng-etal2024a, Mastroserio-etal2024}, during the spectral fitting. In Fig. \ref{fig:res}, we depict the variation of the ratio of data to fitted model (\texttt{M1} and \texttt{M3}) for these sources, which indicates significant improvement of the respective fits by incorporating reflection component. We state that, \texttt{M3} fails to assert its relevance to the spectra of other sources in different states due to the absence of prominent reflection features.

\subsubsection{Spectral Properties}

The spectral energy distribution of the individual sources are generally described by the model combination comprising thermal Comptonization, standard disc and occasionally reflection components. Although several models developed based on different coronal geometries are found to describe the observed spectra satisfactorily, we examine the spectral states of the respective sources using model \texttt{M1} and \texttt{M3} from the extracted spectral parameters. We present the best-fitted and estimated model parameters for each source in Table \ref{table:spec_para}. The variation of the spectral parameters over different observational epochs indicates the presence of distinct spectral states of the sources under consideration.

\begin{figure*}
    \begin{center}
         \includegraphics[trim={0 4cm 0 2cm},clip, width=1\textwidth]{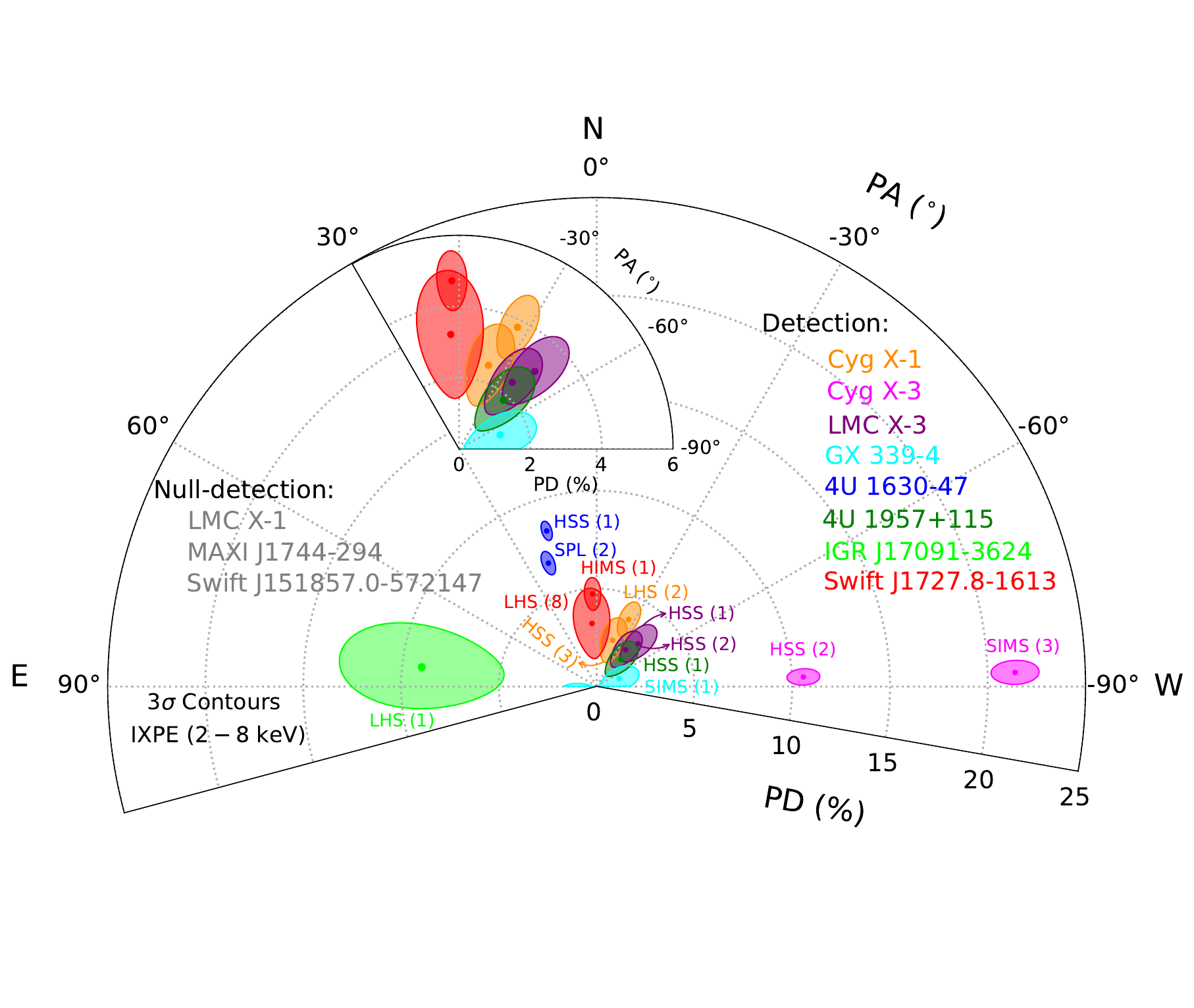}
    \end{center}
    \vskip -0.5cm
    \caption{Confidence contours ($3\sigma$) in the PD–PA plane obtained from the model-independent polarization measurements using $2-8$ keV {\it IXPE} data, combining all DUs across various spectral states of the sources. The epoch of each observation is indicated in parentheses next to the corresponding spectral state. For Swift J$151857.0-572147$, LMC X$-1$ and MAXI J$1744-294$, null-detection of polarization is observed, whereas all other sources exhibit significant polarization. Inset highlights a zoomed-in region of PD–PA space for better clarity. 
    }
    \label{fig:contour}
\end{figure*}

We find that the effects of Comptonization are noticeably high during the LHS of Cyg X$-$1 with covering fraction $cov_{\rm frac} \sim 0.61_{-0.04}^{+0.04}-0.77_{-0.02}^{+0.02}$, indicating Comptonization of up to $\sim 77\%$ of the soft photons in the LHS spectra. In addition, a high Comptonized flux contribution of $73-80\%$ is also observed in the spectra (see Table \ref{table:spec_para}). Indeed, such a high $cov_{\rm frac}$ results in low photon index ($\Gamma_{\rm th} \lesssim 1.64_{-0.02}^{+0.02}$) as observed during the LHS observations of Cyg X$-1$. In the HIMS and LHS of Swift J1727.8$-$1613, we find $cov_{\rm frac} \sim 0.35_{-0.01}^{+0.01}-0.82_{-0.02}^{+0.02}$ and $\Gamma_{\rm th}\sim 1.71_{-0.02}^{+0.03}-2.11_{-0.04}^{+0.05}$ with up to $\sim 64\%$ Comptonized emission. Similarly, LHS of IGR J17091$-$3624 which is dominated by the Comptonized emission ($\sim 87\%$ flux contribution) yields harder ($\Gamma_{\rm th}\sim 1.62_{-0.01}^{+0.02}$) spectral characteristics. On the other hand, the HSS/SIMS spectra of Cyg X$-$1 and Swift J1727.8$-$1613 results in low Comptonization with $cov_{\rm frac}$ in the range $0.005_{-0.002}^{+0.002}-0.38_{-0.03}^{+0.04}$ and thermal disc emission with $39-98\%$ flux contribution along with a steeper spectral index ($\Gamma_{\rm th}\sim 2.32_{-0.04}^{+0.06}-5.14_{-0.39}^{+0.36}$). Moreover, HSS with $76-98\%$ disc contribution are observed for LMC X$-$1, 4U 1630$-$47, LMC X$-$3 and 4U 1957$+$115 with negligible Comptonized emission ($cov_{\rm frac} < 0.1$) as shown in Table \ref{table:spec_para}. We find that GX 339$-$4 (Swift J151857.0$-$572147) exhibits intermediate spectral characteristics with $cov_{\rm frac} \sim 0.34$ ($0.21$) and  $\Gamma_{\rm th}\sim 2$ ($2.5$) with $\sim 50\%$ ($78\%$) disc emission during SIMS (HSS). 

The reflection spectroscopy of Cyg X$-1$, Swift J$1727.8-1613$ and GX $339-4$ with model \texttt{M3} during the HSS/SIMS results in disc ionization parameter ($\log \xi$) in the range of  $3.31_{-0.05}^{+0.05}-4.29_{-0.21}^{+0.14}$ erg cm $\rm s^{-1}$, iron abundance ($A_{\rm fe}$) between $\sim 4.61_{-0.31}^{+0.36}$ and $6.35_{-1.46}^{+2.08}$ $A_\odot$, and reflection fraction ($R_{\rm f}$) ranging $\sim 0.59_{-0.11}^{+0.05}$ to $3.16_{-0.33}^{+0.35}$. Here, $A_\odot$ represents the solar iron abundance. The best-fitted parameters of the \texttt{relxill} component obtained from model \texttt{M3} are tabulated in Table \ref{tab:reflect}, whereas the parameters associated with other spectral components (\texttt{diskbb} and \texttt{thcomp}) of \texttt{M3} are presented in Table \ref{table:spec_para} for the respective sources. Notably, a high ionization ($\log \xi \sim 4$) observed in HSS/SIMS generally suggests the reflection of significant coronal emission producing strong reflection signatures.

In addition, we find that the model component \texttt{relxilllp} \citep{Dauser-etal2013} also delineates the reflection features assuming the primary source emissivity profile from a lamppost corona. Moreover, another reflection flavor \texttt{relxillCp}, in which the spectrum of the coronal emission is computed from the \texttt{nthComp} \citep{Zdziarski-etal1996} model of spherical geometry, also fits the data for Cyg X$-1$ (E6 and E7) and Swift J$1727.8-1613$ (E4). Therefore, it is evident that the reflection spectroscopy also results in degeneracy among several models relying on contemporary coronal geometries.

Overall, the wide-band spectro-temporal analysis (see \S4.1 and \S4.2) confirms the presence of distinct spectral states in BH-XRBs, but falls short of resolving the inherent degeneracy among the possible coronal geometries. This highlights the importance of incorporating X-ray polarimetric observations into the study of BH-XRBs. Subsequently, we explore how polarization measurements offer deeper insights to this issue.

\subsection{Polarization Measurements}

We study the in-depth polarimetric properties of the eleven BH-XRBs observed with {\it IXPE} during multiple campaigns. In general, the polarization measurements are performed using two distinct methods. The first one relies on the model-independent approach based on the analysis of the polarization cube \citep{Kislat-etal2015} using the dedicated \texttt{python} package \texttt{IXPEOBSSIMv31.0.3}\footnote{\url{https://ixpeobssim.readthedocs.io/en/latest/}} \citep{Baldini-etal2022}. This provides the polarimetric properties by computing the polarization degree (PD) and polarization angle (PA) from the normalized Stokes parameters without any prior assumption on the emission mechanism. The alternative method depends on the simultaneous modeling of the I, Q, and U Stokes spectra in \texttt{XPSEC}. In this approach, the source intensity spectrum (I) is modeled using the appropriate physical components responsible for the emission mechanism and a polarization model accounting for the Q and U spectra \citep{Strohmayer-etal2017}. Accordingly, we carry out both model-independent and model-dependent analyses to ensure the consistency of the polarimetric measurements. 

We perform model-independent polarization cube analysis using \texttt{XPBIN} tool inside the software routine with \texttt{PCUBE} algorithm. Following \citealt[]{Kushwaha-etal2023a, Majumder-etal2024a, Majumder-etal2024b}, we consider all events from the three DUs of {\it IXPE} in the analysis. Subsequently, the normalized Q and U Stokes parameters, the polarization degree $({\rm PD} =\sqrt{(Q/I)^2+(U/I)^2})$ and polarization angle $({\rm PA} =\frac{1}{2} \tan^{-1}(U/Q))$ are computed in $2-8$ keV energy band for all observations of the respective sources. Following the guidelines of IAU\footnote{\url{https://aas.org/posts/news/2015/12/iau-calls-consistency-use-polarization-angle}} \citep{Contopoulos-etal1974}, we adopt the convention that PA increases counter-clockwise from North to East direction in the sky. We also calculate the minimum detectable polarization at $99\%$ confidence level ($\rm MDP_{99}\%$) and the significance (SIGNIF in units of $\sigma$) for each measurement. It may be noted that an observed $\rm PD > MDP_{99}\%$ indicates a polarization level that would arise from random statistical fluctuations with only $1\%$ chance probability \citep{Kislat-etal2015}. All the calculated polarimetric parameters and the measurement statistics are tabulated in Table \ref{tab:en-pol}. In Fig. \ref{fig:contour}, we present $3\sigma$ confidence contours associated with the measurements of PD and PA for the respective sources in different spectral states. 

We find significant degree of polarized emission ($\rm PD >\rm MDP_{\rm 99}\%$) in all sources except LMC X$-1$, Swift J$151857.0-572147$ and MAXI J1744$-$294. More precisely, we find that PD varies within $1.22\pm0.35\%-21.41\pm0.41\%$ including all sources in different spectral states within $2-8$ keV energy band. We note that Cyg X$-3$ shows the maximum degree of polarization, whereas GX 339$-$4 exhibits the lowest PD ($> \rm MDP_{99}\%$) among all sources under consideration. Further, a significant variation of PD is observed over different spectral states of the sources. For example, Cyg X$-1$ manifests a maximum PD of $\sim 4.8\%$ in the LHS, which decreases to $\sim 1.4\%$ in the HSS. Interestingly, a similar variation of PD is observed over several spectral states during the outburst phase of Swift J$1727.8-1613$ in which the polarization degree drops down to $\sim 0.4\%$ (HSS) from $\sim 4.7\%$ (HIMS) and again increases to $\sim 3.2\%$ in the LHS. Most astonishingly, 4U $1630-47$ exhibits exceedingly high PD in the HSS ($\sim 8.3\%$) and SPL ($\sim 6.8\%$), respectively. Moreover, we find an unprecedentedly high PD of $\sim 9\%$ in the LHS of the recent outburst of IGR J$17091-3624$, which is the maximum reported for any confirmed BH-XRB observed with {\it IXPE} to date (see also \citealt[]{Ewing-etal2025}). In contrast, 4U 1957$+$115 and LMC X$-$3 are found to show $\sim 1.9\%$ and $\sim 2.4-3\%$ polarization in the HSS, respectively. It may be noted that, for GX 339$-$4, significant PD ($\sim 1.2\%$, E1) is obtained in 3$-$8 keV energy range \cite[see also][]{Mastroserio-etal2024}, whereas null-detection is seen considering the entire {\it IXPE} energy band of 2$-$8 keV. Moreover, we find that PA of the sources in different spectral states shows marginal variation (see Table \ref{tab:en-pol}). 

\noindent {\it \textbf{Null-detection of Polarization}}: It is worth mentioning that we re-confirm the null-detection of polarized emission ($\rm PD <\rm MDP_{\rm 99}\%$) for a few sources under consideration. In particular, LMC X$-1$ remains the first source that fails to manifest significant polarization with {\it IXPE} (see Table \ref{tab:en-pol} and \citealt[]{Podgorny-etal2023}). Further, Swift J$151857.0-572147$ exhibits a PD of $\sim 0.25\%$ in $2-8$ keV which remains significantly low from the corresponding $\rm MDP_{\rm 99} \sim 0.78\%$, confirming a null-detection. This finding contradicts the polarization detection with PD $\sim 1.3\%$ as reported by \cite{Mondal-etal2024}. This discrepancy possibly arises as the present analysis is carried out using the latest version of the software \texttt{IXPEOBSSIMv31.0.3}, fixing the bugs related to the visualization of polarization cubes in the previous releases\footnote{\url{https://ixpeobssim.readthedocs.io/en/latest/release_notes.html}}. We emphasize that our results are consistent with the recent polarization measurements of Swift J$151857.0-572147$ \cite[]{Ling-etal2024}. Further, the null-detection of polarization is also observed in a few observations of Swift J$1727.8-1613$ (E6, E7), GX $339-4$ (E2) and LMC X$-3$ (E3). In addition, for the first time, we report the null-detection of X-ray polarization with PD ($\sim 0.7\%$) $< \rm MDP_{\rm 99} (1.3\%)$ in the recently discovered BH-XRB candidate MAXI J$1744-294$ during HSS.

\vspace{-0.5cm}
\subsection{Energy-dependent Polarization}

\begin{figure*}
	\begin{center}
		\includegraphics[width=0.35\textwidth]{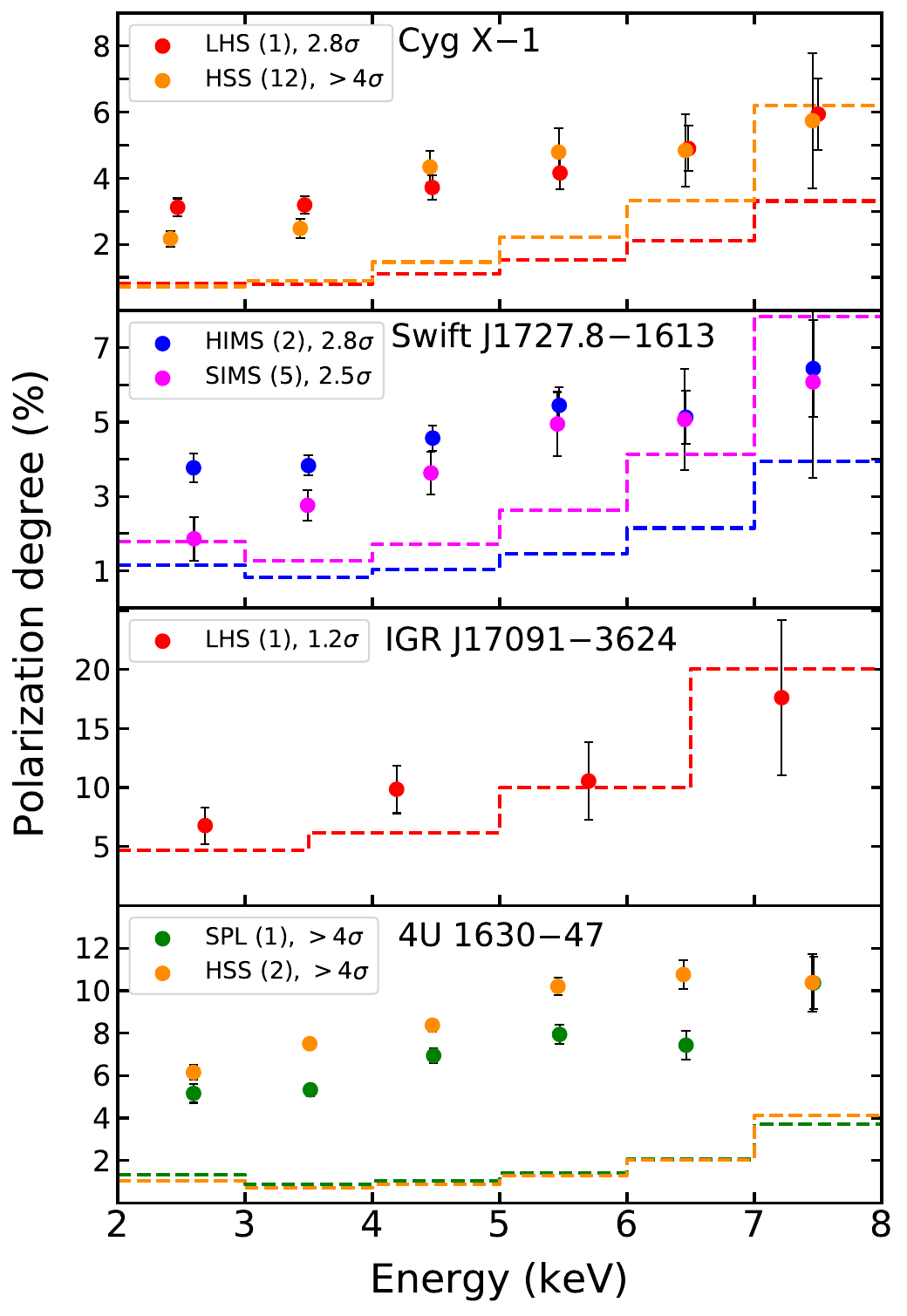}
		\includegraphics[width=0.35\textwidth]{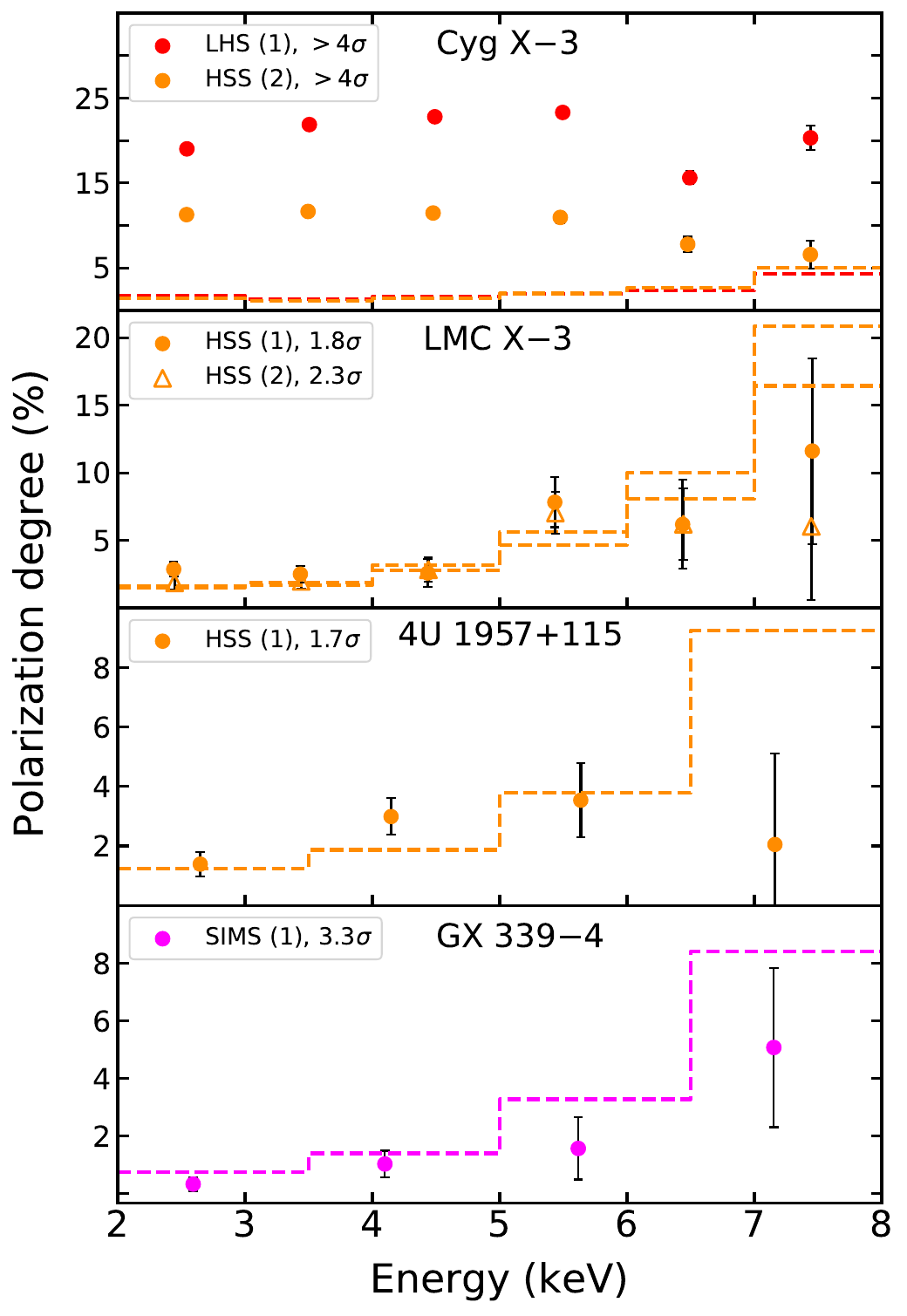} \\
	\end{center}
    \vskip -0.5cm
	\caption{Variation of polarization degree (PD) with energy. In each panel, source name and spectral state (observed during different epoch) are marked. The significance of energy variation of PD for the respective cases are mentioned in units of $\sigma$ in the plot legends. The histograms represent the minimum detectable polarization at $99\%$ confidence ($\rm MDP_{99}\%$). See the text for details.}
    \label{fig:pol1}
\end{figure*}

We further explore energy-dependent polarization properties of the selected sources that confirm the presence of significant polarized emission in $2-8$ keV energy range of {\it IXPE}. For each source except 4U $1957+115$, IGR J$17091-3624$ and GX $339-4$, we divide $2-8$ keV energy range into six linearly spaced energy bins and estimate polarization using the model-independent approach (see also \citealt[]{Svoboda-etal2024a, Veledina-etal2023}). For 4U $1957+115$, IGR J$17091-3624$ and GX $339-4$, only four energy bins are considered for the energy-dependent study to ensure relatively improved photon statistics per energy bin (see also \citealt[]{Marra-etal2024}). Further, we estimate the significance of the energy variation of PD over these bins following \cite{Krawczynski-etal2022b, Majumder-etal2024b}. In doing so, we consider PD in $2-8$ keV energy range (see Table \ref{tab:en-pol}) as the null hypothesis against which the significance of energy variations in PD is computed for each epoch of a given source. With this, following \cite{Majumder-etal2024b}, we calculate the probability of rejecting the null hypothesis using $(N-1)$ degrees of freedom, taking into account the variation of PD over $N$ energy bins for the respective cases. This essentially indicates the confidence level at which the PD variation is significant over different energy bins. The obtained significance levels of the energy variation of PD are mentioned in Table \ref{tab:en-pol} for the respective cases. We observe that this significance varies from $\lesssim 1\sigma$ to $> 4\sigma$ across different epochs of the sources in various spectral states. In Fig. \ref{fig:pol1}, we present the variation of PD with energy over different observational epochs using filled circles in the respective panels. The histograms in colors represent the $\rm MDP_{99}\%$ associated with the measurements in the selected energy bins for the corresponding epochs. Note that, for a given source, when multi-epoch observations are available, we present the results for two epochs of different spectral states only, except LMC X$-$3.

Interestingly, we observe that Cyg X$-1$ and Swift J1727.8$-1613$ show similar energy variation of PD within $\sim 2-6\%$ in different spectral states (LHS, HSS, HIMS and SIMS). Moreover, for Cyg X$-1$, the variation in PD with energy remains most significant at a confidence level exceeding $4\sigma$ in HSS, whereas it remains $2.8\sigma$ for Swift J$1727.8-1613$ in HIMS, when all observational epochs are considered (see Fig. \ref{fig:pol1}). Note that, PD $< \rm MDP_{99}\%$ at higher energy bins in a few epochs of several sources, indicates a null-detection. Interestingly, unlike Cyg X$-1$ and Swift J$1727.8-1613$, a weak energy dependency of PD is noticed for IGR $17091-3624$ in the LHS, for which PD corresponding to two out of four energy bins remain below the $\rm MDP_{99}\%$ (see Fig. \ref{fig:pol1}). We find a significant increase in PD with energy up to $\sim 11\%$ in HSS ($>4\sigma$) and SPL ($>4\sigma$) state of 4U 1630$-47$ (see Fig. \ref{fig:pol1}). Similarly, distinct variations of PD over several energy bins are observed for Cyg X$-3$ during LHS ($>4\sigma$) and SIMS ($>4\sigma$). LMC X$-3$ shows constant polarization over the initial three energy bins with a sudden jump in PD up to $\sim 7.8\%$ at $\sim 5.4$ keV in both epochs (see Fig. \ref{fig:pol1}). Further, we only observe a marginal variation ($1.7\sigma$) of PD in 4U 1957$+115$ up to $\sim 3\%$. Interestingly, significant ($3.3\sigma$) energy variation of polarization degree is observed for GX $339-4$. Although, PD $< \rm MDP_{99}\%$ is obtained in all four energy bins (see Fig. \ref{fig:pol1}). Overall, we state that a significant ($>3\sigma$) variation of polarization degree with energy is observed only for Cyg X$-1$, 4U 1630$-47$ and Cyg X$-3$, whereas a moderate ($2\sigma$ to $3\sigma$) variation is seen for Swift J1727.8$-1613$ and LMC X$-3$. It is worth mentioning that the energy variation of PA for all sources under consideration remains statistically insignificant ($< 2\sigma$). However, we note that PA drops from $-30^{\circ}$ to $-60^{\circ}$ at $\sim 6$ keV in 4U $1957+115$, which is marginally significant at $1.6\sigma$ confidence level. 

\vspace{-0.5cm}
\subsection{Spectro-polarimetric Modeling}

We explore the model-dependent spectro-polarimetric properties of the sources from the simultaneous fitting of the I, Q and U Stokes spectra of {\it IXPE} in $2-8$ keV energy range. Due to the calibration issues \cite[]{Krawczynski-etal2022b, Veledina-etal2024a, Steiner-etal2024}, only DU1 data is considered for the spectro-polarimetric study of Cyg X$-1$, Cyg X$-3$ and Swift J1727.8$-1613$, whereas data combining all the DUs are used for the remaining sources. For simplicity, we first adopt the model combination \texttt{M4: Tbabs$\times$polconst$\times$(diskbb + powerlaw)} for the spectro-polarimetric fitting. Here, \texttt{polconst} represents a constant polarization model, where the degree and angle of polarization are treated as the model parameters. The best-fit results yield $\chi_{\rm red}^2$ ranging from $0.91-1.79$ across all selected sources, with PD varying as $\sim 1.16-13.06\%$ and PA spanning from  $-89.13^{\circ}$ to $85.15^{\circ}$. Note that the spectro-polarimetric modeling of Swift J1727.8$-$1613 does not constrain hydrogen column density, likely due to low-energy threshold of {\it IXPE} \cite[$\sim 2$ keV; see][]{Veledina-etal2023, Ingram-etal2024}. 

Next, we investigate the energy-dependent polarization properties using model-dependent approach for all the sources that demonstrate the signature of X-ray polarization. In doing so, we replace the \texttt{polconst} component with the energy-dependent polarization model \texttt{polpow}. Empirically, \texttt{polpow} assumes a power-law variation in the polarization parameters of the form PD($E$) $= A_{\rm norm}\times E^{-A_{\rm index}}$ (in fraction) and PA($E$) $= \psi_{\rm norm}\times E^{-\psi_{\rm index}}$ (in $^\circ$) while modeling the Q and U Stokes spectra. We find that the best-fitted $\psi_{\rm index}$ remains consistent with zero within its $1\sigma$ errors and hence, we freeze it to zero during the spectral fitting of all observations. Accordingly, we obtain the best spectral fit for all sources with $\chi_{\rm red}^2 \sim 0.97-1.46$. Furthermore, integrating PD($E$) in $2-8$ keV energy range using the best-fitted model parameters ($A_{\rm index}$ and $A_{\rm norm}$), we obtain PD as $3.23-14.33\%$. Moreover, with $\psi_{\rm index}$ frozen to zero, the polarization angles are obtained as the best-fitted $\psi_{\rm norm}$ values that vary from $-88.92^{\circ}$ to $85.11^{\circ}$. Note that, in a few observations of Cyg X$-1$, 4U 1957$+115$ and GX $339-4$ (see Table \ref{tab:pol_para}), the data is unable to constrain the parameters of \texttt{polpow} component even at $1 \sigma$ level, and hence, we refrain from the modeling with \texttt{polpow} for these observations. In Table \ref{tab:pol_para}, we present the best-fitted model parameters obtained from the spectro-polarimetric study of all sources. We find that the model-dependent polarimetric results are in agreement with the findings of the model-independent study except Cyg X$-3$, confirming the robustness of the polarization measurements for the individual sources. However, for Cyg X$-$3, we obtain significantly low PD compared to the results of the model-independent study, likely due to multiple low-energy unpolarized emission lines including strong iron complex around $\sim 6.4$ keV \cite[]{Veledina-etal2024a}, dominating Stokes spectra.

\vspace{-0.7cm}
\subsection{Spectro-polarimetric Correlation}

\begin{figure}
    \begin{center}
	\includegraphics[width=\columnwidth]{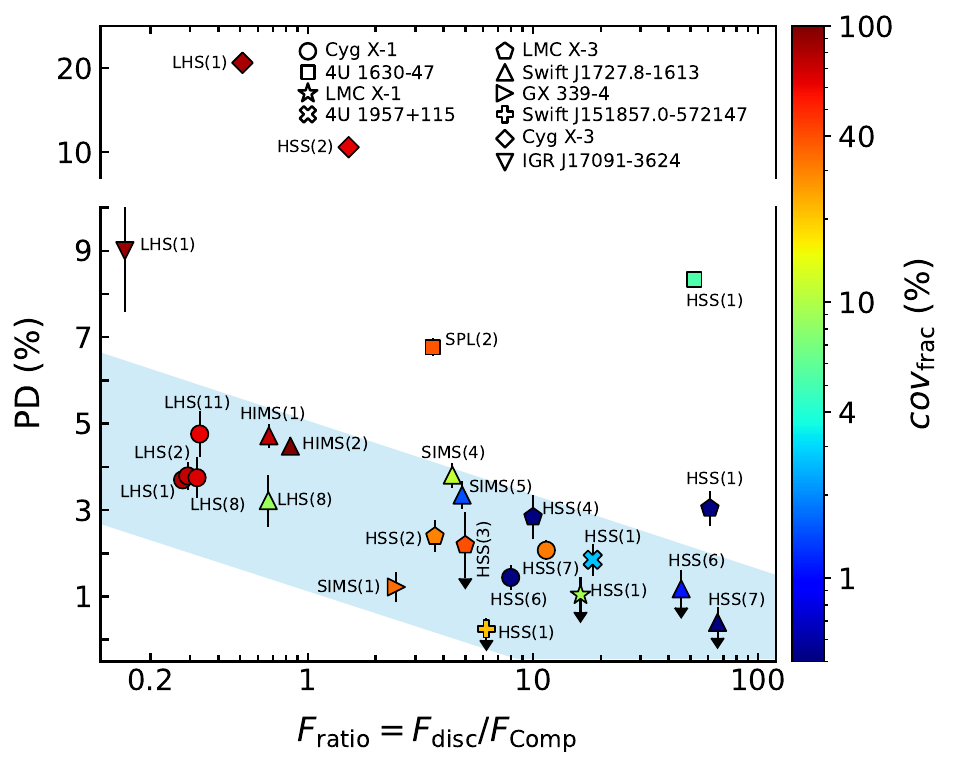}
    \end{center}
    \vskip -0.5cm
    \caption{Variation of PD ($2-8$ keV), obtained from model-independent analysis, as a function of the disc-to-Comptonized flux ratio ($F_{\rm ratio}$) derived in $1-100$ keV energy range from the wide-band spectral modeling. The color bar represents the covering fraction ($cov_{\rm frac}$) obtained from \texttt{thcomp} model, while each data point is labeled with the corresponding spectral state and observational epoch. Data points within the shaded region (light blue) are used to compute the Pearson correlation coefficient. Downward arrows indicate null-detections of polarization for the respective sources.
    }
    \label{fig:corr_plt}
\end{figure}

We examine the spectro-polarimetric correlation properties for each source based on the results obtained from wide-band spectral modeling using model \texttt{M1} (see \S4.2) and the polarimetric measurements (see \S4.3). Specifically, we explore the correlation between the observed PD obtained from model-independent study, and best-fitted spectral parameters, namely the ratio of disc to Comptonized spectral flux ($F_{\rm ratio}$) in $1-100$ keV energy range, and the covering fraction ($cov_{\rm frac}$) obtained from the \texttt{thcomp} model. In Fig. \ref{fig:corr_plt}, we present the variation of PD with $F_{\rm ratio}$ (see Table \ref{table:spec_para}) using different markers, where the covering fraction ($cov_{\rm frac}$, in percent) is represented by the colors of respective markers. The color bar on the right indicates the range of $cov_{\rm frac}$. The spectral state and the corresponding epoch (in parentheses) for each observation are marked in the figure. 

We observe an apparent anti-correlation between PD and $F_{\rm ratio}$, and a correlation between PD and $cov_{\rm frac}$ for the sources under consideration. However, 4U 1630$-47$, Cyg X$-3$, and IGR J$17091-3624$ deviate significantly from this trend in the PD$-F_{\rm ratio}$ plane (see Fig. \ref{fig:corr_plt}) and therefore, these sources are excluded from the present correlation study. To ascertain the firmness of the apparent correlations, we compute the Pearson correlation coefficient ($\rho$) based on the variations in spectro-polarimetric parameters. Excluding the aforementioned outliers, we find $\rho = 0.7$ for the PD$-cov_{\rm frac}$ correlation and $\rho = -0.5$ for the PD$-F_{\rm ratio}$ anti-correlation. Furthermore, excluding epoch E1 of LMC X$-3$, which appears as a marginal outlier in the PD$-F_{\rm ratio}$ plane, along with 4U 1630$-47$, Cyg X$-3$ and IGR J$17091-3624$ as before, the correlations become relatively stronger yielding $\rho = 0.8$ and $\rho = -0.6$ for PD$-cov_{\rm frac}$ and PD$-F_{\rm ratio}$, respectively.

Furthermore, we find a strong correlation (anti-correlation) between PD and $cov_{\rm frac}$ ($F_{\rm ratio}$) with $\rho \sim 0.9$ ($-0.9$) for Swift J$1727.8-1613$. In particular, a sharp drop in PD is noticed from $\sim 4.7\%$ (HIMS) to $\sim 0.4\%$ (HSS) (see also \citealt[]{Svoboda-etal2024b}) during which $F_{\rm ratio}$ increases from $\sim 0.7$ to $\sim 66$ and $cov_{\rm frac}$ noticeably reduces down to $13\%$ from a maximum value of $82\%$. A similar trend is observed for Cyg X$-1$ with $\rho = 0.9$ ($-0.9$) for the respective positive (negative) correlations between PD and $cov_{\rm frac}$ ($F_{\rm ratio}$). As before, a low (high) $F_{\rm ratio} \sim 0.3$ ($11.5$) and high (low) $cov_{\rm frac}\sim 61-77\%$ ($0.5-31\%$) are observed in the LHS (HSS) of Cyg X$-1$ with PD $\sim 3.7-4.8\%$ ($1.4-2.1\%$) (see Fig. \ref{fig:corr_plt}). Moreover, we also find a similar behavior in the LHS of IGR J$17091-3624$, for which a very high PD ($\sim 9\%$) is seen with $cov_{\rm frac}\sim 87\%$ and $F_{\rm ratio} \sim 0.2$. These findings suggest that a significant contribution in the polarization degree comes from the Comptonized emission for Swift J$1727.8-1613$, Cyg X$-1$ and IGR J$17091-3624$ in the HIMS and/or LHS. Furthermore, we notice a low polarization degree (PD $\lesssim 2\%$) for all the sources except 4U 1630$-47$ and LMC X$-3$ in the HSS/SIMS with $F_{\rm ratio} \gtrsim 2$ and $cov_{\rm frac} \lesssim 30\%$. Interestingly, 4U 1630$-47$ being the source of extreme soft nature ($F_{\rm ratio} \sim 52$, $cov_{\rm frac} \sim 5\%$) manifests a high polarization degree of $\sim 8.3\%$ in the HSS, whereas it reduces to $6.8\%$ in the SPL state with $F_{\rm ratio} \sim 3.6$ and $cov_{\rm frac} \sim 38\%$. We observe that 4U $1957+115$ (GX $339-4$) shows PD $\sim 1.2\%$ ($1.9\%$) with $cov_{\rm frac}\sim 2.7\%$ ($34\%$) and $F_{\rm ratio}\sim 18$ ($2.4$) in the HSS (SIMS). 
Moreover, LMC X$-3$ exhibits moderate polarization with PD in the range of $2.4-3\%$ for $F_{\rm ratio} \gtrsim 4$ and $cov_{\rm frac} \sim 0.5\%$ during the HSS.

\vspace{-0.7cm}
\section{Discussion}

In this study, we present a comprehensive spectro-polarimetric investigation of eleven BH-XRBs using quasi-simultaneous observations from \textit{IXPE}, \textit{NICER}, \textit{AstroSat} and \textit{NuSTAR}. \textit{IXPE} data are analyzed to examine the polarization properties in $2-8$ keV energy range, while combined \textit{NICER}, \textit{AstroSat} and \textit{NuSTAR} observations provide insights into the spectro-temporal behavior across the wide energy range ($0.5-100$ keV). Subsequently, the spectro-polarimetric findings are used to investigate the accretion-ejection dynamics and geometry surrounding the sources.

\vspace{-0.6cm}
\subsection{Spectro-temporal Properties and Spectral States}

Spectro-temporal studies using {\it NICER}, {\it NuSTAR} and {\it AstroSat} observations in $0.5-100$ keV energy range impart distinct variability properties in various spectral states of the sources (see \S4.1 and \S4.2). The obtained results are in line with the spectro-temporal characteristics generally seen in the respective spectral states of BH-XRBs \citep{Belloni-etal2005, McClintock-etal2009, Nandi-etal2012, Sreehari-etal2019, Mendez-etal2022, Aneesha-etal2024}. These findings reinforce the presence of distinct spectral states as observed during the multi-mission campaigns of the sources.

In particular, we observe a high fractional variability amplitude $rms_{\rm tot}\% \sim 30-52$ in the PDS of LHS, whereas significantly low variability with $rms_{\rm tot}\% \sim 6-14$ are seen in the intermediate states (HIMS and SIMS) of the sources (see Table \ref{table:spec_para}). Negligible power distribution at lower frequencies results in marginal variability ($rms_{\rm tot}\% \sim 1-10$) for the sources in the HSS (see Table \ref{table:spec_para} and Fig. \ref{fig:spec_temp}). In addition, we find the presence of a strong type-C QPO feature at $\sim 1.4$ Hz (E1), which evolves to $\sim 6.7$ Hz (E2-E5), during the HIMS and SIMS of Swift J1727.8$-1613$ over a wide energy range of $0.5-100$ keV with {\it NICER}, {\it NuSTAR} and {\it AstroSat} observations (see Fig. \ref{fig:spec_temp}). The origin of this QPO is attributed to a strong Comptonizing corona surrounding the central source \citep{Chakrabarti-etal1995,Chakrabarti-etal2008, Nandi-etal2024}; however, based on the correlation between the evolution of QPO frequency and quasi-simultaneous radio jet ejections, \cite{Liao-etal2024} suggest a jet-like, vertically elongated corona for Swift J$1727.8-1613$. A similar type-C QPO at $\sim 0.2$ Hz is detected in LHS of IGR J$17091-3624$ (see Fig. \ref{fig:spec_temp}). Moreover, we also observe a type-B QPO in the disc-dominated SIMS of GX $339-4$ in the $0.5-80$ keV energy range from {\it NICER} and {\it AstroSat} data. Similar type-B QPOs have been observed earlier in GX $339-4$, indicating the presence of a compact corona located very close to the black hole, accompanied by dominant disc emission \citep{Nandi-etal2012, Aneesha-etal2024}. 

The wide-band ($0.7-60$ keV) energy spectral analysis reveals distinct spectral states of each source. The observed spectra of all sources are well described by a standard accretion disc component (\texttt{diskbb}) combined with thermal Comptonization of soft photons by a spherical corona (\texttt{thcomp}) (see Table \ref{table:spec_para} and Fig. \ref{fig:spec_temp}). The spectral properties exhibit a strong correlation with the spectral states of the sources. In particular, during LHS and HIMS, the spectra are dominated by Comptonized emission, contributing $\sim 64-87\%$ of the total flux, accompanied by a high covering fraction ($cov_{\rm frac} \sim 61-87$). In contrast, the HSS and SIMS are characterized by a prominent disc component, contributing up to $39-98\%$ with marginal Comptonization effects ($cov_{\rm frac} \sim 0.5-38\%$) across all sources. We observe the present findings to be broadly consistent with the previously reported results for the respective sources. For example, \citealt[]{Svoboda-etal2024b} found a steeper $\Gamma_{\rm th} \sim 4.9$ and low $cov_{\rm frac} \sim 20\%$ in the HSS of Swift J$1727.8-1613$, which agrees well with our findings of $\Gamma_{\rm th} \sim 4.6-5.1$ and $cov_{\rm frac} \sim 17\%$. In addition, the LHS results for Cyg X$-$1 (E1) are consistent with those of \cite{Krawczynski-etal2022b}, except for their measurement of a high electron temperature, which is likely because of the differences in spectral energy coverage and/or the choice of model components. 

We also notice that the Comptonization model with a disc-like coronal geometry (\texttt{comptt}) yields a comparable fit to the observed spectra similar to the spherical corona model (\texttt{thcomp}). This evidently indicates the degeneracy in determining the corona geometry (see \S4.2). Furthermore, strong reflection signatures detected in the soft state observations (HSS and SIMS) of Cyg X$-1$, Swift J$1727.8-1613$, and GX $339-4$ allow the detailed investigation of alternative coronal geometries using various reflection models (see \S4.2). We also observe that the reflected emission, either from a lamppost corona, modeled with \texttt{relxilllp}, or from a spherical corona as described in \texttt{relxillCp}, can be reproduced satisfactorily considering a highly ionized accretion disc ($\log \xi \sim 4$). All these findings highlight the inherent degeneracy among different disc-corona geometry models in explaining the observed spectral characteristics.

\vspace{-0.6cm}
\subsection{Model Prescriptions and Limitations of X-ray Polarization in BH-XRBs}

X-ray polarization measurements with {\it IXPE} reveal significant polarized emission across several spectral states of the sources in $2-8$ keV energy range, except for LMC X$-$1 (see \S4.3 and \citealt[]{Podgorny-etal2023}), Swift J$151857.0-57214$ (see \S4.3 and \citealt[]{Ling-etal2024}) and MAXI J$1744-294$ (see \S4.3). We find that Cyg X$-$1 exhibits PD within $3-4.8\%$ in LHS, which decreases to $1.4-2.5\%$ in HSS. A similar noticeable reduction in the PD from $4.7\%$ (HIMS) to $0.4\%$ (HSS) is also observed for the exceptionally bright transient Swift J1727.8$-$1613. The most recent {\it IXPE} observation of this source confirms the recovery of PD $\sim 3.2\%$ in the faint LHS. We mention that these results are in agreement with the previously reported polarimetric findings of Cyg X$-$1 \citep{Krawczynski-etal2022b, Jana-etal2024, Steiner-etal2024} and Swift J1727.8$-$1613 \citep{Veledina-etal2023, Ingram-etal2024, Svoboda-etal2024b, Podgorny-etal2024}.

Intriguingly, an exceptionally high PD of $\sim 8.3\%$ is observed in the HSS of 4U 1630$-47$, which drops to $\sim 6.8\%$ during the SPL state exhibiting strong ($> 4\sigma$) energy dependency (see Table \ref{tab:en-pol} and Fig. \ref{fig:pol1}; also \citealt[]{Kushwaha-etal2023a, Rodriguez-etal2023, Rawat-etal2023a, Rawat-etal2023b, Ratheesh-etal2024}). In contrast, low PD values are observed in several sources: LMC X$-3$ ($\sim 2.4-3\%$), 4U 1957$+115$ ($\sim 1.9\%$) and GX $339-4$ ($\sim 1.2\%$), being consistent with the previous measurements \citep{Majumder-etal2024a, Svoboda-etal2024a, Garg-etal2024, Marra-etal2024, Mastroserio-etal2024}. In the present work, we report polarization detections analyzing the most recent {\it IXPE} observations of Cyg X$-$3 (PD $\sim 21.4\%$, SIMS), LMC X$-$3 (PD $\sim 2.4\%$, HSS) and IGR J17091$-$3624 (PD $\sim 9\%$, LHS), apart from the earlier measurements of these sources. In addition, for the first time to the best of our knowledge, we report the null-detection of X-ray polarization in LMC X$-3$ (E3) and a newly discovered BH-XRB candidate MAXI J$1744-294$. We note that, subsequent to our report, null-detection of polarization for MAXI J$1744-294$ is also confirmed by \cite{Marra-etal2025}.

Meanwhile, several models have been proposed to account for the observed polarization in BH-XRBs. For instance, remarkably high PD ($\sim 8.3\%$) of 4U 1630$-$47 is attributed to the emission from a partially ionized thin/slim accretion disc viewed at inclination $i \gtrsim 60^{\circ}$ \citep{Ratheesh-etal2024}. Similar explanation is also suggested for LMC X$-3$ in the HSS that exhibits PD $\sim 3\%$ at $i \sim 70^{\circ}$ \citep{Majumder-etal2024a, Svoboda-etal2024a}. Further, scattering of photons in strong disc winds of different opening angles ($\alpha_{\rm w} \sim 30^{\circ}-40^{\circ}$) has also been suggested as another possible mechanism to explain high PD values of 4U 1630$-$47 \citep{Kushwaha-etal2023a, Rawat-etal2023a, Nitindala-etal2025}. However, \cite{Tomaru-etal2024} demonstrated that electron scattering in an optically thin wind produces polarization in a direction opposite to that obtained from the intrinsic emission from an optically thick disc. As a result, the wind component is expected to depolarize the total emission, including the intrinsic disc emission. Therefore, the presence of disc-wind from the optically thick disc appears to be insufficient to account for the observed high PD in 4U 1630$-$47. On the other hand, in high-spin black hole systems, strong relativistic effects cause intense gravitational lensing that bends emitted photons back onto the disc. This enhances the reflection features and significantly affects the observed polarization \cite[see \S4.2,][]{Schnittman-etal2009}. Such an effect appears consistent with the polarimetric observations (PD $\sim 1.2-2.5\%$) in the soft state as reported for rapidly spinning systems like Cyg X$-1$ ($a_{*} > 0.998$) and 4U $1957+115$ ($a_{*} \sim 0.988$) \citep{Marra-etal2024, Steiner-etal2024}.

Furthermore, it has been suggested that $\sim 4\%$ (Cyg X$-1$, LHS) and $\sim 9\%$ (IGR J$17091-3624$, LHS) polarization can be obtained from a wedge-shaped corona with mildly relativistic outflowing plasma model of \cite{Beloborodov-etal1998} having inclination in the range $30^{\circ} \lesssim i \lesssim 60^{\circ}$ \citep{Poutanen-etal2023, Ewing-etal2025}. A similar model of bulk Comptonization involving a mildly relativistic outflow predicts an even higher polarization degree up to $\lesssim 10\%$ \citep{Dexter-etal2024}. Notably, observed $\sim 4.7\%$ PD at $i \sim 40^{\circ}$ in HIMS of Swift J$1727.8-1613$ also seems to be consistent with the predictions from a wedge-shaped corona with outflowing plasma velocity $\lesssim 0.2c$ \citep{Poutanen-etal2023}. However, QPOs are detected in Swift J$1727.8-1613$ ($0.5-100$ keV) and IGR J$17091-3624$ ($3-79$ keV) during observations simultaneous with {\it IXPE}, although their physical origin remains unclear within the framework of the outflowing corona geometry. On the other hand, \cite{Kumar-etal2024} predicted PD $\sim 0.5-4\%$ assuming a simple static spherical corona for $30^{\circ} \lesssim i \lesssim 60^{\circ}$. This appears broadly consistent with the low PD of $\sim 1.2\%$ observed at $i\sim 30^{\circ}$ in the SIMS of GX $339-4$. 

All the above interpretations of the polarimetric findings point toward a persistent degeneracy in understanding the different accretion-ejection dynamics. More precisely, predictions regarding the disc-corona-jet geometry become even more challenging when the simultaneous origin of temporal features (QPOs), spectral distributions and polarization (PD) is considered. This emphasizes the need for a unified framework of accretion-ejection processes in BH-XRBs that can consistently explain both timing and spectro-polarimetric observations.

\vspace{-0.5cm}
\subsection{Accretion-Ejection Scenarios in BH-XRBs}

\begin{figure}
	\begin{center}
		\includegraphics[width=0.8\columnwidth]{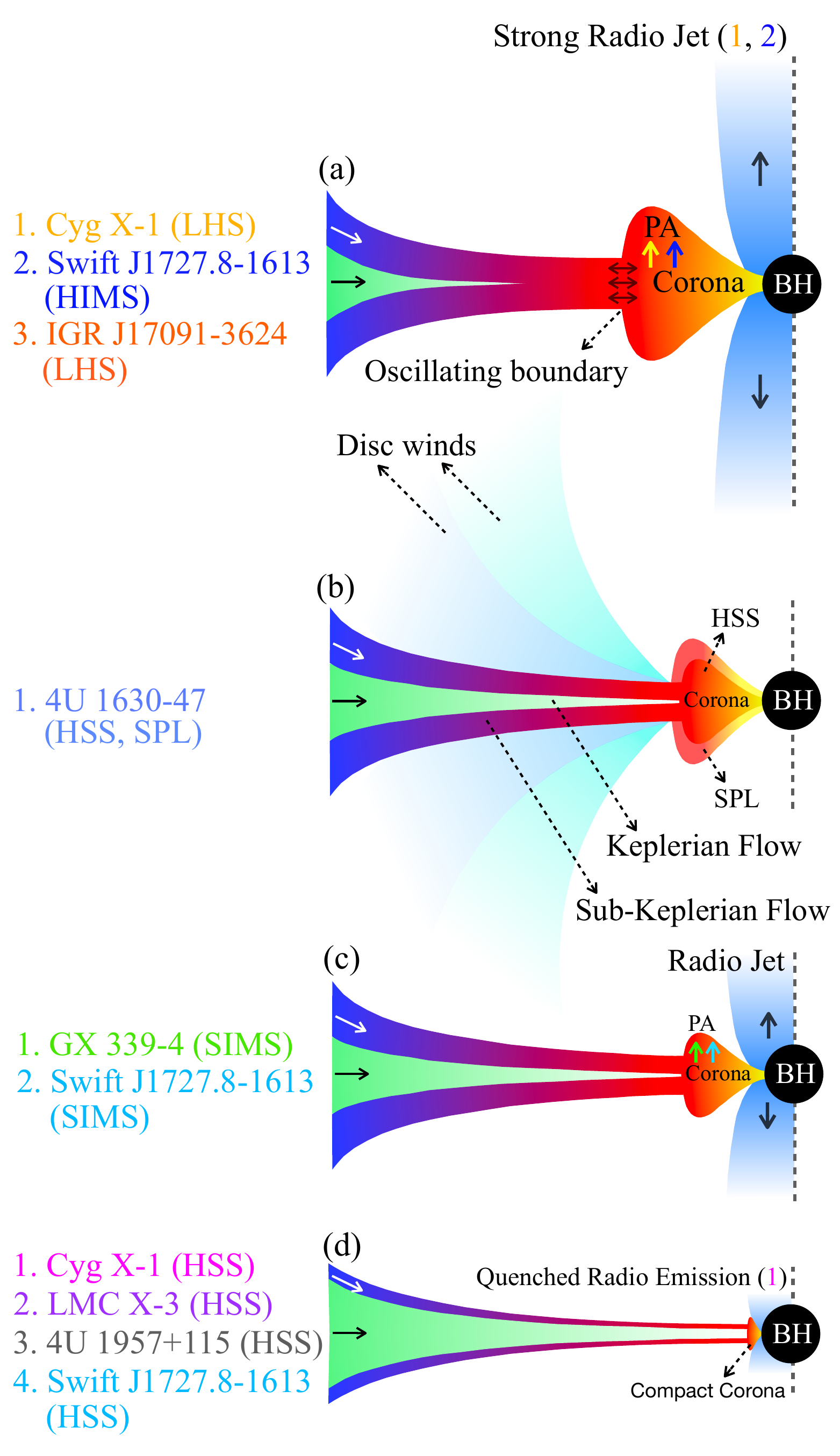}
	\end{center}
    \vskip -0.4cm
	\caption{Schematic representation of the possible disc–corona–jet configurations for seven BH-XRB sources within the two-component accretion flow framework. The Keplerian and sub-Keplerian components shown in panel (b) are also present in the other three configurations. Colored vertical arrows in the corona regions of panels (a) and (c) denote the polarization angle for the respective sources. In panels (a) and (d), the jet/radio emission configuration applies only to sources in distinct spectral states, indicated by their serial numbers. See the text for details.
    }
	\label{fig:config}
\end{figure}

Following the discussion of model degeneracies and the need for a unified model framework, we next explore physically motivated accretion-ejection geometries capable of explaining the commonly observed spectro-temporal features, while also offering a promising basis for interpreting the spectro-polarimetric signatures in BH-XRBs. Towards this, a plausible disc-corona-jet configuration based on the two-component accretion flow model \cite[]{Chakrabarti-etal1995,Chakrabarti-etal1996} is illustrated in Fig.~\ref{fig:config}. In this model, the accretion flow around black holes comprise with Keplerian disc flanked between the sub-Keplerian components. This sub-Keplerian component generally forms the post-shock corona (PSC, equivalently Comptonizing corona) in the inner region of the disc \cite[]{Chakrabarti-etal1995, Molteni-etal1996, Chakrabarti-etal2008, Nandi-etal2012, Das-etal2014}. Soft photons from the Keplerian disc are up-scattered by hot electrons present in the corona and produce the hard spectral tail seen in BH-XRBs. The inner edge of the Keplerian disc smoothly connects to the outer boundary of the coronal region.
 
The spectro-polarimetric correlation study (see \S4.6 and Fig. \ref{fig:corr_plt}) suggests that the polarization signatures (PD $\sim 3-9\%$) in the LHS/HIMS of Cyg X$-1$, Swift J$1727.8-1613$ and IGR J$17091-3624$ are primarily governed by the dominant Comptonization process characterized by $F_{\rm ratio} < 1$ and $cov_{\rm frac} \sim 35-87\%$. Interestingly, the radio jet position angle of Cyg X$-1$ and Swift J$1727.8-1613$ roughly aligns along the observed PA throughout the respective {\it IXPE} campaigns \citep{Miller-Jones-etal2021, Wood-etal2024}. This suggests that PA is possibly oriented perpendicular to the disc-corona geometry. Further, it may be noted that PA resulting from multiple Compton up-scatterings generally tends to align along the minor axis of the corona \citep{Ingram-etal2024}. These pieces of evidence suggest a disc-corona configuration comprising a radial corona located close to the black hole, along with a truncated accretion disc at larger radius during LHS/HIMS (see Fig. \ref{fig:config}a) of Swift J$1727.8-1613$ and Cyg X$-1$. The above configuration potentially explains the QPOs in Swift J$1727.8-1613$ (see \S4.2 and Fig. \ref{fig:spec_temp}c) in terms of the oscillating boundary of the Comptonizing region (see Fig. \ref{fig:config}a and \citealt[]{Molteni-etal1996, Nandi-etal2024}). It is worth mentioning that the coronal geometry in IGR J$17091-3624$ remains inconclusive from the polarimetric results due to the unresolved position angle of the radio jet. Although the presence of QPO in LHS of IGR J$17091-3624$ (see Fig. \ref{fig:spec_temp}g) perhaps indicates a similar disc-corona configuration presented in Fig. \ref{fig:config}a. However, the absence of QPO in Cyg X$-1$ implies that either the coronal boundary fails to satisfy the necessary resonance condition for modulation, or oscillation is too subtle to be detected as predicted by \cite{Majumder-etal2025}. 

For 4U $1630-47$, an absorption feature around $\sim 7$ keV is observed, suggesting the presence of strong disc winds in the HSS ($F_{\rm ratio}\sim 52$ and $cov_{\rm frac}\sim 5\%$) spectra (see \S4.2). Accordingly, a disc-dominated accretion configuration along with a weak corona, accompanied by strong disc winds with varying opening angles (see also \citealt[]{Nitindala-etal2025}), is likely favored for this source (see Fig. \ref{fig:config}b). However, a relatively strong corona ($cov_{\rm frac}\sim 38\%$) may be present in the SPL state as compared to the HSS (see Fig. \ref{fig:config}b).

For GX $339-4$, a low PD ($\sim 1.2\%$) is observed within $3-8$ keV energy range during SIMS ($cov_{\rm frac} \sim 34\%$ and $F_{\rm ratio} \sim 2.4$), while null-detection of PD is seen in the HSS (Table \ref{tab:en-pol}). This possibly associates the polarization signature of the source with Comptonized emission. We notice $\rm PD < \rm MDP_{99}\%$ in the entire {\it IXPE} energy range of $2-8$ keV during SIMS, which possibly resulted from the depolarization of the overall radiation by unpolarized and/or oppositely polarized (as compared to corona) disc emission \citep{Ingram-etal2024} at lower energies ($2-3$ keV). The polarization angle of GX $339-4$ aligns with the direction of the discrete radio jet knot \citep{Mastroserio-etal2024}. These findings predict the GX $339-4$ configuration comprising a radially extended weak compact corona coexisting with a strong disc ($F_{\rm ratio}\sim 2.4$, $\Gamma_{\rm th}\sim 2$), as illustrated in Fig. \ref{fig:config}c. Moreover, a type-B QPO (see Fig. \ref{fig:spec_temp}e) observed in this source could be explained from this configuration using an analogous argument presented above for Fig. \ref{fig:config}a (see also \citealt[]{Nandi-etal2012, Aneesha-etal2024}). Notably, the SIMS of Swift J$1727.8-1613$ also corresponds to the configuration shown in Fig. \ref{fig:config}c.

Furthermore, we conjecture that a similar disc-corona configuration (see Fig. \ref{fig:config}d) is also preferred in the disc-dominated HSS of other sources (LMC X$-3$, 4U $1957+115$, Cyg X$-1$ and Swift J$1727.8-1613$) with major contributions from thermal disc and/or various general relativistic effects \citep{Majumder-etal2024a, Svoboda-etal2024a, Marra-etal2024, Steiner-etal2024, Svoboda-etal2024b} in the observed PD ($1.4-3\%$). Among these, quenched radio emission with flux densities of $\sim 2.4-11.7$ mJy is observed in the HSS of Cyg X$-1$ (see Table \ref{tab:log} and Fig. \ref{fig:config}d), which is found to be consistent with previous reports of radio quenching in this source \citep{Tigelaar-etal2004}. It is noteworthy that although the presence of a radial corona is favored for 4U $1630-47$, LMC X$-3$, and 4U $1957+115$ in Fig. \ref{fig:config}, the coronal geometry remains uncertain due to the lack of resolved radio jet orientation for these sources. Moreover, high PD (up to $\sim 21\%$) dominated by reflected emission in Cyg X$-3$ appears inconsistent with the disc-corona configurations presented in Fig. \ref{fig:config}. It has been suggested that a disc-corona geometry featuring an optically thick funnel in the innermost region of the accretion disc, which obscures most of the primary emission, is favored for Cyg X$-3$ \cite[]{Veledina-etal2024a, Veledina-etal2024b}. Finally, we mention that, although the two-component accretion flow configuration aligns with the disc-corona geometry inferred from polarimetric studies for most of the BH-XRBs, a detailed radiative transfer computation is needed within this framework for the predictions on X-ray polarization. Therefore, the schematic in Fig.~\ref{fig:config} illustrates a possible disc-corona-jet geometry in BH-XRBs; however, it does not yet provide a quantitative explanation for the polarimetric results.

\vspace{-0.5cm}
\subsection{Frontiers of X-ray Polarimetry: A Promising Era Ahead}

\begin{figure}
    \begin{center}
        \includegraphics[width=1\columnwidth]{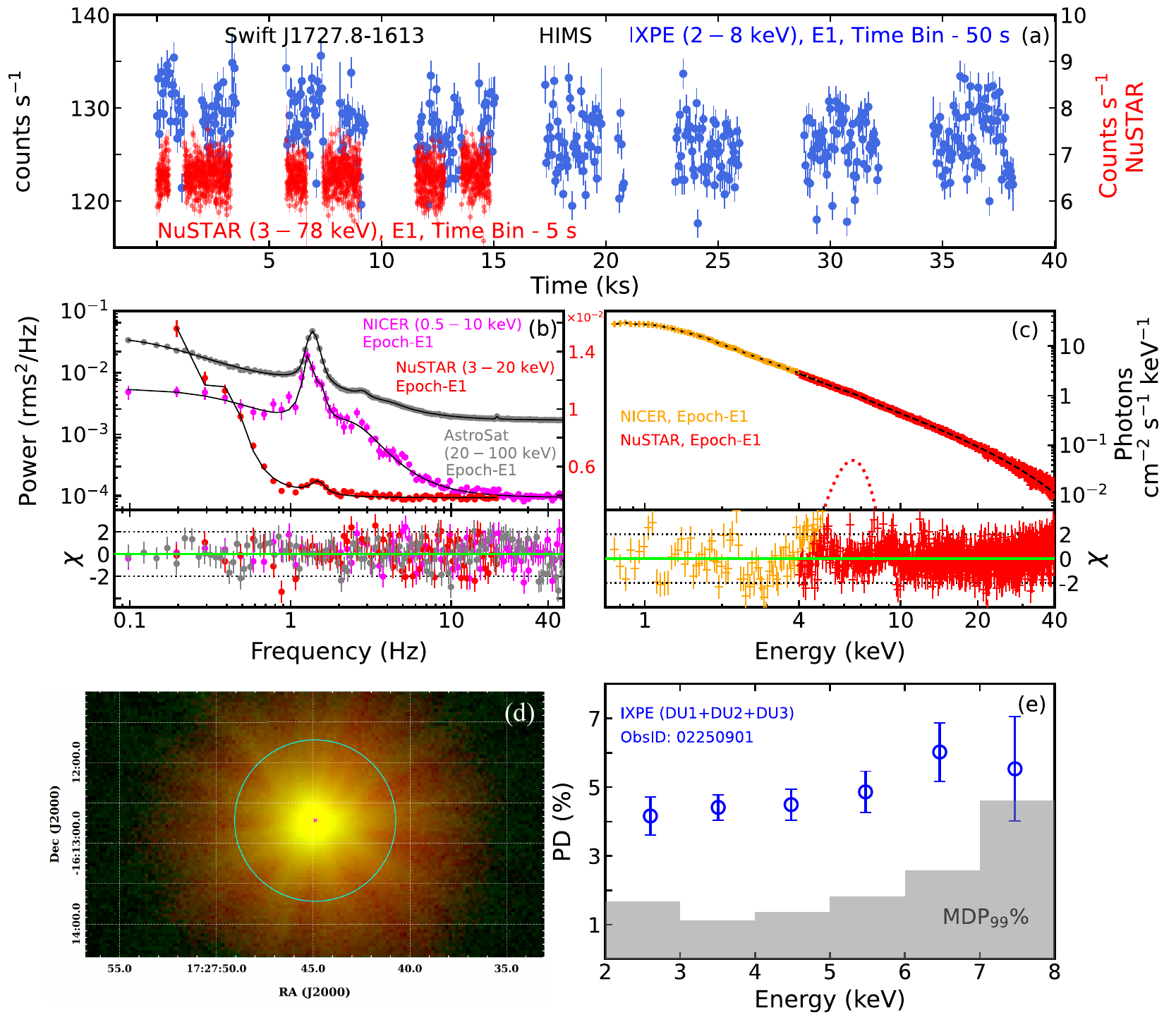}
    \end{center}
    \vskip -0.5cm
    \caption{(a) Time binned ($50$ s) light curve of Swift J$1727.8-1613$ observed with {\it IXPE} in the hard-intermediate state (HIMS). (b) Best-fitted power density spectra ($0.1-50$ Hz) obtained from the quasi-simultaneous {\it NICER} ($0.5-10$ keV), {\it NuSTAR} ($3-20$ keV) and {\it AstroSat} ($20-100$ keV) observations. (c) Best-fitted energy spectra from the quasi-simultaneous {\it NICER} and {\it NuSTAR} observations in $0.7-40$ keV energy range. (d) Colour composite X-ray image ({\it NuSTAR}: reddish and {\it IXPE}: greenish) of Swift J$1727.8-1613$ with the overlap in yellow colour. The magenta cross marks the source location in X-ray and the cyan circle of radius 60 arcsec around it denotes the region considered for polarimetric analysis. (e) Variation of the polarization degree (PD) with energy obtained from the model-independent polarimetric analysis. The gray histograms represent the $\rm MDP_{\rm 99}\%$ level. 
    }
   \label{fig:master}
\end{figure}

Indeed, the first X-ray polarization measurements were conducted in the 1970s by \textit{OSO-8}, revealing significant polarization in Cyg X$-1$ with PD $\sim 2.4\%$ at $2.6$ keV and $\sim 5.3\%$ at $5.2$ keV \citep{Long-etal1980}. Thereafter, higher polarization was observed in hard X-rays for Cyg X$-1$ with \textit{INTEGRAL} as $\lesssim 20\%$ in the $130-230$ keV band \citep{Jourdain-etal2012} and $\sim 75\%$ above $400$ keV \citep{Rodriguez-etal2015}. Recently, \citet{Chattopadhyay-etal2024} detected $\sim 23\%$ polarization in the HIMS of Cyg X$-1$ using \textit{AstroSat/CZTI}. Despite these advances, low-energy X-ray polarimetry remained relatively underexplored until the launch of \textit{IXPE}, which has made substantial progress in this domain during its first three and a half years of operation. Nevertheless, several questions remain unanswered, paving the way for future research.

At present, Swift J$1727.8-1613$ is the only BH-XRB transient observed during its outburst with {\it IXPE} exhibiting polarization over four spectral states (HIMS, SIMS, HSS and LHS) along with QPO features. To establish robust spectro-polarimetric correlations across different spectral states, coordinated {\it IXPE} observations along with simultaneous spectro-temporal coverage from other missions are indispensable. Further, since the effects of Comptonization become prominent beyond the coverage of {\it IXPE} up to $8$ keV, complementary wide-band X-ray spectro-polarimetry and timing studies are crucial to constrain the coronal geometry. Thus, the four windows of X-ray astronomy (imaging, timing, spectroscopy and polarimetry) with wide energy coverage is crucial to unravel the complete understanding of BH-XRBs (see Fig. \ref{fig:master}).

In this context, the recently launched {\it XPoSat}\footnote{\url{https://www.isro.gov.in/XPoSat.html}} mission and the upcoming {\it XL-Calibur} mission \citep{Abarr-etal2021} are set to play a crucial role in investigating hard X-ray polarimetric properties of BH-XRBs. In particular, {\it POLIX} ($8-30$ keV) and {\it XSPECT} ($0.8-15$ keV) onboard on {\it XPoSat} are suitable for mid-energy polarimetry and simultaneous spectro-temporal studies, respectively. The balloon-borne {\it XL-Calibur} mission, operating in $15-80$ keV energy range, completed a successful flight\footnote{\url{https://sscspace.com/nasa-xl-calibur-balloon-launched/}} in July 2024, aiming to measure polarization in Cyg X$-1$. This eventually led to the recent measurement of a polarization degree of $\sim 5\%$ in LHS of Cyg X$-1$ in $\sim 19-64$ keV energy range \citep{Awaki-etal2025}, which is comparable to the low energy ($2-8$ keV) polarimetric results of Cyg X$-1$ (PD $\sim 4.8\%$, LHS) with {\it IXPE} (see Fig. \ref{fig:contour} and Table \ref{tab:en-pol}). The consistency of the measured PDs and PAs across the $2-64$ keV range observed with {\it IXPE} and {\it XL-Calibur} indicates that the same underlying physical processes likely govern the observed polarization in both energy bands \citep{Awaki-etal2025}. In addition, the future soft X-ray spectro-polarimetric mission {\it eXTP}\footnote{\url{https://heasarc.gsfc.nasa.gov/docs/heasarc/missions/extp.html}} ($2-10$ keV), expected to launch in $2027$, will offer improved sensitivity for low-energy polarization measurements. Indeed, these dedicated missions will significantly advance our understanding of the accretion dynamics and coronal geometries in BH-XRBs. However, simultaneous observations combining wide-band timing and spectro-polarimetric measurement are still lacking, which emphasizes the need for future missions with broader energy ranges and enhanced capabilities. 

\vspace{-0.7cm}
\section{Conclusions}

In this paper, we perform a detailed timing and spectro-polarimetric study of eleven BH-XRBs using quasi-simultaneous {\it IXPE}, {\it NICER}, {\it NuSTAR} and {\it AstroSat} observations. Our analyses provide valuable insights into the accretion-ejection dynamics and the geometry of BH-XRBs under consideration. The key findings from our study along with their implications are summarized below.

\begin{itemize}

\item The combined spectro-temporal results in the wide-band energy range ($0.5-100$ keV) reveal the presence of distinct canonical spectral states of BH-XRBs, which are closely connected to the temporal characteristics and emission mechanism of the sources. The timing and spectral features result in degeneracy among the different viable disc-corona geometries of the sources.

\item The detection of X-ray polarization is reconfirmed in eight out of eleven BH-XRBs with moderate to strong energy dependence in PD. A comprehensive spectro-polarimetric correlation study reveals significant positive (negative) correlations between PD and $cov_{\rm frac}$ ($F_{\rm ratio}$) indicating contributions from various surrounding components (disc and corona) to the observed polarization.

\item Timing and spectro-polarimetric results, combined with known radio-jet angles, suggest that a two-component disc-corona model may be plausible. In this framework, a radially extended strong corona is likely to be present at the inner part of a truncated accretion disc during harder states, while a comparatively weaker corona persists in softer states for Swift J1727.8$-$1613, Cyg X$-$1, GX 339$-$4, and IGR J17091$-$3624.

\item For 4U 1630$-$47, LMC X$-$3, and 4U 1957$+$115, a thermally dominated accretion disc with a weak corona seems to be preferred. However, the coronal geometry remains unclear due to the minimal contribution of Comptonized emission in the {\it IXPE} band and the lack of complementary radio observations, though this analysis does support the presence of disc winds in 4U 1630$-$47. We also note that the disc-corona geometry outlined in this work bears limitations to explain the observed features of Cyg X$-$3, for which an alternative configuration involving an optically thick inner funnel has been proposed \citep{Veledina-etal2024a, Veledina-etal2024b}.

\end{itemize}

\vspace{-0.8cm}
\section{Acknowledgments}

Authors thank the anonymous reviewer for constructive comments and useful suggestions that helped to improve the quality of the manuscript. SM and SD thank the Department of Physics, IIT Guwahati, India for providing the facilities to complete this work. SM also thanks Luca Baldini for the discussion in identifying the bugs in \texttt{IXPEOBSSIM} software and releasing the latest version \texttt{IXPEOBSSIM V31.0.3}. AK, SS, KJ and AN thank GH, SAG; DD, PDMSA, and Director, URSC for encouragement and continuous support to carry out this research. This publication uses data from the {\it IXPE}, {\it NICER}, {\it NuSTAR} and {\it AstroSat} missions. This research has made use of the {\it MAXI} data provided by {\it RIKEN}, {\it JAXA} and the {\it MAXI} team \citep{Matsuoka-etal2009}. The {\it Swift/BAT} transient monitor results provided by the {\it Swift/BAT} team are also used \citep{Krimm-etal2013}. We thank each instrument team for processing the data and providing the necessary software tools for the analysis.

\vspace{-0.7cm}
\section{Data Availability}

Data used for this publication from {\it IXPE}, {\it NICER} and {\it NuSTAR} missions are currently available at the HEASARC browse (\url{https://heasarc.gsfc.nasa.gov/db-perl/W3Browse/w3browse.pl}). The {\it AstroSat} data is available at the ISSDC data archive (\url{https://webapps.issdc.gov.in/astro_archive/archive/Home.jsp}). The {\it MAXI/GSC} on-demand data is available at \url{http://maxi.riken.jp/mxondem/} and {\it Swift/BAT} data is taken from \url{https://swift.gsfc.nasa.gov/results/transients/}.


\begin{table*}
	\centering
	\caption{Details of the multi-mission observations of the selected sources over different epochs. In the table, blue shades represent the epochs for which quasi-simultaneous {\it IXPE}, {\it NICER}, and {\it NuSTAR} observations are available. The gray shades denote the quasi-simultaneous observations with {\it IXPE} and {\it NICER} only. The magenta colored shade indicates the epoch for which only quasi-simultaneous {\it IXPE} and {\it NuSTAR} observations are available. The orange colored shade denotes the available {\it AstroSat} observations close to the {\it IXPE} epochs. See the text for details.}
	
	\renewcommand{\arraystretch}{1.47}
	\resizebox{1.0\textwidth}{!}{%
		\begin{tabular}{lcllcccccccc}
			\hline
			Source & Epoch  &  Instrument & Obs. ID & Start Date & MJD& Exposure & Mean Rate & {\it MAXI/GSC} & {\it Swift/BAT} & Radio Flux & Spectral \\
			
			&             &                &  &   &  Start & (ks) & (cts/s) & Flux (mCrab) & Flux (mCrab) & (mJy) & State \\
			&             &                &  &   & &  & & ($2-20$ keV) & ($15-50$ keV) &  &  \\
			
			\hline
			
			\rowcolor{Lightblue}
			Cyg X$-1$& E1  & IXPE        & 01002901       & 2022-05-15 & 59714.64 & 242 & $9.82\pm0.21$ &  $401\pm11$  & $928\pm34$ & $-$ & LHS \\
			\rowcolor{Lightblue}
			& & NICER       & 5100320101 & 2022-05-15 & 59714.26 & 8.2  & $2938 \pm 25$ & $-$ & $-$ & $-$ & $-$ \\
			\rowcolor{Lightblue}
			& & NuSTAR   & 30702017002 & 2022-05-18 & 59717.57 & 16  & $324\pm15$ & $-$ & $-$ & $-$ &  $-$ \\
			\rowcolor{orange}
			& & AstroSat  & A11\_080T01\_9000005146 & 2022-05-15 & 59714.03 & $47.3$ &  $1505 \pm 37$ & $-$ & $-$ & $-$ &  $-$ \\

			\rowcolor{Lightblue} 
			& E2  & IXPE        & 01250101       & 2022-06-18 & 59748.86 &  86  & $10.32\pm0.12$ & $420\pm14$   & $885\pm33$ & $-$ &  LHS  \\
			\rowcolor{Lightblue}
			& & NICER       & 5100320108     & 2022-06-20 & 59750.60 &  3  & $3001 \pm 25$ & $-$ & $-$ & $-$ &  $-$ \\
			\rowcolor{Lightblue}
			& & NuSTAR      & 90802013002    & 2022-06-20 & 59750.50 &  13  & $338\pm14$ & $-$ & $-$ & $-$ & $-$ \\

			& E3  & IXPE        & 02008201       & 2023-05-02 & 	60066.96 &  21  & $26.93\pm0.18$ & $888\pm12$  & $546\pm21$ &  $9.1\pm0.3^{\boxdot}$ & HSS \\

			& E4  & IXPE        & 02008301       & 2023-05-09 & 	60073.44 &  31  & $33.93\pm0.23$ & $701\pm12$  & $613\pm23$ & $11.7\pm0.2^{\boxdot}$ & HSS \\

			& E5  & IXPE        & 02008401       & 2023-05-24 & 	60088.82 &  25  & $44.26\pm0.30$ & $809\pm20$  & $569\pm35$ & $2.4\pm0.2^{\boxdot}$ & HSS \\
			\rowcolor{lightgray}
			\rowcolor{orange}
			&    & AstroSat       & T05\_105T01\_9000005662  & 2023-05-24 & 60088.79 & 60.5  & $1354\pm36$ & $-$ & $-$ & $-$ & $-$  \\
			
		\rowcolor{magenta2}	
		& E6  & IXPE        & 02008501       & 2023-06-13 & 	60108.96 &  29  & $39.33\pm0.28$ & $765\pm27$  & $593\pm45$ & $8.9\pm0.2^{\boxdot}$ & HSS\\

            \rowcolor{magenta2}
            &   &  NuSTAR    & 80902318004      & 2023-06-14 & 	60109.02 &  9.8  & $694\pm22$ & $-$  & $-$ & $-$ & HSS\\

			\rowcolor{Lightblue}
			& E7  & IXPE        & 02008601       & 2023-06-20 & 60115.04 &  34  & $48.09 \pm 0.24$ & $803\pm24$  & $859\pm46$ & $9.4\pm0.4^{\boxdot}$ & HSS \\
			\rowcolor{Lightblue}
			&    & NICER       & 6643010104     & 2023-06-20 & 	60115.97 &  7.6  & $23227\pm 201$ & $-$ & $-$ & $-$ & $-$ \\
            \rowcolor{Lightblue}
            &    & NuSTAR       & 80902318006     & 2023-06-20 & 	60115.82 &  10.2  & $878\pm 26$ & $-$ & $-$ & $-$ & $-$ \\

            \rowcolor{lightgray}
            & E8  & IXPE   &   03002201  & 2024-04-12 & 60412.02 & 55.8 & $7.28\pm0.12$ & $333\pm9$  & $736\pm31$ & $-$ & LHS \\

            \rowcolor{lightgray}
		&    & NICER       & 7100320104  & 2024-04-11 & 60411.10 & 0.7 & $1508 \pm 18$ & $-$ & $-$ & $-$ & $-$  \\

            & E9  & IXPE   &   03003101  & 2024-05-06 & 60436.37 & 53.9 & $7.14\pm0.11$ & $283\pm8$  & $919\pm42$ & $-$ & LHS \\

            & E10  & IXPE   &   03010001  & 2024-05-26 & 	60456.04 & 57.5 & $7.29\pm0.12$ & $286\pm18$  & $747\pm28$ & $-$ & LHS \\

            \rowcolor{lightgray}
            & E11  & IXPE   &   03010101  & 2024-06-14 &  60475.65 & 55.8  & $5.72\pm0.09$ & $293\pm11$  & $725\pm27$ & $-$ & LHS \\

            \rowcolor{lightgray}
		&    & NICER       & 7706010104	 & 2024-06-14 & 60475.99 & 6.1 & $1698 \pm 54$ & $-$ & $-$ & $-$ & $-$  \\

         & E12  & IXPE   &   03002599  & 2024-10-10 & 60593.22  & 110 & $22.1\pm2.2$ & $600\pm13$  & $795\pm27$ & $-$ & HSS \\
   
			\hline
			
			\rowcolor{Lightblue}
			4U 1630$-47$ & E1  & IXPE        & 01250401   & 2022-08-23 & 59814.95 & 459 & $4.56\pm0.08$ & $247\pm8$ & $10\pm2$ & $-$ & HSS \\
			\rowcolor{Lightblue}
			& & NICER       & 5501010102  & 2022-08-23 & 59814.01 &  2 & $479\pm10$ & $-$ & $-$ & $-$ & $-$ \\
			\rowcolor{Lightblue}
			& & NuSTAR      & 80802313002 & 2022-08-25 & 59816.19 &  17 & $348\pm14$ & $-$ & $-$ & $-$ & $-$ \\
			
			\rowcolor{Lightblue}
			& E2  & IXPE     &  02250601   & 2023-03-10 & 60013.78 &  138  & $10.16\pm0.12$ & $550\pm16$ & $246\pm20$ &  $-$ & SPL  \\
			\rowcolor{Lightblue}
			& & NICER       & 6557010101  & 2023-03-10 & 60013.76 & 5  & $1009\pm15$ & $-$ & $-$ & $-$ & $-$ \\
			\rowcolor{Lightblue}
			& & NuSTAR      & 80801327002 & 2023-03-09 & 60012.36 & 12  & $884\pm29$ & $-$ & $-$ & $-$ & $-$ \\	
                \rowcolor{orange}
            &    & AstroSat     & A12\_056T02\_9000005538  & 2023-03-10 & 60013.08 & 29.4  & $1938\pm43$ & $-$ & $-$ & $-$ & $-$  \\
			
			\hline
			
			\rowcolor{Lightblue}
			Cyg X$-3$ & E1  & IXPE        & 02001899    & 2022-10-14 & 59866.04 & 538  & $1.33 \pm 0.04$ &  $106\pm20$  & $129\pm26$ & $106 \pm 24^*$ & LHS \\
			\rowcolor{Lightblue}
			& & NICER  & 5142010105  & 2022-10-14 & 59866.28 &  5  & $44.79\pm3.12$ & $-$ & $-$ & $-$ & $-$ \\
			\rowcolor{Lightblue}
			& & NuSTAR      & 90802323002 & 2022-10-13 & 59865.62  &  18  & $197\pm1$ & $-$ & $-$ & $-$ & $-$ \\
			
			\rowcolor{magenta2}
			& E2  & IXPE &  02250301  & 2022-12-25 & 59938.41 &  199  & $4.55 \pm 0.09$ & $295\pm31$ & $144\pm35$ & $107 \pm 36^*$ & HSS \\
			
			\rowcolor{magenta2}
			& & NuSTAR  & 90801336002 & 2022-12-25 & 59938.33 & 36  & $467\pm2$ & $-$ & $-$ & $-$ & $-$ \\ 
			
			\rowcolor{lightgray}
			& E3  & IXPE  &  02009101   & 2023-11-17 & 60265.83 &  291  & $1.43 \pm 0.05$ & $131\pm8$ & $159\pm13$ & $-$ & SIMS \\
			\rowcolor{lightgray}
			& & NICER       & 6692010101  & 2023-11-17 & 60265.77 &  1.9  & $125 \pm 7$ & $-$ & $-$ & $-$ & $-$ \\

                & E4  & IXPE  &   03250301  & 2024-06-02 & 60463.77 &  50  & $7.45 \pm 0.12$ & $411\pm16$ & $14\pm9$ & $-$ & HSS \\
			\hline
			
			\rowcolor{Lightblue}
			LMC X$-1$ & E1  & IXPE        & 02001901       & 2022-10-19  &  59871.61 &  563 & $0.96\pm0.03$ & $14\pm9$ & $13\pm3$ & $-$ & HSS \\
			\rowcolor{Lightblue}
			& & NICER       & 	5100070101	  & 2022-10-19 & 59871.79 & 2.8  & $193\pm6$ & $-$ & $-$ & $-$ & $-$ \\
			\rowcolor{Lightblue}
			& & NuSTAR      & 90801324002 & 2022-10-24 & 59876.16 &  19 & $15\pm2$ & $-$ & $-$ & $-$ & $-$ \\
			
			\hline
			
			\rowcolor{Lightblue}
			4U 1957$+115$ & E1  & IXPE   & 02006601   & 2023-05-12 & 60076.10 &  572 & $1.09\pm0.04$ & $19\pm4$  & $2\pm1$ & $-$ &  HSS \\
			\rowcolor{Lightblue}
			& & NICER  & 6100400101  & 2023-05-12 & 60076.55 & 0.6  & $265\pm 8$ & $-$ & $-$ & $-$ & $-$ \\
			\rowcolor{Lightblue}
			& & NuSTAR  & 30902042002 & 2023-05-15 & 60079.33 & 19  & $29\pm4$ & $-$ & $-$ & $-$ & $-$ \\
			
			\hline
			
			\rowcolor{Lightblue}
			LMC X$-3$& E1  & IXPE        & 02006599       & 2023-07-07 & 60132.77 &  562 & $0.97\pm0.03$ & $16\pm4$  & $18\pm3$ & $-$ & HSS \\
			\rowcolor{Lightblue}
			& & NICER       & 6101010117  & 2023-07-08 & 60133.42 &  0.1  & $358\pm23$ & $-$ & $-$ & $-$ & $-$ \\
			\rowcolor{Lightblue}
			& & NuSTAR      & 30902041002 & 2023-07-09 &  60134.51 &  28  & $15\pm 3$ & $-$  & $-$ & $-$ & $-$ \\ 

            \rowcolor{lightgray}
		& E2  & IXPE    & 03004899   & 2024-10-03 & 60586.51 & 385  & $1.62\pm0.04$ & $40\pm5$  & $14\pm9$ & $-$ & HSS \\
		\rowcolor{lightgray}
		& & NICER       & 7704010101  & 2024-10-03 & 60586.63 &  2.4  & $606\pm11$ & $-$ & $-$ & $-$ & $-$ \\

            \rowcolor{lightgray}
            & E3  & IXPE  & 03004901  & 2024-11-21 & 60635.94 & 382  & $0.42\pm0.02$ & $8\pm5$  & $-$ & $-$ & HSS \\

            \rowcolor{lightgray}
            &   & NICER  & 	7704010201  & 2024-11-23 & 60637.10 & 1.7 & $246\pm7$ & $-$ & $-$ & $-$ & $-$ \\

            \rowcolor{magenta2}
            & E4  & IXPE  & 03005001  & 2025-01-11 & 60686.18 & 389  & $0.86\pm0.41$ & $11\pm4$  & $9\pm7$ & $-$ & HSS \\

            \rowcolor{magenta2}
            &  & NuSTAR  & 31002032006  & 2025-01-11 & 60686.68 & 35 & $16 \pm 5$ & $-$  & $-$ & $-$ &  \\

			\hline
			
		\end{tabular}
	}
	
	\label{tab:log}
	
	\begin{list}{}{}
              \item $^{\dagger}$Data are not available in public domain. $^{\boxdot}${\it AMI-LA} ($15.5$ GHz, \citealt{Steiner-etal2024}). $^*${\it RATAN} ($4.7$ GHz, \citealt[]{Veledina-etal2024a}).
		\end{list}

\end{table*}


\begin{table*}
	\centering
	\caption{Same as Table \ref{tab:log}.}
	
	\renewcommand{\arraystretch}{1.47}
	\resizebox{1.0\textwidth}{!}{%
		\begin{tabular}{lcllcccccccc}
			\hline
			Source & Epoch  &  Instrument & Obs. ID & Start Date & MJD& Exposure & Mean Rate & {\it MAXI/GSC} & {\it Swift/BAT} & Radio Flux & Spectral \\
			
			&             &                &  &   &  Start & (ks) & (cts/s) & Flux (mCrab) & Flux (mCrab) & (mJy) & State \\
			&             &                &  &   & &  & & ($2-20$ keV) & ($15-50$ keV) &  &  \\
			
			\hline

	       	\rowcolor{Lightblue}
			Swift J1727.8$-1613$ & E1  & IXPE        & 02250901      & 2023-09-07 & 60194.81 &  19 & $42.21\pm0.28$ & $6053\pm71$  & $4898\pm173$ & $120 \pm 12^{**}$ & HIMS \\
			\rowcolor{Lightblue}
			&    & NICER       &  6203980115     &  2023-09-08 &  60195.61 &  0.2  &  $64955\pm274$ & $-$ & $-$ & $-$ & $-$ \\

                \rowcolor{Lightblue}
                &    & NuSTAR       &  80902333006     &  2023-09-07 &  60194.78 &  0.7  &  $6536\pm8$ & $-$ & $-$ & $-$ & $-$ \\

                \rowcolor{orange}
                &    & AstroSat       &  T05\_145T01\_9000005836    & 2023-09-08  & 60195.07  &  $32$  &  $541\pm23$ & $-$ & $-$ & $-$ & $-$ \\
			
			\rowcolor{Lightblue}
			& E2  & IXPE        & 02251001       & 2023-09-16 & 	60203.70 & 37  & $41.29\pm0.25$ & $5058\pm31$ & $2472\pm98$ & $144 \pm 4.32^*$ & HIMS \\
			\rowcolor{Lightblue}
			&    & NICER       & 6203980119    & 2023-09-13 &  60200.18 &  4.9 &  $65392\pm 122$ & $-$ & $-$ & $-$ & $-$ \\

               \rowcolor{Lightblue}
                &    & NuSTAR       &  80902313002     &  2023-09-16 &  60203.81 &  0.6  &  $5478\pm26$ & $-$ & $-$ & $-$ & $-$ \\
			
			& E3  & IXPE        & 02251101     &  2023-09-27 & 60214.92 &  21  & $36.85\pm0.22$ & $3395\pm49$  &  $300\pm10$ & $88 \pm 2.64^*$ &  HIMS  \\ 
			
			\rowcolor{Lightblue}
			& E4  & IXPE        & 02251201       & 2023-10-04 & 	60221.54  & 17  & $41.31\pm0.24$ & $4689\pm46$  & $206\pm38$ & $11 \pm 0.33^*$ & SIMS \\
			\rowcolor{Lightblue}
			&    & NICER       & 6557020401     & 2023-10-04 & 60221.18 &  6.5  & $66263\pm62$ & $-$ & $-$ & $-$ & $-$ \\

                \rowcolor{Lightblue}
                &    & NuSTAR       &  80902313008  &  2023-10-04 &  60221.54 &  0.5  &  $4111\pm4$ & $-$ & $-$ & $-$ & $-$ \\

			\rowcolor{Lightblue}
			& E5  & IXPE        & 02251301       & 2023-10-10 & 	60227.47 &  18  & $34.62\pm0.20$ & $3446\pm38$  & $206\pm38$ & $35\pm1.05^*$ & SIMS \\
			\rowcolor{Lightblue}
			&    & NICER       & 6203980136     & 2023-10-09 & 60226.02 &  1.9  & $57789\pm149$ & $-$ & $-$ & $-$ & $-$ \\

                \rowcolor{Lightblue}
                &    & NuSTAR       &  80902313016  &  2023-10-10 &  60227.82 &  1.1  &  $3196\pm4$ & $-$ & $-$ & $-$ & $-$ \\

			\rowcolor{lightgray}
			& E6  & IXPE        & 03005701       & 2024-02-11 & 60351.39 & 67 & $8.09\pm0.13$ & $80\pm7$  & $21\pm12$ & $-$ & HSS \\
			
			\rowcolor{lightgray}
			&   & NICER        & 7708010101       & 2024-02-11 & 60351.39 & 3.1 & $6415\pm37$ & $-$  & $-$ & $-$ & $-$ \\
			
			\rowcolor{lightgray}
			& E7  & IXPE        & 03006001       & 2024-02-20 & 60360.07 & 151 & $5.59\pm0.09$ & $67\pm8$  & $26\pm22$ & $-$ & HSS \\
			
			\rowcolor{lightgray}
			&   & NICER        & 7708010106       & 2024-02-19 & 60359.07 & 0.6 & $5354\pm34$ & $-$  & $-$ & $-$ & $-$ \\
   
                \rowcolor{lightgray}
			& E8  & IXPE        &  03005801  & 2024-04-03 & 60403.66 & 202 & $7.87\pm0.11$ & $76\pm11$ & $109\pm14$ & $-$ & LHS \\
   
	        \rowcolor{lightgray}
			&   & NICER        &  7708010109  & 2024-04-03 & 60403.29 & 2.5 & $675\pm12$ & $-$  & $-$ & $-$ & $-$ \\
			
			\hline

             \rowcolor{Lightblue}
		GX $339-4$ & E1  & IXPE   &  03005101 & 2024-02-14 & 60354.93 & 95 & $17.44\pm0.21$ & $-$  & $79\pm25$ & $-$ & SIMS \\
		\rowcolor{Lightblue}
		&    & NICER       &   7702010112    & 2024-02-14  & 60354.98 &  3.8  &  $5263\pm34$ & $-$ & $-$ & $-$ & $-$ \\
            \rowcolor{Lightblue}
		&    & NuSTAR       &   91002306002    & 2024-02-14  & 	60354.70 & 16  &  $308\pm13$ & $-$ & $-$ & $-$ & $-$ \\

            \rowcolor{orange}
         &    & AstroSat  & A05\_166T01\_9000006070  & 2024-02-14 & 60354.31 & $50.7$  & $629\pm26$ & $-$ & $-$ & $-$ & $-$  \\

            & E2  & IXPE   &  03005301 & 2024-03-08 & 60377.68 & 98 & $7.37\pm0.11$ & $103\pm6$  & $12\pm8$ & $-$ & HSS \\

            \rowcolor{orange}
         &    & AstroSat$^{\dagger}$   & A13\_028T01\_9000006122  & 2024-03-10 & 60379.44 & $-$  & $-$ & $-$ & $-$ & $-$ & $-$  \\

            \hline

            \rowcolor{Lightblue}
		Swift J$151857.0-572147$ & E1  & IXPE   & 03250201  & 2024-03-18 & 60387.15 & 96 & $12.18\pm0.12$ & $422\pm8$  & $152\pm17$ & $-$ & HSS \\
		\rowcolor{Lightblue}
		&    & NICER       &   7204220111    & 2024-03-18  & 60387.55 &  4.2  &  $1468\pm18$ & $-$ & $-$ & $-$ & $-$ \\
            \rowcolor{Lightblue}
		&    & NuSTAR       &  91001311004     & 2024-03-18  & 	60387.64 & 9.2  &  $444\pm17$ & $-$ & $-$ & $-$ & $-$ \\

        \hline

            \rowcolor{magenta2}
		IGR J$17091-3624$ & E1  & IXPE   & 04250201  & 2025-03-07 & 60741.30 & 163 & $0.27\pm0.02$ & $755\pm11$  & $82\pm27$ & $-$ & LHS \\

            \rowcolor{magenta2}
		&    & NuSTAR       &  81002342008    & 2025-03-07  & 	60741.58 & 21  &  $27\pm15$ & $-$ & $-$ & $-$ & $-$ \\

            \hline

            MAXI J$1744-294$ & E1  & IXPE   & 04250301  & 2025-04-05 & 60770.09 & 149 & $0.17\pm0.11$ & $45\pm28$  & $-$ & $-$ & HSS \\

            \hline

            GRS $1915+105$ & E1  & IXPE   & 04003501  & 2025-05-10 & 60805.69 & 142 & $-$ & $6\pm3$  & $-$ & $-$ & $-$ \\

            \hline
            
		\end{tabular}
	}
	
	\label{tab:log2}
	
	\begin{list}{}{}
        \item $^{\dagger}$Data are not available in public domain. $^*${\it RATAN} ($4.7$ GHz, \citealt[]{Veledina-etal2023, Ingram-etal2024}); $^{**}${\it E-MERLIN} ($5.1$ GHz, \citealt[]{Williams-Baldwin-etal2023}).
		\end{list}
\end{table*}


\begin{table*}
	\caption{Results from the wide-band spectral analysis of ten BH-XRBs using {\it NICER} and {\it NuSTAR} data, modeled with \texttt{M1: constant$\times$Tbabs$\times$(thcomp$\otimes$diskbb)}, unless stated otherwise. Here, $n_{\rm H}$, $kT_{\rm in}$, $kT_{\rm e}$, $\Gamma_{\rm th}$, and $cov_{\rm frac}$ denote the column density, inner disc temperature, electron temperature, photon index, and covering fraction, respectively. $F_{\rm disc}$, $F_{\rm Comp}$, and $F_{\rm bol}$ represent the disc, Comptonized, and total bolometric fluxes, respectively. $L_{\rm bol}$ indicates the bolometric luminosity. $rms_{\rm tot}$ corresponds to the rms amplitude derived from the respective PDS. The energy ranges used for each spectrum and the associated spectral states are also mentioned. See text for details.}
	
	\renewcommand{\arraystretch}{1.7}
	
	\label{table:spec_para}
	
	\resizebox{1.0\textwidth}{!}{%
		\begin{tabular}{l @{\hspace{0.2cm}} l @{\hspace{0.2cm}} c @{\hspace{0.2cm}} c @{\hspace{0.2cm}} c @{\hspace{0.4cm}} c @{\hspace{0.4cm}} c @{\hspace{0.4cm}} c @{\hspace{0.4cm}} c @{\hspace{0.1cm}} c @{\hspace{0.1cm}} c @{\hspace{0.3cm}} c @{\hspace{0.3cm}} c @{\hspace{0.3cm}} c @{\hspace{0.3cm}} c @{\hspace{0.3cm}} c @{\hspace{0.3cm}} c}
			\hline

			Source & Epoch & $n_{\rm H}$ & $kT_{\rm in}$ & $kT_{\rm e}$ & $\Gamma_{\rm th}$ & $cov_{\rm frac}$ & $\chi^2/d.o.f$ & $F_{\rm disc} ^{\boxtimes}$ & $F_{\rm Comp}^{\boxtimes}$ & $F_{\rm bol}^{\boxtimes}$ & $L_{\rm bol}^{\boxtimes}$ & Spectral & Energy & $rms_{\rm tot}^{\dagger}$\\
			
			& & ($10^{22}$ ${\rm cm}^{-2}$) & (keV) & (keV) & &  &  & & ($\tiny{10^{-8}}$ erg ${\rm cm}^{-2}$ ${\rm s}^{-1}$) & & ($\% L_{\rm Edd}$) & State  & range (keV) & ($\%$) \\
			
			\hline
			
		Cyg X$-1$ &  E1 & $0.44_{-0.02}^{+0.02}$ & $0.41_{-0.03}^{+0.03}$ & $30.35_{-2.22}^{+2.87}$ & $1.61_{-0.01}^{+0.02}$ & $0.77_{-0.02}^{+0.02}$ & $1517/1468$ & $0.68$ & $2.45$ & $3.07$ & $1.4$ & LHS  & $0.7-60$ & $32.5$ \\

		($d=2.2$ kpc) & E2 & $0.34_{-0.02}^{-0.02}$ & $0.38_{-0.01}^{+0.01}$ & $22.57_{-1.13}^{+1.50}$ & $1.61_{-0.01}^{+0.01}$ & $0.73_{-0.03}^{+0.04}$ & $1479/1403$ & $0.74$ & $2.53$ & $3.29$ & $1.5$ & LHS  & $0.7-60$ & $31$ \\

         &  E6$^{\boxdot}$ & $0.51^{*}$ & $0.52_{-0.02}^{+0.02}$ & $10^{*}$ & $1.01^{*}$ & $0.005_{-0.002}^{+0.002}$ & $2122/1913$ & $1.75$ & $0.22$ & $4.43$ & $2$ & HSS & $3-60$ & $-$ \\

		& E7$^{\boxdot}$ & $0.52_{-0.01}^{+0.01}$ & $0.43_{-0.01}^{+0.02}$ & $10^{*}$ & $3.21_{-0.14}^{+0.11}$ & $0.31_{-0.02}^{+0.03}$ & $2185/2167$ & $3.78$ & $0.33$ & $6.18$ & $2.8$ & HSS  & $0.7-60$ & $9.1$ \\

            & E8 & $0.43_{-0.03}^{+0.03}$ & $0.27_{-0.03}^{+0.03}$ & $20^{*}$ & $1.64_{-0.02}^{+0.02}$ & $0.65_{-0.07}^{+0.06}$ & $115/133$ & $0.37$ & $1.15$ & $1.53$ & $0.7$ & LHS & $0.7-10$ & $41.5$ \\

            & E11 & $0.39_{-0.04}^{+0.02}$ & $0.32_{-0.04}^{+0.03}$ & $20^{*}$ & $1.62_{-0.03}^{+0.03}$ & $0.61_{-0.04}^{+0.04}$ & $152/131$ & $0.39$ & $1.18$ & $1.58$ & $0.7$ & LHS & $0.7-10$ & $52.4$ \\
			
			\hline
			
			4U 1630$-47$ & E1 & $6.69_{-0.30}^{+0.24}$ & $1.36_{-0.01}^{+0.01}$ & $20^{*}$ & $3.34_{-0.11}^{+0.12}$ & $0.052_{-0.007}^{+0.008}$ & $1134/928$ & $1.56$ & $0.03$ & $1.59$ & $15.1$ & HSS & $0.7-40$ & $6.7$ \\

			($d=10$ kpc) & E2 & $4.55_{-0.05}^{+0.07}$ & $1.46_{-0.02}^{+0.02}$ & $20^{*}$ & $2.47_{-0.04}^{+0.05}$ & $0.38_{-0.01}^{+0.02}$ & $2366/1980$ & $2.76$ & $0.77$ & $3.65$ & $34.7$ & SPL  & $0.7-60$ & $4.8$ \\
			
			\hline
			
			Cyg X$-3$ & E1 & $4.04_{-0.11}^{+0.26}$ & $1.21_{-0.02}^{+0.02}$ & $6.14_{-0.34}^{+0.28}$ & $1.55_{-0.04}^{+0.03}$ & $0.83_{-0.01}^{+0.02}$ & $2921/1842$ & $0.21$ & $0.41$ & $0.64$ & $5.7$ & LHS  & $3-50$ & $-$ \\

            ($d=9.7$ kpc) & E2 & $5.54_{-0.13}^{+0.07}$ & $1.01_{-0.02}^{+0.01}$ & $7.95_{-0.14}^{+0.22}$ & $2.21_{-0.01}^{+0.01}$ & $0.63_{-0.01}^{+0.01}$ & $2709/1903$ & $0.94$ & $0.62$ & $1.58$ & $14.1$ & HSS & $3-50$ & $-$ \\

			\hline
			
			LMC X$-1$ & E1 & $0.50_{-0.01}^{+0.01}$ & $0.98_{-0.01}^{+0.01}$ & $10^{*}$ & $2.58_{-0.09}^{+0.10}$ & $0.091_{-0.010}^{+0.012}$ & $744/613$ & $0.065$ & $0.004$ & $0.074$ & $16$ & HSS  & $0.8-30$ & $10.1$ \\
			
			($d=48.1$ kpc)& & & & & & & & & & & &  & &  \\
			
			\hline
			
			4U 1957$+115$ & E1 & $0.08_{-0.01}^{+0.01}$ & $1.40_{-0.01}^{+0.01}$ & $10^{*}$ & $1.86_{-0.10}^{+0.12}$ & $0.027_{-0.004}^{+0.004}$ & $728/701$ & $0.092$ & $0.005$ & $0.096$ & $0.9$ & HSS & $0.7-40$ & $5$ \\
			
			($d=10$ kpc)& & & & & & & & & & & & &  &  \\
			
			\hline
			
			LMC X$-3$ & E1 & $0.04^{*}$ & $1.10_{-0.01}^{+0.01}$ & $10^{*}$ & $1.78_{-0.35}^{+0.45}$ & $0.005_{-0.002}^{+0.003}$ & $567/543$ & $0.061$ & $0.001$ & $0.062$ & $13.4$ & HSS & $0.7-20$ & $-$ \\

            ($d=48.1$ kpc) & E2 & $0.04^{*}$ & $1.09_{-0.01}^{+0.01}$ & $10^{*}$ & $2.51_{-0.07}^{+0.08}$ & $0.3^{*}$ & $153/128$ & $0.11$ & $0.03$ & $0.14$ & $30.8$ & HSS & $1-10$ & $-$ \\

            & E3 & $0.04^{*}$ & $0.6_{-0.03}^{+0.04}$ & $1.04_{-0.05}^{+0.08}$ & $1.72_{-0.08}^{+0.11}$ & $0.4^{*}$ & $118/117$ & $0.03$ & $0.006$ & $0.035$ & $7.7$ & HSS & $1-10$ & $-$ \\

            & E4 & $0.04^{*}$ & $1.07_{-0.01}^{+0.01}$ & $10^{*}$ & $2.05_{-0.04}^{+0.05}$ & $0.005_{-0.003}^{+0.003}$ & $771/706$ & $0.06$ & $0.006$ & $0.07$ & $15.4$ & HSS & $3-30$ & $-$ \\
			
			\hline

        Swift J1727.8$-1613$ & E1 & $0.23_{-0.01}^{+0.01}$ & $0.29_{-0.02}^{+0.01}$ & $8.44_{-0.11}^{+0.16}$ & $1.98_{-0.02}^{+0.04}$ & $0.74_{-0.01}^{+0.01}$ & $1181/931$ & $10.07$ & $15.01$ & $25.24$ &  $17.5$ &  HIMS & $0.7-40$ & $12$ \\

        ($d=2.7$ kpc) & E2 & $0.20_{-0.01}^{+0.01}$ & $0.43_{-0.02}^{+0.02}$ & $9.34_{-0.23}^{+0.22}$ & $2.11_{-0.04}^{+0.05}$ & $0.82_{-0.02}^{+0.02}$ & $912/847$ & $11.35$ & $13.58$ & $25.11$ & $17.4$ & HIMS  & $0.7-40$ & $-$ \\

        & E4$^{\boxdot}$ & $0.25_{-0.01}^{+0.01}$ & $0.85_{-0.02}^{+0.03}$ & $10^{*}$ & $2.74_{-0.04}^{+0.04}$ & $0.38_{-0.03}^{+0.04}$ & $1373/1447$ & $15.89$ & $3.63$ & $25.04$ & $17.3$ & SIMS & $0.7-40$ & $6.1$ \\

	& E5$^{\boxdot}$ & $0.29_{-0.01}^{+0.02}$ & $0.88_{-0.02}^{+0.02}$ & $10^{*}$ & $2.32_{-0.04}^{+0.06}$ & $0.19_{-0.02}^{+0.02}$ & $1572/1601$ & $10.12$ & $2.09$ & $16.78$ & $11.6$ & SIMS & $0.7-60$ & $6.4$ \\

	& E6 & $0.17_{-0.01}^{+0.01}$ & $0.48_{-0.01}^{+0.01}$ & $10^{*}$ & $5.14_{-0.39}^{+0.36}$ & $0.17_{-0.05}^{+0.03}$ & $127/132$ & $0.91$ & $0.02$ & $0.93$ & $0.6$ & HSS & $0.7-10$ & $1.4$ \\

		& E7 & $0.18_{-0.01}^{+0.01}$ & $0.45_{-0.01}^{+0.01}$ & $10^{*}$ & $4.57_{-0.60}^{+0.65}$ & $0.13_{-0.05}^{+0.09}$ & $102/111$ & $0.66$ & $0.01$ & $0.67$ & $0.5$ & HSS & $0.7-10$ & $1.3$ \\

            & E8 & $0.12^{*}$ & $0.24_{-0.01}^{+0.01}$ & $20^{*}$ & $1.71_{-0.02}^{+0.03}$ & $0.35_{-0.01}^{+0.01}$ & $126/130$ & $0.08$ & $0.12$ & $0.19$ & $0.1$ & LHS & $0.7-10$ & $30.2$ \\
			
			\hline

        GX $339-4$ & E1$^{\boxdot}$ & $0.51_{-0.01}^{+0.01}$ & $0.60_{-0.02}^{+0.02}$ & $0.97_{-0.01}^{+0.02}$ & $2.02_{-0.52}^{+0.44}$ & $0.34_{-0.05}^{+0.08}$ & $1603/1538$ & $0.84$ & $0.36$ & $1.71$ & $11.5$ & SIMS & $0.7-50$ & $14.2$ \\

        ($d=8.4$ kpc)& & & & & & & & & & & &  &  & \\

        \hline

        Swift J$151857.0-572147$ & E1 & $3.83_{-0.01}^{+0.02}$ & $0.94_{-0.01}^{+0.01}$ & $20^{*}$ & $2.50_{-0.01}^{+0.01}$ & $0.21_{-0.01}^{+0.01}$ & $1572/1367$ & $2.11$ & $0.34$ & $2.71$ & $8.7$ & HSS & $0.7-50$ & $12.6$ \\

        ($d=5.8$ kpc)& & & & & & & & & & & &  &  & \\

        \hline

         IGR J$17091-3624$ & E1 & $1.1^{*}$ & $0.88_{-0.38}^{+0.27}$ & $23.54_{-2.91}^{+4.01}$ & $1.62_{-0.01}^{+0.02}$ & $0.87_{-0.03}^{+0.03}$ & $1465/1486$ & $0.02$ & $0.13$  & $0.15$ & $1.7$ & LHS & $3-60$ & $24.3$ \\

         ($d=11$ kpc)& & & & & & & & & & & &  &  & \\

        \hline
		\end{tabular}%
	}
	\begin{list}{}{}
		\item $L_{\rm Edd} = 1.26 \times 10^{39}$ erg $\rm s^{-1}$ for $10 M_\odot$ BH. $^{\boxtimes}$Calculated in $1-100$ keV energy range. $^{\dagger}$Computed in $0.1-50$ Hz using {\it NICER} observations in $0.5-10$ keV energy range. $^{*}$Frozen values. $^{\boxdot}$Results are obtained using model \texttt{M3}.
		
	\end{list}
\end{table*}

\begin{table*}
	\caption{Best-fitted parameters of the reflection model component \texttt{relxill} used in the model combination \texttt{M3: constant$\times$Tbabs$\times$(thcomp$\otimes$diskbb + relxill)}. The parameters of other components in \texttt{M3} are listed in Table \ref{table:spec_para} for the respective epochs. Here, $\Gamma$ is the power-law index of the source spectrum, $A_{\rm fe}$ is the iron abundance in solar units, $\log \xi$ is the ionization of the accretion disc, and $R_{\rm f}$ is the reflection fraction parameter. $F_{\rm relxill}$ is the flux associated with the \texttt{relxill} model component. The references for the spin and inclination of the sources are mentioned in the footnote. See text for details.}
	
	\renewcommand{\arraystretch}{1.7}

	\resizebox{1.0\textwidth}{!}{%
		\begin{tabular}{l @{\hspace{0.4cm}} c @{\hspace{0.2cm}} c @{\hspace{0.2cm}} c @{\hspace{0.2cm}} c @{\hspace{0.4cm}} c @{\hspace{0.4cm}} c @{\hspace{0.4cm}} c @{\hspace{0.4cm}} c @{\hspace{0.4cm}} c }
			\hline

			Source & Epoch & Spin & Inclination ($i$) & $\Gamma$ & $A_{\rm fe}$ & $\log \xi$ & $R_{\rm f}$ & $F_{\rm relxill}^{\boxdot}$ & Spectral \\
			
			& & ($a_{*}$) & ($^{\circ}$) &  & ($A_\odot$) & ($\log$ erg ${\rm cm}$ ${\rm s}^{-1}$) & & ($\tiny{10^{-8}}$ erg ${\rm cm}^{-2}$ ${\rm s}^{-1}$) & State \\
			
			\hline

            $^{a}$Cyg X$-1$ & E6 & $0.98^{*}$ & $27^{*}$  & $2.32_{-0.06}^{+0.04}$ & $6.35_{-1.46}^{+2.08}$ & $3.86_{-0.08}^{+0.09}$ & $0.67_{-0.04}^{+0.04}$ & $2.33$ & HSS \\

            & E7 & $0.98^{*}$ & $27^{*}$ & $2.04_{-0.01}^{+0.01}$ & $4.61_{-0.31}^{+0.36}$ & $3.31_{-0.05}^{+0.05}$ & $1.11_{-0.03}^{+0.03}$ & $2.14$ & HSS \\

            $^{b}$GX $339-4$ & E1 & $0.85^{*}$ & $50^{*}$ & $2.27_{-0.03}^{+0.03}$ & $0.5^{\boxtimes}$ & $3.92_{-0.05}^{+0.12}$ & $3.16_{-0.33}^{+0.35}$ & $0.54$ & SIMS \\

            $^{c}$Swift J$1727.8-1613$ & E4 & $0.98^{*}$ & $40^{*}$ & $2.19_{-0.02}^{+0.03}$ & $5.02^{\boxtimes}$ & $4.11_{-0.13}^{+0.12}$ & $0.68_{-0.14}^{+0.17}$ & $5.44$ & SIMS \\

           & E5 & $0.98^{*}$ & $40^{*}$ & $2.48_{-0.03}^{+0.02}$ & $5.1^{*}$ & $4.29_{-0.21}^{+0.14}$ & $0.59_{-0.11}^{+0.05}$ & $4.38$ & SIMS \\

            \hline

		\end{tabular}%
	}
	\begin{list}{}{}
		\item $^{*}$Frozen values. $^{\boxtimes}$Fixed to the best-fitted values. $^{\boxdot}$Calculated in $1-100$ keV energy range. $A_\odot$ is the solar iron abundance. $r_{\rm g} = GM/\rm c^{2}$. $^{a}$\cite{Kushwaha-etal2021}, $^{b}$\cite{Mastroserio-etal2024}, $^{c}$\cite{Peng-etal2024a}.
		
	\end{list}

    \label{tab:reflect}
\end{table*}

\begin{table*}
	\centering
	\caption{Model-independent polarimetric results obtained using the \texttt{PCUBE} algorithm in the $2-8$ keV energy band from \textit{IXPE} observations of eleven BH-XRBs. PD and PA represent the polarization degree and angle, respectively. Q$/$I and U$/$I are the normalized Stokes parameters. MDP denotes the minimum detectable polarization, and SIGNIF is the detection significance in units of $\sigma$. PD$_{\rm E}$ represents the significance of the energy variation of PD in units of $\sigma$. Gray-shaded entries indicate null-detection of polarization. See text for details.}
	\renewcommand{\arraystretch}{1.2}
	\resizebox{1.0\textwidth}{!}{%
		\begin{tabular}{l @{\hspace{0.2cm}} c @{\hspace{0.2cm}} c @{\hspace{0.2cm}} c @{\hspace{0.2cm}} c @{\hspace{0.2cm}} c @{\hspace{0.2cm}} c @{\hspace{0.2cm}} c @{\hspace{0.2cm}} c @{\hspace{0.2cm}} c @{\hspace{0.2cm}} c @{\hspace{0.2cm}} c}
			\hline
			Source & Epoch & Obs. ID & PD & PA & Q$/$I & U$/$I & $\rm MDP_{99}$ & SIGNIF & Spectral & Remarks & PD$_{\rm E}$$^{\boxtimes}$ \\
			
			&  &  & (\%) & ($^{\circ}$) & (\%) & (\%) & (\%) & ($\sigma$) & State & & Sig. ($\sigma$) \\
			
			\hline
			
			Cyg X$-1$ & E1 & 01002901 & $3.70\pm0.19$ & $-20.75\pm1.50$ & $2.77\pm0.19$ & $-2.45\pm0.19$ & $0.59$ & $19$ & LHS & Detection & $2.8$ \\
			
			& E2 & 01250101 & $3.79\pm0.32$ & $-25.54\pm2.39$ & $2.38\pm0.32$ & $-2.95\pm0.32$ & $0.96$ & $12$  & LHS  & Detection & $< 1$\\
			
			& E3 & 02008201 & $2.50\pm0.40$ & $-19.23\pm4.63$ & $1.95\pm0.40$ & $-1.55\pm0.40$ & $1.22$ & $5.7$  & HSS &  Detection & $< 1$ \\
			
			& E4 & 02008301 & $2.46\pm0.30$ & $-22.89\pm3.45$ & $1.71\pm0.30$ & $-1.76\pm0.30$ & $0.90$ & $8$  & HSS  & Detection & $3$\\
			
			& E5 & 02008401 & $2.01\pm0.29$ & $-25.82\pm4.11$ & $1.25\pm0.29$ & $-1.58\pm0.29$ & $0.88$ & $6.6$  & HSS & Detection  & $> 4$ \\
			
			& E6 & 02008501 & $1.44\pm0.28$ & $-25.38\pm5.66$ & $0.91\pm0.28$ & $-1.12\pm0.28$ & $0.86$ & $4.5$  & HSS & Detection & $1.2$  \\
			
			& E7 & 02008601 & $2.07\pm0.23$ & $-36.37\pm3.19$ & $0.61\pm0.23$ & $-1.98\pm0.23$ & $0.70$ & $9$  &  HSS  & Detection & $1.1$ \\

                &  E8 & 03002201 & $3.75\pm0.47$ & $-24.22\pm3.61$ & $2.49\pm0.47$ & $-2.80\pm0.47$ & $1.43$ & $7.6$  &  LHS  & Detection & $1.8$ \\

                &  E9 & 03003101 & $3.04\pm0.48$ & $-18.35\pm4.54$ & $2.44\pm0.48$ & $-1.82\pm0.48$ & $1.46$ & $5.9$ &  LHS  &  Detection & $< 1$ \\

                &  E10 &  03010001 & $4.65\pm0.46$ & $-27.77\pm2.85$ & $2.63\pm0.46$ & $-3.83\pm0.46$ & $1.4$ & $10.1$  &  LHS  & Detection & $< 1$ \\

                &  E11 &  03010101 & $4.76\pm0.53$ & $-32.64\pm3.18$ & $1.99\pm0.53$ & $-4.32\pm0.53$ & $1.60$ & $9$ &  LHS  & Detection & $1.5$ \\

                &  E12 &  03002599 & $2.79\pm0.19$ & $-21.97\pm1.97$ & $2.01\pm0.19$ & $-1.94\pm0.19$ & $0.58$ & $14.5$ &  HSS  & Detection & $> 4$ \\
			
			&&&&&&&& \\
			
			4U 1630$-47$ & E1 & 01250401 & $8.34\pm0.17$ & $17.80\pm0.60$ & $6.79\pm0.17$ & $4.86\pm0.17$ & $0.53$ & $48$  & HSS  & Detection & $> 4$ \\
			
			& E2 & 02250601 & $6.77\pm0.21$ & $21.36\pm0.90$ & $4.97\pm0.21$ & $4.59\pm0.21$ & $0.65$ & $31.7$  & SPL  & Detection & $> 4$ \\
			
			&&&&&&&& \\
			
			Cyg X$-3$ & E1 & 02001899 & $20.60\pm0.31$ & $-89.79\pm0.43$ & $-20.60\pm0.31$ & $-0.15\pm0.31$ & $0.94$ & $66.3$  & LHS  & Detection & $> 4$ \\
			
			& E2 & 02250301 & $10.58\pm0.28$ & $-87.39\pm0.75$ & $-10.54\pm0.28$ & $-0.96\pm0.28$ & $0.85$ & $38$  &  HSS  & Detection & $> 4$ \\
			
			& E3 & 02009101 & $21.41\pm0.41$ & $-88.09\pm0.55$ & $-21.36\pm0.41$ & $-1.43\pm0.41$ & $1.24$ & $52.7$  & SIMS  & Detection & $> 4$ \\

            &  E4 &  03250301 & $11.98\pm0.43$ & $-85.22\pm1.03$ & $-11.82\pm0.43$ & $-1.99\pm0.43$ & $1.31$ & $27.8$  &  HSS & Detection & $> 4$ \\
			
			&&&&&&&& \\
			
			\rowcolor{lightgray}
			LMC X$-1$ & E1 & 02001901 & $1.04\pm0.40$ & $53.97\pm11.09$ & $-0.32\pm0.40$ & $0.99\pm0.40$ & $1.23$ & $1.8$ & HSS  &  Null-detection & $-$ \\

			&&&&&&&& \\
			
			4U 1957$+115$ & E1 & 02006601 & $1.85\pm0.37$ & $-42.09\pm5.75$ & $0.19\pm0.37$ & $-1.84\pm0.37$ & $1.13$ & $4.5$  & HSS & Detection  & $1.7$ \\
			
			&&&&&&&& \\
			
			LMC X$-3$ & E1 & 02006599 & $3.04\pm0.40$ & $-44.24\pm3.77$ & $0.08\pm0.40$ & $-3.04\pm0.40$ & $1.21$ & $7.2$  & HSS  & Detection  & $1.8$ \\

             & E2 & 03004899 & $2.39\pm0.37$ & $-38.43\pm4.45$ & $0.54\pm0.37$ & $-2.33\pm0.37$ & $1.13$ & $6$  & HSS  & Detection  & $2.3$ \\

             \rowcolor{lightgray}
             & E3 & 03004901 & $2.19\pm0.77$ & $-39.42\pm10.01$ & $0.42\pm0.77$ & $-2.15\pm0.77$ & $2.32$ & $2.4$ & HSS & Null-detection  &  $-$\\

             & E4 & 03005001 & $2.84\pm0.52$ & $-47.32\pm5.24$ & $-0.23\pm0.52$ & $-2.83\pm0.52$ & $1.57$ & $5$ & HSS & Detection  &  $<1$\\
			
			&&&&&&&& \\
			
			Swift J1727.8$-1613$ & E1 & 02250901 & $4.72\pm0.28$ & $2.50\pm1.69$ & $4.71\pm0.28$ & $0.41\pm0.28$ & $0.84$ & $17$  & HIMS  & Detection & $< 1$ \\
			
			& E2 & 02251001 & $4.48\pm0.20$ & $2.28\pm1.31$ & $4.46\pm0.20$ & $0.36\pm0.20$ & $0.62$ & $22$  & HIMS  & Detection  & $2.8$ \\
			
			& E3 & 02251101 & $4.39\pm0.29$ & $0.95\pm1.87$ & $4.38\pm0.29$ & $0.15\pm0.29$ & $0.87$ & $15.3$  & HIMS  & Detection  & $2.6$ \\
			
			& E4 & 02251201 & $3.81\pm0.29$ & $-0.49\pm2.22$ & $3.81\pm0.29$ & $-0.07\pm0.29$ & $0.89$ & $12.8$  & SIMS  & Detection & $1.1$ \\
			
			& E5 & 02251301 & $3.35\pm0.32$ & $-2.01\pm2.75$ & $3.34\pm0.32$ & $-0.23\pm0.32$ & $0.98$ & $10.4$  & SIMS  & Detection & $2.5$ \\
			
			\rowcolor{lightgray}
			& E6 & 03005701 & $1.18\pm0.44$ & $-2.74\pm10.55$ & $1.18\pm0.44$ & $-0.11\pm0.44$ & $1.32$ & $2$  & HSS  & Null-detection & $-$ \\
			
			\rowcolor{lightgray}
			& E7 & 03006001 & $0.40\pm0.36$ & $6.05\pm25.83$ & $0.39\pm0.36$ & $0.08\pm0.36$ & $1.08$ & $<1$  & HSS  & Null-detection & $-$ \\

                & E8 & 03005801 & $3.22\pm0.60$ & $4.18\pm5.37$ & $3.19\pm0.60$ & $0.47\pm0.60$ & $1.83$ & $4.8$  & LHS  & Detection & $1.6$ \\

            &&&&&&&& \\

            GX $339-4$ & E$1^{\dagger}$ & 03005101 & $1.22\pm0.35$ & $-71.03\pm8.17$ & $-0.96\pm0.35$ & $-0.75\pm0.35$ & $1.05$ & $3.1$  & SIMS  & Detection  & $3.3$ \\

            \rowcolor{lightgray}
            & E2 & 03005301 & $0.47\pm0.36$ & $-25.15\pm22.07$ & $0.30\pm0.36$ & $-0.36\pm0.36$ & $1.09$ & $0.8$  & HSS  & Null-detection & $-$ \\

            &&&&&&&& \\

            \rowcolor{lightgray}
            Swift J$151857.0-572147$ & E1 & 03250201 & $0.25\pm0.26$ & $-22.35\pm29.98$ & $0.17\pm0.26$ & $-0.17\pm0.26$ & $0.78$ & $0.5$  & HSS  &  Null-detection & $-$ \\

            &&&&&&&& \\

            IGR J$17091-3624$ & E$1$ & 04250201 & $9\pm1.41$ & $83.71\pm4.48$ & $-8.78\pm1.41$ & $1.96\pm1.41$ & $4.27$ & $6.1$  & LHS  & Detection & $1.2$ \\

            &&&&&&&& \\

            \rowcolor{lightgray}
            MAXI J$1744-294$ & E$1$ & 04250301 & $0.71\pm0.42$ & $-15.37\pm17.21$ & $-0.60\pm0.42$ & $-0.36\pm0.42$ & $1.27$ & $1.2$  & HSS  & Null-Detection & $-$ \\
            
            \hline
			
		\end{tabular}
	}

    \begin{list}{}{}
		\item $^{\dagger}$Measurements in $3-8$ keV. Null-detection in $2-8$ keV. $^{\boxtimes}$Computed considering $6$ linear energy bins for all sources except GX $339-4$, 4U 1957$+115$ and IGR J$17091-3624$, where $4$ linear bins are used.
\end{list}
 
	\label{tab:en-pol}
\end{table*}


\begin{table*}
	\centering
	\caption{Results from the spectro-polarimetric modeling of \textit{IXPE} Stokes spectra across different observation epochs of the sources in $2-8$ keV energy band. The parameters, with their standard definitions, are obtained from the constant polarization model (\texttt{polconst}) and the energy-dependent polarization model (\texttt{polpow}). See text for details.}
	\renewcommand{\arraystretch}{1.5}
	\resizebox{1.0\textwidth}{!}{%
		\begin{tabular}{l c c c c c c c c c c}
			\hline
			
			Source & Epoch & $A_{\rm norm}$  & $A_{\rm index}$ & $\psi_{\rm norm}$ & $\chi^{2}_{\rm red}$ & $\rm PD_{polpow}$ & $\rm PA_{polpow}$ & $\rm PD_{polconst}$ & $\rm PA_{polconst}$ & $\chi^{2}_{\rm red}$ \\
			
			& & & &  ($^\circ$) & (polpow) & ($\%$) & ($^\circ$) & ($\%$) & ($^\circ$) & (polconst) \\ \hline
			
			Cyg X$-1$$^{a}$ & E1 & $0.020_{-0.006}^{+0.008}$ & $-0.43 \pm 0.25$ & $-19.83 \pm 2.22$ & $1.01$ &  $3.93 \pm 0.95$ & $-19.83 \pm 2.22$ & $3.40 \pm 0.26$ & $-19.73 \pm 2.24$ & $0.91$ \\ 
			
			& E2 & $-$ & $-$ & $-$ & $-$ & $-$ & $-$ & $3.41 \pm 0.44$ & $-25.13 \pm 3.73$ & $1.09$ \\ 
			
			& E3 & $-$ & $-$ & $-$ & $-$ & $-$ & $-$ & $2.24 \pm 0.61$ & $-11.87 \pm 7.95$ & $1.15$ \\
			
			& E4 & $0.008_{-0.004}^{+0.005}$ & $-0.90 \pm 0.56$ & $-17.74 \pm 5.45$ & $1.13$ &  $3.39 \pm 0.40$ & $-17.74 \pm 5.45$ & $2.29 \pm 0.45$ & $-18.33 \pm 5.68$ & $1.13$ \\

			& E5 & $0.003_{-0.002}^{+0.003}$ & $-1.57 \pm 0.57$ & $-8.63 \pm 6.16$ & $1.10$ &  $3.95 \pm 0.31$ & $-8.63 \pm 6.16$ & $1.91 \pm 0.47$ & $-9.48 \pm 7.18$ & $1.10$ \\

			& E6 & $0.003 \pm 0.002$ & $-1.52 \pm 0.68$ & $-16.26 \pm 5.74$ & $1.12$ &  $3.63 \pm 0.47$ & $-16.26 \pm 5.74$ & $2.06 \pm 0.45$ & $-16.64 \pm 6.27$ & $1.13$ \\

		& E7 & $0.009 \pm 0.005$ & $-0.80 \pm 0.47$ & $-25.03 \pm 4.56$ & $1.11$ &  $3.23 \pm 0.99$ & $-25.03 \pm 4.56$ & $2.19 \pm 0.37$ & $-25.77 \pm 4.69$ & $1.11$ \\

            & E8$^{b}$ & $-$ & $-$ & $-$ & $-$ &  $-$ & $-$ & $3.04\pm 0.66$ & $-27.09\pm6.24$ & $1.07$ \\

            & E9$^{b}$ & $-$ & $-$ & $-$ & $-$ &  $-$ & $-$ & $3.35\pm0.67$ & $-25.73\pm5.79$ & $1.25$ \\

            & E10$^{b}$ & $-$ & $-$ & $-$ & $-$ &  $-$ & $-$ & $5.83\pm0.64$ & $-32.72\pm3.15$ & $0.96$ \\

            & E11$^{b}$ & $-$ & $-$ & $-$ & $-$ &  $-$ & $-$ & $4\pm 0.7$ & $-34.98\pm 5.27$ & $1.05$ \\

            & E12 & $0.009_{-0.004}^{+0.005}$ & $-0.9_{-0.4}^{+0.4}$ & $-24.1\pm3.26$ & $1.03$ &  $3.38\pm0.48$ & $-24.1\pm3.26$ & $2.52\pm 0.48$ & $-23.03\pm3.34$ & $1.05$ \\
			
			\hline
			
			4U 1630$-47$ & E1$^{c}$ & $0.028 \pm 0.003$ & $-0.75 \pm 0.07$ & $17.90 \pm 0.52$ & $1.28$ &  $9.25 \pm 0.02$ & $17.90 \pm 0.52$ & $7.89 \pm 0.14$ & $17.94 \pm 0.53$ & $1.36$ \\ 
			& E2 & $0.022 \pm 0.004$ & $-0.74 \pm 0.11$ & $21.45 \pm 0.77$ & $1.13$ & $8.02 \pm 0.02$ & $21.45 \pm 0.77$ & $6.27 \pm 0.17$ & $21.30 \pm 0.78$ & $1.17$ \\
			\hline

                Cyg X$-3$$^{a}$  & E1$^{c}$ & $0.028\pm0.004$ & $-0.73\pm0.09$ & $-88.92\pm0.93$ & $1.46$ & $8.89 \pm 0.11$ & $-88.92\pm0.93$ & $7.09 \pm 0.24$ & $-89.13 \pm 0.95$ & $1.79$ \\

                & E2$^{c}$ & $0.018\pm0.002$ & $-0.58\pm0.11$ & $-88.48\pm0.97$ & $1.31$ & $4.43 \pm 0.16$ & $-88.48\pm0.97$ & $3.58 \pm 0.21$ & $-88.58 \pm 1.01$ & $1.36$ \\

                & E3$^{c}$ & $0.024\pm0.005$ & $-0.87\pm0.12$ & $-86.98\pm3.45$ & $1.32$ & $9.75\pm0.21$ & $-86.98\pm3.45$ & $7.33 \pm 0.31$ & $-87.84 \pm 1.19$ & $1.63$ \\

                & E4$^{c}$ & $0.11\pm0.03$ & $-0.17\pm0.11$ & $-86.35\pm1.34$ & $1.31$ & $14.33\pm0.44$ & $-86.35\pm1.34$ & $13.06 \pm 1.01$ & $-86.29 \pm 2.19$ & $1.31$ \\

                \hline
			4U 1957$+115$ & E1 & $-$ & $-$ & $-$ & $-$ & $-$ & $-$ & $1.83 \pm 0.33$ & $-43.73 \pm 5.19$ & $1.09$ \\
			\hline
			LMC X$-3$ & E1 & $0.011_{-0.006}^{+0.007}$ & $-0.82 \pm 0.5$ & $-44.29 \pm 3.59$ & $1.12$ & $4.08 \pm 0.26$ & $-44.29 \pm 3.59$ & $2.94 \pm 0.37$ & $-44.94 \pm 3.63$ & $1.12$ \\

            & E2 & $0.003_{-0.001}^{+0.002}$ & $-1.67 \pm 0.51$ & $-40.47 \pm 4.02$ & $1.07$ & $4.71 \pm 0.58$ & $-40.47 \pm 4.02$ & $2.24\pm0.56$ & $-38.21\pm7.17$ & $1.08$ \\

            & E4 & $-$ & $-$ & $-$ & $-$ & $-$ & $-$ & $2.85\pm0.83$ & $-44.18\pm8.08$ & $1.03$ \\
			
			\hline
			
			Swift J1727.8$-1613$$^{a}$ & E1 & $0.022_{-0.009}^{+0.015}$ & $-0.42 \pm 0.36$ & $2.21 \pm 2.74$ & $1.08$ & $4.26 \pm 0.21$ & $2.21 \pm 2.74$ & $3.94\pm0.38$ & $2.02\pm2.75$ & $1.08$ \\

			& E2 & $0.018_{-0.006}^{+0.008}$ & $-0.55 \pm 0.26$ & $1.53 \pm 2.08$ & $1.38$ & $4.29 \pm 0.85$ & $1.53 \pm 2.08$ & $3.85 \pm 0.28$ & $1.41 \pm 2.10$ & $1.39$ \\
			
			& E3 & $0.021_{-0.010}^{+0.018}$ & $-0.37 \pm 0.23$ & $-3.55 \pm 3.24$ & $1.04$ & $3.75 \pm 0.88$ & $-3.55 \pm 3.24$ & $3.53 \pm 0.39$ & $-3.63 \pm 3.25$ & $1.04$ \\
			
			& E4 & $0.007_{-0.004}^{+0.005}$ & $-1.08 \pm 0.54$ & $-1.21 \pm 3.76$ & $1.05$ & $4.00 \pm 0.36$ & $-1.21 \pm 3.76$ & $3.11 \pm 0.41$ & $*$ & $1.06$ \\
			
			& E5 & $0.004_{-0.002}^{+0.003}$ & $-1.49 \pm 0.54$ & $-7.21 \pm 4.11$ & $0.97$ & $4.59 \pm 0.41$ & $-7.21 \pm 4.11$ & $2.94 \pm 0.45$ & $-8.38 \pm 4.44$ & $0.98$ \\

                & E8 & $0.006_{-0.003}^{+0.005}$ & $-1.28 \pm 0.47$ & $0.76 \pm 4.21$ & $0.97$ & $4.81 \pm 0.17$ & $0.76 \pm 4.21$ & $3.22 \pm 0.51$ & $2.68 \pm 4.54$ & $0.97$ \\
                			
			\hline

			GX $339-4$$^{d}$ & E1 & $-$ & $-$ & $-$ & $-$ & $-$ & $-$ & $1.16\pm0.31$ & $-72.52\pm7.78$ & $1.03$ \\
			
			\hline

            IGR J$17091-3624$$^{b}$ & E1 & $0.025_{-0.013}^{+0.016}$ & $-0.89\pm0.43$ & $85.11\pm3.88$ & $1.04$ & $10.41\pm0.51$ & $85.11\pm3.88$ & $8.24\pm1.16$ & $85.15\pm4.03$ & $1.05$ \\

            \hline

		\end{tabular}
	}
	
	\label{tab:pol_para}
	
	\begin{list}{}{}
		\item $^{a}$Only {\it IXPE} spectra of DU1 is fitted. $^*$Not constrained. $^{b}$\texttt{diskbb} is not required. $^{c}$\texttt{powerlaw} is not required.  $^{d}$Spectral fitting performed within $3-8$ keV range.
	\end{list}
	
\end{table*}


\begin{thebibliography}{}
\makeatletter
\relax
\def\mn@urlcharsother{\let\do\@makeother \do\$\do\&\do\#\do\^\do\_\do\%\do\~}
\def\mn@doi{\begingroup\mn@urlcharsother \@ifnextchar [ {\mn@doi@}
  {\mn@doi@[]}}
\def\mn@doi@[#1]#2{\def\@tempa{#1}\ifx\@tempa\@empty \href
  {http://dx.doi.org/#2} {doi:#2}\else \href {http://dx.doi.org/#2} {#1}\fi
  \endgroup}
\def\mn@eprint#1#2{\mn@eprint@#1:#2::\@nil}
\def\mn@eprint@arXiv#1{\href {http://arxiv.org/abs/#1} {{\tt arXiv:#1}}}
\def\mn@eprint@dblp#1{\href {http://dblp.uni-trier.de/rec/bibtex/#1.xml}
  {dblp:#1}}
\def\mn@eprint@#1:#2:#3:#4\@nil{\def\@tempa {#1}\def\@tempb {#2}\def\@tempc
  {#3}\ifx \@tempc \@empty \let \@tempc \@tempb \let \@tempb \@tempa \fi \ifx
  \@tempb \@empty \def\@tempb {arXiv}\fi \@ifundefined
  {mn@eprint@\@tempb}{\@tempb:\@tempc}{\expandafter \expandafter \csname
  mn@eprint@\@tempb\endcsname \expandafter{\@tempc}}}

\bibitem[\protect\citeauthoryear{{Abarr} et~al.,}{{Abarr}
  et~al.}{2021}]{Abarr-etal2021}
{Abarr} Q.,  et~al., 2021, \mn@doi [Astroparticle Physics]
  {10.1016/j.astropartphys.2020.102529}, \href
  {https://ui.adsabs.harvard.edu/abs/2021APh...12602529A} {126, 102529}

\bibitem[\protect\citeauthoryear{{Agrawal}}{{Agrawal}}{2006}]{Agrawal-etal2006}
{Agrawal} P.~C.,  2006, \mn@doi [Advances in Space Research]
  {10.1016/j.asr.2006.03.038}, \href
  {https://ui.adsabs.harvard.edu/abs/2006AdSpR..38.2989A} {38, 2989}

\bibitem[\protect\citeauthoryear{{Aneesha}, {Das}, {Katoch}  \&
  {Nandi}}{{Aneesha} et~al.}{2024}]{Aneesha-etal2024}
{Aneesha} U.,  {Das} S.,  {Katoch} T.~B.,   {Nandi} A.,  2024, \mn@doi [\mnras]
  {10.1093/mnras/stae1753}, \href
  {https://ui.adsabs.harvard.edu/abs/2024MNRAS.532.4486A} {532, 4486}

\bibitem[\protect\citeauthoryear{{Antia} et~al.,}{{Antia}
  et~al.}{2017}]{Antia-etal2017}
{Antia} H.~M.,  et~al., 2017, \mn@doi [\apjs] {10.3847/1538-4365/aa7a0e}, \href
  {https://ui.adsabs.harvard.edu/abs/2017ApJS..231...10A} {231, 10}

\bibitem[\protect\citeauthoryear{{Antia} et~al.,}{{Antia}
  et~al.}{2021}]{Antia-etal2021}
{Antia} H.~M.,  et~al., 2021, \mn@doi [Journal of Astrophysics and Astronomy]
  {10.1007/s12036-021-09712-8}, \href
  {https://ui.adsabs.harvard.edu/abs/2021JApA...42...32A} {42, 32}

\bibitem[\protect\citeauthoryear{{Antia}, {Agrawal}, {Katoch}, {Manchanda},
  {Mukerjee}  \& {Shah}}{{Antia} et~al.}{2022}]{Antia-etal2022}
{Antia} H.~M.,  {Agrawal} P.~C.,  {Katoch} T.,  {Manchanda} R.~K.,  {Mukerjee}
  K.,   {Shah} P.,  2022, \mn@doi [\apjs] {10.3847/1538-4365/ac6dd0}, \href
  {https://ui.adsabs.harvard.edu/abs/2022ApJS..260...40A} {260, 40}

\bibitem[\protect\citeauthoryear{{Athulya}, {Radhika}, {Agrawal},
  {Ravishankar}, {Naik}, {Mandal}  \& {Nandi}}{{Athulya}
  et~al.}{2022}]{Athulya-etal2022}
{Athulya} M.~P.,  {Radhika} D.,  {Agrawal} V.~K.,  {Ravishankar} B.~T.,  {Naik}
  S.,  {Mandal} S.,   {Nandi} A.,  2022, \mn@doi [\mnras]
  {10.1093/mnras/stab3614}, \href
  {https://ui.adsabs.harvard.edu/abs/2022MNRAS.510.3019A} {510, 3019}

\bibitem[\protect\citeauthoryear{{Awaki} et~al.,}{{Awaki}
  et~al.}{2025}]{Awaki-etal2025}
{Awaki} H.,  et~al., 2025, \mn@doi [arXiv e-prints]
  {10.48550/arXiv.2507.23126}, \href
  {https://ui.adsabs.harvard.edu/abs/2025arXiv250723126A} {p. arXiv:2507.23126}

\bibitem[\protect\citeauthoryear{{Baby}, {Agrawal}, {Ramadevi}, {Katoch},
  {Antia}, {Mandal}  \& {Nandi}}{{Baby} et~al.}{2020}]{Baby-etal2020}
{Baby} B.~E.,  {Agrawal} V.~K.,  {Ramadevi} M.~C.,  {Katoch} T.,  {Antia}
  H.~M.,  {Mandal} S.,   {Nandi} A.,  2020, \mn@doi [\mnras]
  {10.1093/mnras/staa1965}, \href
  {https://ui.adsabs.harvard.edu/abs/2020MNRAS.497.1197B} {497, 1197}

\bibitem[\protect\citeauthoryear{{Baldini} et~al.,}{{Baldini}
  et~al.}{2022}]{Baldini-etal2022}
{Baldini} L.,  et~al., 2022, \mn@doi [SoftwareX] {10.1016/j.softx.2022.101194},
  \href {https://ui.adsabs.harvard.edu/abs/2022SoftX..1901194B} {19, 101194}

\bibitem[\protect\citeauthoryear{{Barillier}, {Grinberg}, {Horn}, {Nowak},
  {Remillard}, {Steiner}, {Walton}  \& {Wilms}}{{Barillier}
  et~al.}{2023}]{Barillier-etal2023}
{Barillier} E.,  {Grinberg} V.,  {Horn} D.,  {Nowak} M.~A.,  {Remillard} R.~A.,
   {Steiner} J.~F.,  {Walton} D.~J.,   {Wilms} J.,  2023, \mn@doi [\apj]
  {10.3847/1538-4357/acaeaf}, \href
  {https://ui.adsabs.harvard.edu/abs/2023ApJ...944..165B} {944, 165}

\bibitem[\protect\citeauthoryear{{Bellavita}, {Garc{\'\i}a}, {M{\'e}ndez}  \&
  {Karpouzas}}{{Bellavita} et~al.}{2022}]{Bellavita-etal2022}
{Bellavita} C.,  {Garc{\'\i}a} F.,  {M{\'e}ndez} M.,   {Karpouzas} K.,  2022,
  \mn@doi [\mnras] {10.1093/mnras/stac1922}, \href
  {https://ui.adsabs.harvard.edu/abs/2022MNRAS.515.2099B} {515, 2099}

\bibitem[\protect\citeauthoryear{{Bellavita}, {M{\'e}ndez}, {Garc{\'\i}a}, {Ma}
   \& {K{\"o}nig}}{{Bellavita} et~al.}{2025}]{Bellavita-etal2025}
{Bellavita} C.,  {M{\'e}ndez} M.,  {Garc{\'\i}a} F.,  {Ma} R.,   {K{\"o}nig}
  O.,  2025, \mn@doi [\aap] {10.1051/0004-6361/202453092}, \href
  {https://ui.adsabs.harvard.edu/abs/2025A&A...696A.128B} {696, A128}

\bibitem[\protect\citeauthoryear{{Belloni}, {M{\'e}ndez}, {van der Klis},
  {Lewin}  \& {Dieters}}{{Belloni} et~al.}{1999}]{Belloni-etal1999}
{Belloni} T.,  {M{\'e}ndez} M.,  {van der Klis} M.,  {Lewin} W.~H.~G.,
  {Dieters} S.,  1999, \mn@doi [\apjl] {10.1086/312130}, \href
  {https://ui.adsabs.harvard.edu/abs/1999ApJ...519L.159B} {519, L159}

\bibitem[\protect\citeauthoryear{Belloni, Psaltis  \& van~der Klis}{Belloni
  et~al.}{2002}]{Belloni-etal2002}
Belloni T.,  Psaltis D.,   van~der Klis M.,  2002, \mn@doi [The Astrophysical
  Journal] {10.1086/340290}, 572, 392

\bibitem[\protect\citeauthoryear{{Belloni}, {Homan}, {Casella}, {van der Klis},
  {Nespoli}, {Lewin}, {Miller}  \& {M{\'e}ndez}}{{Belloni}
  et~al.}{2005}]{Belloni-etal2005}
{Belloni} T.,  {Homan} J.,  {Casella} P.,  {van der Klis} M.,  {Nespoli} E.,
  {Lewin} W.~H.~G.,  {Miller} J.~M.,   {M{\'e}ndez} M.,  2005, \mn@doi [\aap]
  {10.1051/0004-6361:20042457}, \href
  {https://ui.adsabs.harvard.edu/abs/2005A&A...440..207B} {440, 207}

\bibitem[\protect\citeauthoryear{{Beloborodov}}{{Beloborodov}}{1998}]{Beloborodov-etal1998}
{Beloborodov} A.~M.,  1998, \mn@doi [\apjl] {10.1086/311260}, \href
  {https://ui.adsabs.harvard.edu/abs/1998ApJ...496L.105B} {496, L105}

\bibitem[\protect\citeauthoryear{{Bhuvana}, {Radhika}, {Agrawal}, {Mandal}  \&
  {Nandi}}{{Bhuvana} et~al.}{2021}]{Bhuvana-etal2021}
{Bhuvana} G.~R.,  {Radhika} D.,  {Agrawal} V.~K.,  {Mandal} S.,   {Nandi} A.,
  2021, \mn@doi [\mnras] {10.1093/mnras/staa4012}, \href
  {https://ui.adsabs.harvard.edu/abs/2021MNRAS.501.5457B} {501, 5457}

\bibitem[\protect\citeauthoryear{{Capitanio}, {Del Santo}, {Bozzo}, {Ferrigno},
  {De Cesare}  \& {Paizis}}{{Capitanio} et~al.}{2012}]{Capitanio-etal2012}
{Capitanio} F.,  {Del Santo} M.,  {Bozzo} E.,  {Ferrigno} C.,  {De Cesare} G.,
   {Paizis} A.,  2012, \mn@doi [\mnras] {10.1111/j.1365-2966.2012.20834.x},
  \href {https://ui.adsabs.harvard.edu/abs/2012MNRAS.422.3130C} {422, 3130}

\bibitem[\protect\citeauthoryear{Casella, Belloni  \& Stella}{Casella
  et~al.}{2005}]{Casella-etal2005}
Casella P.,  Belloni T.,   Stella L.,  2005, \mn@doi [The Astrophysical
  Journal] {10.1086/431174}, 629, 403

\bibitem[\protect\citeauthoryear{{Chakrabarti}}{{Chakrabarti}}{1996}]{Chakrabarti-etal1996}
{Chakrabarti} S.~K.,  1996, \mn@doi [\apj] {10.1086/177354}, \href
  {https://ui.adsabs.harvard.edu/abs/1996ApJ...464..664C} {464, 664}

\bibitem[\protect\citeauthoryear{{Chakrabarti} \& {Manickam}}{{Chakrabarti} \&
  {Manickam}}{2000}]{Chakrabarti-etal2000}
{Chakrabarti} S.~K.,  {Manickam} S.~G.,  2000, \mn@doi [\apjl]
  {10.1086/312512}, \href
  {https://ui.adsabs.harvard.edu/abs/2000ApJ...531L..41C} {531, L41}

\bibitem[\protect\citeauthoryear{{Chakrabarti} \& {Titarchuk}}{{Chakrabarti} \&
  {Titarchuk}}{1995}]{Chakrabarti-etal1995}
{Chakrabarti} S.,  {Titarchuk} L.~G.,  1995, \mn@doi [\apj] {10.1086/176610},
  \href {https://ui.adsabs.harvard.edu/abs/1995ApJ...455..623C} {455, 623}

\bibitem[\protect\citeauthoryear{{Chakrabarti}, {Debnath}, {Nandi}  \&
  {Pal}}{{Chakrabarti} et~al.}{2008}]{Chakrabarti-etal2008}
{Chakrabarti} S.~K.,  {Debnath} D.,  {Nandi} A.,   {Pal} P.~S.,  2008, \mn@doi
  [\aap] {10.1051/0004-6361:200810136}, \href
  {https://ui.adsabs.harvard.edu/abs/2008A&A...489L..41C} {489, L41}

\bibitem[\protect\citeauthoryear{{Chandrasekhar}}{{Chandrasekhar}}{1960}]{Chandrasekhar-1960}
{Chandrasekhar} S.,  1960, {Radiative transfer}.
New York: Dover Publications

\bibitem[\protect\citeauthoryear{{Chatterjee}, {Debnath}, {Bhowmick}, {Nath}
  \& {Chatterjee}}{{Chatterjee} et~al.}{2022}]{Chatterjee-etal2022}
{Chatterjee} K.,  {Debnath} D.,  {Bhowmick} R.,  {Nath} S.~K.,   {Chatterjee}
  D.,  2022, \mn@doi [\mnras] {10.1093/mnras/stab3570}, \href
  {https://ui.adsabs.harvard.edu/abs/2022MNRAS.510.1128C} {510, 1128}

\bibitem[\protect\citeauthoryear{{Chatterjee}, {Pujitha Suribhatla}, {Mondal}
  \& {Singh}}{{Chatterjee} et~al.}{2024}]{Chatterjee-etal2024}
{Chatterjee} K.,  {Pujitha Suribhatla} S.,  {Mondal} S.,   {Singh} C.~B.,
  2024, \mn@doi [arXiv e-prints] {10.48550/arXiv.2406.17629}, \href
  {https://ui.adsabs.harvard.edu/abs/2024arXiv240617629C} {p. arXiv:2406.17629}

\bibitem[\protect\citeauthoryear{{Chattopadhyay} et~al.,}{{Chattopadhyay}
  et~al.}{2024}]{Chattopadhyay-etal2024}
{Chattopadhyay} T.,  et~al., 2024, \mn@doi [\apjl] {10.3847/2041-8213/ad118d},
  \href {https://ui.adsabs.harvard.edu/abs/2024ApJ...960L...2C} {960, L2}

\bibitem[\protect\citeauthoryear{{Contopoulos} \& {Jappel}}{{Contopoulos} \&
  {Jappel}}{1974}]{Contopoulos-etal1974}
{Contopoulos} G.,  {Jappel} A.,  1974, {Transactions of the International
  Astronomical Union, Volume\_XVB: Proceedings of the Fifteenth General
  Assembly, Sydney 1973 and Extraordinary Assembly, Poland 1973.}

\bibitem[\protect\citeauthoryear{{Das}, {Chattopadhyay}, {Nandi}  \&
  {Molteni}}{{Das} et~al.}{2014}]{Das-etal2014}
{Das} S.,  {Chattopadhyay} I.,  {Nandi} A.,   {Molteni} D.,  2014, \mn@doi
  [\mnras] {10.1093/mnras/stu864}, \href
  {https://ui.adsabs.harvard.edu/abs/2014MNRAS.442..251D} {442, 251}

\bibitem[\protect\citeauthoryear{{Dauser}, {Garcia}, {Wilms}, {B{\"o}ck},
  {Brenneman}, {Falanga}, {Fukumura}  \& {Reynolds}}{{Dauser}
  et~al.}{2013}]{Dauser-etal2013}
{Dauser} T.,  {Garcia} J.,  {Wilms} J.,  {B{\"o}ck} M.,  {Brenneman} L.~W.,
  {Falanga} M.,  {Fukumura} K.,   {Reynolds} C.~S.,  2013, \mn@doi [\mnras]
  {10.1093/mnras/sts710}, \href
  {https://ui.adsabs.harvard.edu/abs/2013MNRAS.430.1694D} {430, 1694}

\bibitem[\protect\citeauthoryear{{Debnath}, {Srimani}  \& {Chang}}{{Debnath}
  et~al.}{2025}]{Debnath-etal2025}
{Debnath} D.,  {Srimani} S.,   {Chang} H.-K.,  2025, \mn@doi [\apj]
  {10.3847/1538-4357/adf1e6}, \href
  {https://ui.adsabs.harvard.edu/abs/2025ApJ...989..165D} {989, 165}

\bibitem[\protect\citeauthoryear{{Dexter} \& {Begelman}}{{Dexter} \&
  {Begelman}}{2024}]{Dexter-etal2024}
{Dexter} J.,  {Begelman} M.~C.,  2024, \mn@doi [\mnras]
  {10.1093/mnrasl/slad182}, \href
  {https://ui.adsabs.harvard.edu/abs/2024MNRAS.528L.157D} {528, L157}

\bibitem[\protect\citeauthoryear{{Done} \& {Zycki}}{{Done} \&
  {Zycki}}{1999}]{Done-etal1999}
{Done} C.,  {Zycki} P.~T.,  1999, \mn@doi [\mnras]
  {10.1046/j.1365-8711.1999.02439.x}, \href
  {https://ui.adsabs.harvard.edu/abs/1999MNRAS.305..457D} {305, 457}

\bibitem[\protect\citeauthoryear{{Done}, {Gierli{\'n}ski}  \& {Kubota}}{{Done}
  et~al.}{2007}]{Done-etal2007}
{Done} C.,  {Gierli{\'n}ski} M.,   {Kubota} A.,  2007, \mn@doi [\aapr]
  {10.1007/s00159-007-0006-1}, \href
  {https://ui.adsabs.harvard.edu/abs/2007A&ARv..15....1D} {15, 1}

\bibitem[\protect\citeauthoryear{{Dov{\v{c}}iak} et~al.,}{{Dov{\v{c}}iak}
  et~al.}{2024}]{Dovciak-etal2024}
{Dov{\v{c}}iak} M.,  et~al., 2024, \mn@doi [Galaxies]
  {10.3390/galaxies12050054}, \href
  {https://ui.adsabs.harvard.edu/abs/2024Galax..12...54D} {12, 54}

\bibitem[\protect\citeauthoryear{{Eardley}, {Lightman}  \& {Shapiro}}{{Eardley}
  et~al.}{1975}]{Eardley-etal1975}
{Eardley} D.~M.,  {Lightman} A.~P.,   {Shapiro} S.~L.,  1975, \mn@doi [\apjl]
  {10.1086/181871}, \href
  {https://ui.adsabs.harvard.edu/abs/1975ApJ...199L.153E} {199, L153}

\bibitem[\protect\citeauthoryear{{Ewing} et~al.,}{{Ewing}
  et~al.}{2025}]{Ewing-etal2025}
{Ewing} M.,  et~al., 2025, \mn@doi [\mnras] {10.1093/mnras/staf859}, \href
  {https://ui.adsabs.harvard.edu/abs/2025MNRAS.541.1774E} {541, 1774}

\bibitem[\protect\citeauthoryear{{Fender}, {Homan}  \& {Belloni}}{{Fender}
  et~al.}{2009}]{Fender-etal2009}
{Fender} R.~P.,  {Homan} J.,   {Belloni} T.~M.,  2009, \mn@doi [\mnras]
  {10.1111/j.1365-2966.2009.14841.x}, \href
  {https://ui.adsabs.harvard.edu/abs/2009MNRAS.396.1370F} {396, 1370}

\bibitem[\protect\citeauthoryear{{Garc{\'\i}a} et~al.,}{{Garc{\'\i}a}
  et~al.}{2014}]{Garcia-etal2014}
{Garc{\'\i}a} J.,  et~al., 2014, \mn@doi [\apj] {10.1088/0004-637X/782/2/76},
  \href {https://ui.adsabs.harvard.edu/abs/2014ApJ...782...76G} {782, 76}

\bibitem[\protect\citeauthoryear{{Garc{\'\i}a}, {M{\'e}ndez}, {Karpouzas},
  {Belloni}, {Zhang}  \& {Altamirano}}{{Garc{\'\i}a}
  et~al.}{2021}]{Garcia-etal2021}
{Garc{\'\i}a} F.,  {M{\'e}ndez} M.,  {Karpouzas} K.,  {Belloni} T.,  {Zhang}
  L.,   {Altamirano} D.,  2021, \mn@doi [\mnras] {10.1093/mnras/staa3944},
  \href {https://ui.adsabs.harvard.edu/abs/2021MNRAS.501.3173G} {501, 3173}

\bibitem[\protect\citeauthoryear{{Garg}, {Rawat}  \& {M{\'e}ndez}}{{Garg}
  et~al.}{2024}]{Garg-etal2024}
{Garg} A.,  {Rawat} D.,   {M{\'e}ndez} M.,  2024, \mn@doi [\mnras]
  {10.1093/mnras/stae1198}, \href
  {https://ui.adsabs.harvard.edu/abs/2024MNRAS.531..585G} {531, 585}

\bibitem[\protect\citeauthoryear{{Gendreau} et~al.,}{{Gendreau}
  et~al.}{2016}]{Gendreau-etal2016}
{Gendreau} K.~C.,  et~al., 2016, in {den Herder} J.-W.~A.,  {Takahashi} T.,
  {Bautz} M.,  eds,  Society of Photo-Optical Instrumentation Engineers (SPIE)
  Conference Series Vol. 9905, Space Telescopes and Instrumentation 2016:
  Ultraviolet to Gamma Ray. p. 99051H, \mn@doi{10.1117/12.2231304}

\bibitem[\protect\citeauthoryear{{Giacconi}, {Gorenstein}, {Gursky}  \&
  {Waters}}{{Giacconi} et~al.}{1967}]{Giacconi-etal1967}
{Giacconi} R.,  {Gorenstein} P.,  {Gursky} H.,   {Waters} J.~R.,  1967, \mn@doi
  [\apjl] {10.1086/180028}, \href
  {https://ui.adsabs.harvard.edu/abs/1967ApJ...148L.119G} {148, L119}

\bibitem[\protect\citeauthoryear{{Giacconi}, {Murray}, {Gursky}, {Kellogg},
  {Schreier}, {Matilsky}, {Koch}  \& {Tananbaum}}{{Giacconi}
  et~al.}{1974}]{Giacconi-etal1974}
{Giacconi} R.,  {Murray} S.,  {Gursky} H.,  {Kellogg} E.,  {Schreier} E.,
  {Matilsky} T.,  {Koch} D.,   {Tananbaum} H.,  1974, \mn@doi [\apjs]
  {10.1086/190288}, \href
  {https://ui.adsabs.harvard.edu/abs/1974ApJS...27...37G} {27, 37}

\bibitem[\protect\citeauthoryear{{Gomez}, {Mason}  \& {Robinson}}{{Gomez}
  et~al.}{2015}]{Gomez-etal2015}
{Gomez} S.,  {Mason} P.~A.,   {Robinson} E.~L.,  2015, \mn@doi [\apj]
  {10.1088/0004-637X/809/1/9}, \href
  {https://ui.adsabs.harvard.edu/abs/2015ApJ...809....9G} {809, 9}

\bibitem[\protect\citeauthoryear{{Gou} et~al.,}{{Gou}
  et~al.}{2009}]{Gou-etal2009}
{Gou} L.,  et~al., 2009, \mn@doi [\apj] {10.1088/0004-637X/701/2/1076}, \href
  {https://ui.adsabs.harvard.edu/abs/2009ApJ...701.1076G} {701, 1076}

\bibitem[\protect\citeauthoryear{{Gou} et~al.,}{{Gou}
  et~al.}{2014}]{Gou-etal2014}
{Gou} L.,  et~al., 2014, \mn@doi [\apj] {10.1088/0004-637X/790/1/29}, \href
  {https://ui.adsabs.harvard.edu/abs/2014ApJ...790...29G} {790, 29}

\bibitem[\protect\citeauthoryear{{Haardt} \& {Maraschi}}{{Haardt} \&
  {Maraschi}}{1993}]{Haardt-etal1993}
{Haardt} F.,  {Maraschi} L.,  1993, \mn@doi [\apj] {10.1086/173020}, \href
  {https://ui.adsabs.harvard.edu/abs/1993ApJ...413..507H} {413, 507}

\bibitem[\protect\citeauthoryear{Harrison et~al.,}{Harrison
  et~al.}{2013}]{Harrison-etal2013}
Harrison F.~A.,  et~al., 2013, \mn@doi [The Astrophysical Journal]
  {10.1088/0004-637X/770/2/103}, 770, 103

\bibitem[\protect\citeauthoryear{{Homan}, {Wijnands}, {van der Klis},
  {Belloni}, {van Paradijs}, {Klein-Wolt}, {Fender}  \& {M{\'e}ndez}}{{Homan}
  et~al.}{2001}]{Homan-etal2001}
{Homan} J.,  {Wijnands} R.,  {van der Klis} M.,  {Belloni} T.,  {van Paradijs}
  J.,  {Klein-Wolt} M.,  {Fender} R.,   {M{\'e}ndez} M.,  2001, \mn@doi [\apjs]
  {10.1086/318954}, \href
  {https://ui.adsabs.harvard.edu/abs/2001ApJS..132..377H} {132, 377}

\bibitem[\protect\citeauthoryear{{Homan} et~al.,}{{Homan}
  et~al.}{2020}]{Homan-etal2020}
{Homan} J.,  et~al., 2020, \mn@doi [\apjl] {10.3847/2041-8213/ab7932}, \href
  {https://ui.adsabs.harvard.edu/abs/2020ApJ...891L..29H} {891, L29}

\bibitem[\protect\citeauthoryear{{Ingram} \& {Motta}}{{Ingram} \&
  {Motta}}{2019}]{Ingram-etal2019}
{Ingram} A.~R.,  {Motta} S.~E.,  2019, \mn@doi [\nar]
  {10.1016/j.newar.2020.101524}, \href
  {https://ui.adsabs.harvard.edu/abs/2019NewAR..8501524I} {85, 101524}

\bibitem[\protect\citeauthoryear{{Ingram}, {Done}  \& {Fragile}}{{Ingram}
  et~al.}{2009}]{Ingram-etal2009}
{Ingram} A.,  {Done} C.,   {Fragile} P.~C.,  2009, \mn@doi [\mnras]
  {10.1111/j.1745-3933.2009.00693.x}, \href
  {https://ui.adsabs.harvard.edu/abs/2009MNRAS.397L.101I} {397, L101}

\bibitem[\protect\citeauthoryear{{Ingram} et~al.,}{{Ingram}
  et~al.}{2024}]{Ingram-etal2024}
{Ingram} A.,  et~al., 2024, \mn@doi [\apj] {10.3847/1538-4357/ad3faf}, \href
  {https://ui.adsabs.harvard.edu/abs/2024ApJ...968...76I} {968, 76}

\bibitem[\protect\citeauthoryear{Iyer, Nandi  \& Mandal}{Iyer
  et~al.}{2015}]{Iyer-etal2015}
Iyer N.,  Nandi A.,   Mandal S.,  2015, \mn@doi [The Astrophysical Journal]
  {10.1088/0004-637X/807/1/108}, 807, 108

\bibitem[\protect\citeauthoryear{{Jana} \& {Chang}}{{Jana} \&
  {Chang}}{2024}]{Jana-etal2024}
{Jana} A.,  {Chang} H.-K.,  2024, \mn@doi [\mnras] {10.1093/mnras/stad3961},
  \href {https://ui.adsabs.harvard.edu/abs/2024MNRAS.52710837J} {527, 10837}

\bibitem[\protect\citeauthoryear{{Jones}, {Forman}, {Tananbaum}  \&
  {Turner}}{{Jones} et~al.}{1976}]{Jones-etal1976}
{Jones} C.,  {Forman} W.,  {Tananbaum} H.,   {Turner} M.~J.~L.,  1976, \mn@doi
  [\apjl] {10.1086/182291}, \href
  {https://ui.adsabs.harvard.edu/abs/1976ApJ...210L...9J} {210, L9}

\bibitem[\protect\citeauthoryear{Jourdain, Roques, Chauvin  \& Clark}{Jourdain
  et~al.}{2012}]{Jourdain-etal2012}
Jourdain E.,  Roques J.~P.,  Chauvin M.,   Clark D.~J.,  2012, \mn@doi [The
  Astrophysical Journal] {10.1088/0004-637X/761/1/27}, 761, 27

\bibitem[\protect\citeauthoryear{Kalemci, Maccarone  \& Tomsick}{Kalemci
  et~al.}{2018}]{Kalemci-etal2018}
Kalemci E.,  Maccarone T.~J.,   Tomsick J.~A.,  2018, \mn@doi [The
  Astrophysical Journal] {10.3847/1538-4357/aabcd3}, 859, 88

\bibitem[\protect\citeauthoryear{{Karpouzas}, {M{\'e}ndez}, {Ribeiro},
  {Altamirano}, {Blaes}  \& {Garc{\'\i}a}}{{Karpouzas}
  et~al.}{2020}]{Karpouzas-etal2020}
{Karpouzas} K.,  {M{\'e}ndez} M.,  {Ribeiro} E.~M.,  {Altamirano} D.,  {Blaes}
  O.,   {Garc{\'\i}a} F.,  2020, \mn@doi [\mnras] {10.1093/mnras/stz3502},
  \href {https://ui.adsabs.harvard.edu/abs/2020MNRAS.492.1399K} {492, 1399}

\bibitem[\protect\citeauthoryear{{Katoch}, {Baby}, {Nandi}, {Agrawal}, {Antia}
  \& {Mukerjee}}{{Katoch} et~al.}{2021}]{Katoch-etal2021}
{Katoch} T.,  {Baby} B.~E.,  {Nandi} A.,  {Agrawal} V.~K.,  {Antia} H.~M.,
  {Mukerjee} K.,  2021, \mn@doi [\mnras] {10.1093/mnras/staa3756}, \href
  {https://ui.adsabs.harvard.edu/abs/2021MNRAS.501.6123K} {501, 6123}

\bibitem[\protect\citeauthoryear{{Kennea}, {Lien}, {D'Elia}, {Melandri}, {Page}
   \& {Siegel}}{{Kennea} et~al.}{2024}]{Kennea-etal2024}
{Kennea} J.~A.,  {Lien} A.~Y.,  {D'Elia} V.,  {Melandri} A.,  {Page} K.~L.,
  {Siegel} M.~H.,  2024, The Astronomer's Telegram, \href
  {https://ui.adsabs.harvard.edu/abs/2024ATel16500....1K} {16500, 1}

\bibitem[\protect\citeauthoryear{{Kislat}, {Clark}, {Beilicke}  \&
  {Krawczynski}}{{Kislat} et~al.}{2015}]{Kislat-etal2015}
{Kislat} F.,  {Clark} B.,  {Beilicke} M.,   {Krawczynski} H.,  2015, \mn@doi
  [Astroparticle Physics] {10.1016/j.astropartphys.2015.02.007}, \href
  {https://ui.adsabs.harvard.edu/abs/2015APh....68...45K} {68, 45}

\bibitem[\protect\citeauthoryear{{Krawczynski} et~al.,}{{Krawczynski}
  et~al.}{2022}]{Krawczynski-etal2022b}
{Krawczynski} H.,  et~al., 2022, \mn@doi [Science] {10.1126/science.add5399},
  \href {https://ui.adsabs.harvard.edu/abs/2022Sci...378..650K} {378, 650}

\bibitem[\protect\citeauthoryear{{Krawczynski} et~al.,}{{Krawczynski}
  et~al.}{2024}]{Krawczynski-etal2024}
{Krawczynski} H.,  et~al., 2024, \mn@doi [\apjl] {10.3847/2041-8213/ad855c},
  \href {https://ui.adsabs.harvard.edu/abs/2024ApJ...977L..10K} {977, L10}

\bibitem[\protect\citeauthoryear{Krimm et~al.,}{Krimm
  et~al.}{2013}]{Krimm-etal2013}
Krimm H.~A.,  et~al., 2013, \mn@doi [The Astrophysical Journal Supplement
  Series] {10.1088/0067-0049/209/1/14}, 209, 14

\bibitem[\protect\citeauthoryear{{Kudo} et~al.,}{{Kudo}
  et~al.}{2025}]{Kudo-etal2025}
{Kudo} Y.,  et~al., 2025, The Astronomer's Telegram, \href
  {https://ui.adsabs.harvard.edu/abs/2025ATel16975....1K} {16975, 1}

\bibitem[\protect\citeauthoryear{{Kumar}}{{Kumar}}{2024}]{Kumar-etal2024}
{Kumar} N.,  2024, \mn@doi [\pasa] {10.1017/pasa.2024.8}, \href
  {https://ui.adsabs.harvard.edu/abs/2024PASA...41...13K} {41, e013}

\bibitem[\protect\citeauthoryear{{Kushwaha}, {Agrawal}  \& {Nandi}}{{Kushwaha}
  et~al.}{2021}]{Kushwaha-etal2021}
{Kushwaha} A.,  {Agrawal} V.~K.,   {Nandi} A.,  2021, \mn@doi [\mnras]
  {10.1093/mnras/stab2258}, \href
  {https://ui.adsabs.harvard.edu/abs/2021MNRAS.507.2602K} {507, 2602}

\bibitem[\protect\citeauthoryear{{Kushwaha}, {Jayasurya}, {Agrawal}  \&
  {Nandi}}{{Kushwaha} et~al.}{2023}]{Kushwaha-etal2023a}
{Kushwaha} A.,  {Jayasurya} K.~M.,  {Agrawal} V.~K.,   {Nandi} A.,  2023,
  \mn@doi [\mnras] {10.1093/mnrasl/slad070}, \href
  {https://ui.adsabs.harvard.edu/abs/2023MNRAS.524L..15K} {524, L15}

\bibitem[\protect\citeauthoryear{{Kuulkers}, {Lutovinov}, {Parmar},
  {Capitanio}, {Mowlavi}  \& {Hermsen}}{{Kuulkers}
  et~al.}{2003}]{Kuulkers-etal2003}
{Kuulkers} E.,  {Lutovinov} A.,  {Parmar} A.,  {Capitanio} F.,  {Mowlavi} N.,
  {Hermsen} W.,  2003, The Astronomer's Telegram, \href
  {https://ui.adsabs.harvard.edu/abs/2003ATel..149....1K} {149, 1}

\bibitem[\protect\citeauthoryear{{Kylafis}, {Reig}  \& {Papadakis}}{{Kylafis}
  et~al.}{2020}]{Kylafis-etal2020}
{Kylafis} N.~D.,  {Reig} P.,   {Papadakis} I.,  2020, \mn@doi [\aap]
  {10.1051/0004-6361/202038468}, \href
  {https://ui.adsabs.harvard.edu/abs/2020A&A...640L..16K} {640, L16}

\bibitem[\protect\citeauthoryear{{Liao} et~al.,}{{Liao}
  et~al.}{2024}]{Liao-etal2024}
{Liao} J.,  et~al., 2024, \mn@doi [arXiv e-prints] {10.48550/arXiv.2410.06574},
  \href {https://ui.adsabs.harvard.edu/abs/2024arXiv241006574L} {p.
  arXiv:2410.06574}

\bibitem[\protect\citeauthoryear{Ling, Xie, Ge  \& Monaca}{Ling
  et~al.}{2024}]{Ling-etal2024}
Ling Y.-S.,  Xie F.,  Ge M.-Y.,   Monaca F.~L.,  2024, \mn@doi [Research in
  Astronomy and Astrophysics] {10.1088/1674-4527/ad6edf}, 24, 095004

\bibitem[\protect\citeauthoryear{{Liu} et~al.,}{{Liu}
  et~al.}{2022}]{Liu-etal2022}
{Liu} H.~X.,  et~al., 2022, \mn@doi [\apj] {10.3847/1538-4357/ac88c6}, \href
  {https://ui.adsabs.harvard.edu/abs/2022ApJ...938..108L} {938, 108}

\bibitem[\protect\citeauthoryear{{Long}, {Chanan}  \& {Novick}}{{Long}
  et~al.}{1980}]{Long-etal1980}
{Long} K.~S.,  {Chanan} G.~A.,   {Novick} R.,  1980, \mn@doi [\apj]
  {10.1086/158027}, \href
  {https://ui.adsabs.harvard.edu/abs/1980ApJ...238..710L} {238, 710}

\bibitem[\protect\citeauthoryear{{Ma} et~al.,}{{Ma} et~al.}{2021}]{Ma-etal2021}
{Ma} X.,  et~al., 2021, \mn@doi [Nature Astronomy] {10.1038/s41550-020-1192-2},
  \href {https://ui.adsabs.harvard.edu/abs/2021NatAs...5...94M} {5, 94}

\bibitem[\protect\citeauthoryear{{Maitra}, {Miller}, {Reynolds}, {Reis}  \&
  {Nowak}}{{Maitra} et~al.}{2014}]{Maitra-etal2014}
{Maitra} D.,  {Miller} J.~M.,  {Reynolds} M.~T.,  {Reis} R.,   {Nowak} M.,
  2014, \mn@doi [\apj] {10.1088/0004-637X/794/1/85}, \href
  {https://ui.adsabs.harvard.edu/abs/2014ApJ...794...85M} {794, 85}

\bibitem[\protect\citeauthoryear{{Majumder}, {Sreehari}, {Aftab}, {Katoch},
  {Das}  \& {Nandi}}{{Majumder} et~al.}{2022}]{Majumder-etal2022}
{Majumder} S.,  {Sreehari} H.,  {Aftab} N.,  {Katoch} T.,  {Das} S.,   {Nandi}
  A.,  2022, \mn@doi [\mnras] {10.1093/mnras/stac615}, \href
  {https://ui.adsabs.harvard.edu/abs/2022MNRAS.512.2508M} {512, 2508}

\bibitem[\protect\citeauthoryear{{Majumder}, {Kushwaha}, {Das}  \&
  {Nandi}}{{Majumder} et~al.}{2024a}]{Majumder-etal2024a}
{Majumder} S.,  {Kushwaha} A.,  {Das} S.,   {Nandi} A.,  2024a, \mn@doi
  [\mnras] {10.1093/mnrasl/slad148}, \href
  {https://ui.adsabs.harvard.edu/abs/2024MNRAS.527L..76M} {527, L76}

\bibitem[\protect\citeauthoryear{{Majumder}, {Chatterjee}, {Jayasurya}, {Das}
  \& {Nandi}}{{Majumder} et~al.}{2024b}]{Majumder-etal2024b}
{Majumder} S.,  {Chatterjee} R.,  {Jayasurya} K.~M.,  {Das} S.,   {Nandi} A.,
  2024b, \mn@doi [\apjl] {10.3847/2041-8213/ad67e5}, \href
  {https://ui.adsabs.harvard.edu/abs/2024ApJ...971L..21M} {971, L21}

\bibitem[\protect\citeauthoryear{{Majumder}, {Das}  \& {Nandi}}{{Majumder}
  et~al.}{2025a}]{Majumder-etal2025}
{Majumder} S.,  {Das} S.,   {Nandi} A.,  2025a, \mn@doi [arXiv e-prints]
  {10.48550/arXiv.2504.09070}, \href
  {https://ui.adsabs.harvard.edu/abs/2025arXiv250409070M} {p. arXiv:2504.09070}

\bibitem[\protect\citeauthoryear{{Majumder}, {Dutta}  \& {Nandi}}{{Majumder}
  et~al.}{2025b}]{Majumder-etal2025b}
{Majumder} P.,  {Dutta} B.~G.,   {Nandi} A.,  2025b, \mn@doi [arXiv e-prints]
  {10.48550/arXiv.2504.14193}, \href
  {https://ui.adsabs.harvard.edu/abs/2025arXiv250414193M} {p. arXiv:2504.14193}

\bibitem[\protect\citeauthoryear{{Makishima}, {Maejima}, {Mitsuda}, {Bradt},
  {Remillard}, {Tuohy}, {Hoshi}  \& {Nakagawa}}{{Makishima}
  et~al.}{1986}]{Makishima-etal1986}
{Makishima} K.,  {Maejima} Y.,  {Mitsuda} K.,  {Bradt} H.~V.,  {Remillard}
  R.~A.,  {Tuohy} I.~R.,  {Hoshi} R.,   {Nakagawa} M.,  1986, \mn@doi [\apj]
  {10.1086/164534}, \href
  {https://ui.adsabs.harvard.edu/abs/1986ApJ...308..635M} {308, 635}

\bibitem[\protect\citeauthoryear{{Mandel} et~al.,}{{Mandel}
  et~al.}{2025}]{Mandel-etal2025}
{Mandel} S.,  et~al., 2025, The Astronomer's Telegram, \href
  {https://ui.adsabs.harvard.edu/abs/2025ATel17031....1M} {17031, 1}

\bibitem[\protect\citeauthoryear{{Mark}, {Price}, {Rodrigues}, {Seward}  \&
  {Swift}}{{Mark} et~al.}{1969}]{Mark-etal1969}
{Mark} H.,  {Price} R.,  {Rodrigues} R.,  {Seward} F.~D.,   {Swift} C.~D.,
  1969, \mn@doi [\apjl] {10.1086/180322}, \href
  {https://ui.adsabs.harvard.edu/abs/1969ApJ...155L.143M} {155, L143}

\bibitem[\protect\citeauthoryear{{Markert}, {Canizares}, {Clark}, {Lewin},
  {Schnopper}  \& {Sprott}}{{Markert} et~al.}{1973}]{Markert-etal1973}
{Markert} T.~H.,  {Canizares} C.~R.,  {Clark} G.~W.,  {Lewin} W.~H.~G.,
  {Schnopper} H.~W.,   {Sprott} G.~F.,  1973, \mn@doi [\apjl] {10.1086/181290},
  \href {https://ui.adsabs.harvard.edu/abs/1973ApJ...184L..67M} {184, L67}

\bibitem[\protect\citeauthoryear{{Markoff}, {Nowak}  \& {Wilms}}{{Markoff}
  et~al.}{2005}]{Markoff-etal2005}
{Markoff} S.,  {Nowak} M.~A.,   {Wilms} J.,  2005, \mn@doi [\apj]
  {10.1086/497628}, \href
  {https://ui.adsabs.harvard.edu/abs/2005ApJ...635.1203M} {635, 1203}

\bibitem[\protect\citeauthoryear{{Marra} et~al.,}{{Marra}
  et~al.}{2024}]{Marra-etal2024}
{Marra} L.,  et~al., 2024, \mn@doi [\aap] {10.1051/0004-6361/202348277}, \href
  {https://ui.adsabs.harvard.edu/abs/2024A&A...684A..95M} {684, A95}

\bibitem[\protect\citeauthoryear{{Marra} et~al.,}{{Marra}
  et~al.}{2025}]{Marra-etal2025}
{Marra} L.,  et~al., 2025, \mn@doi [arXiv e-prints]
  {10.48550/arXiv.2506.17050}, \href
  {https://ui.adsabs.harvard.edu/abs/2025arXiv250617050M} {p. arXiv:2506.17050}

\bibitem[\protect\citeauthoryear{{Mastroserio} et~al.,}{{Mastroserio}
  et~al.}{2025}]{Mastroserio-etal2024}
{Mastroserio} G.,  et~al., 2025, \mn@doi [\apjl] {10.3847/2041-8213/ad9913},
  \href {https://ui.adsabs.harvard.edu/abs/2025ApJ...978L..19M} {978, L19}

\bibitem[\protect\citeauthoryear{{Mata S{\'a}nchez}, {Mu{\~n}oz-Darias}, {Armas
  Padilla}, {Casares}  \& {Torres}}{{Mata S{\'a}nchez}
  et~al.}{2024}]{Mata-etal2024}
{Mata S{\'a}nchez} D.,  {Mu{\~n}oz-Darias} T.,  {Armas Padilla} M.,  {Casares}
  J.,   {Torres} M.~A.~P.,  2024, \mn@doi [\aap] {10.1051/0004-6361/202348754},
  \href {https://ui.adsabs.harvard.edu/abs/2024A&A...682L...1M} {682, L1}

\bibitem[\protect\citeauthoryear{{Matsuoka} et~al.,}{{Matsuoka}
  et~al.}{2009}]{Matsuoka-etal2009}
{Matsuoka} M.,  et~al., 2009, \mn@doi [\pasj] {10.1093/pasj/61.5.999}, \href
  {https://ui.adsabs.harvard.edu/abs/2009PASJ...61..999M} {61, 999}

\bibitem[\protect\citeauthoryear{{McClintock}, {Remillard}, {Rupen}, {Torres},
  {Steeghs}, {Levine}  \& {Orosz}}{{McClintock}
  et~al.}{2009}]{McClintock-etal2009}
{McClintock} J.~E.,  {Remillard} R.~A.,  {Rupen} M.~P.,  {Torres} M.~A.~P.,
  {Steeghs} D.,  {Levine} A.~M.,   {Orosz} J.~A.,  2009, \mn@doi [\apj]
  {10.1088/0004-637X/698/2/1398}, \href
  {https://ui.adsabs.harvard.edu/abs/2009ApJ...698.1398M} {698, 1398}

\bibitem[\protect\citeauthoryear{{M{\'e}ndez}, {Karpouzas}, {Garc{\'\i}a},
  {Zhang}, {Zhang}, {Belloni}  \& {Altamirano}}{{M{\'e}ndez}
  et~al.}{2022}]{Mendez-etal2022}
{M{\'e}ndez} M.,  {Karpouzas} K.,  {Garc{\'\i}a} F.,  {Zhang} L.,  {Zhang} Y.,
  {Belloni} T.~M.,   {Altamirano} D.,  2022, \mn@doi [Nature Astronomy]
  {10.1038/s41550-022-01617-y}, \href
  {https://ui.adsabs.harvard.edu/abs/2022NatAs...6..577M} {6, 577}

\bibitem[\protect\citeauthoryear{{Miller-Jones} et~al.,}{{Miller-Jones}
  et~al.}{2021}]{Miller-Jones-etal2021}
{Miller-Jones} J. C.~A.,  et~al., 2021, \mn@doi [Science]
  {10.1126/science.abb3363}, \href
  {https://ui.adsabs.harvard.edu/abs/2021Sci...371.1046M} {371, 1046}

\bibitem[\protect\citeauthoryear{{Miyamoto} \& {Kitamoto}}{{Miyamoto} \&
  {Kitamoto}}{1991}]{Miyamoto-etal1991}
{Miyamoto} S.,  {Kitamoto} S.,  1991, \mn@doi [\apj] {10.1086/170158}, \href
  {https://ui.adsabs.harvard.edu/abs/1991ApJ...374..741M} {374, 741}

\bibitem[\protect\citeauthoryear{{Molteni}, {Sponholz}  \&
  {Chakrabarti}}{{Molteni} et~al.}{1996}]{Molteni-etal1996}
{Molteni} D.,  {Sponholz} H.,   {Chakrabarti} S.~K.,  1996, \mn@doi [\apj]
  {10.1086/176775}, \href
  {https://ui.adsabs.harvard.edu/abs/1996ApJ...457..805M} {457, 805}

\bibitem[\protect\citeauthoryear{{Mondal}, {Suribhatla}, {Chatterjee}, {Singh}
  \& {Chatterjee}}{{Mondal} et~al.}{2024}]{Mondal-etal2024}
{Mondal} S.,  {Suribhatla} S.~P.,  {Chatterjee} K.,  {Singh} C.~B.,
  {Chatterjee} R.,  2024, \mn@doi [\apj] {10.3847/1538-4357/ad7d92}, \href
  {https://ui.adsabs.harvard.edu/abs/2024ApJ...975..257M} {975, 257}

\bibitem[\protect\citeauthoryear{{Morgan}, {Remillard}  \& {Greiner}}{{Morgan}
  et~al.}{1997}]{Morgan-etal1997}
{Morgan} E.~H.,  {Remillard} R.~A.,   {Greiner} J.,  1997, \mn@doi [\apj]
  {10.1086/304191}, \href
  {https://ui.adsabs.harvard.edu/abs/1997ApJ...482..993M} {482, 993}

\bibitem[\protect\citeauthoryear{{Nakajima} et~al.,}{{Nakajima}
  et~al.}{2023}]{Nakajima-etal2023}
{Nakajima} M.,  et~al., 2023, The Astronomer's Telegram, \href
  {https://ui.adsabs.harvard.edu/abs/2023ATel16206....1N} {16206, 1}

\bibitem[\protect\citeauthoryear{{Nandi}, {Debnath}, {Mandal}  \&
  {Chakrabarti}}{{Nandi} et~al.}{2012}]{Nandi-etal2012}
{Nandi} A.,  {Debnath} D.,  {Mandal} S.,   {Chakrabarti} S.~K.,  2012, \mn@doi
  [\aap] {10.1051/0004-6361/201117844}, \href
  {https://ui.adsabs.harvard.edu/abs/2012A&A...542A..56N} {542, A56}

\bibitem[\protect\citeauthoryear{{Nandi}, {Das}, {Majumder}, {Katoch}, {Antia}
  \& {Shah}}{{Nandi} et~al.}{2024}]{Nandi-etal2024}
{Nandi} A.,  {Das} S.,  {Majumder} S.,  {Katoch} T.,  {Antia} H.~M.,   {Shah}
  P.,  2024, \mn@doi [\mnras] {10.1093/mnras/stae1208}, \href
  {https://ui.adsabs.harvard.edu/abs/2024MNRAS.531.1149N} {531, 1149}

\bibitem[\protect\citeauthoryear{{Nitindala}, {Veledina}  \&
  {Poutanen}}{{Nitindala} et~al.}{2025}]{Nitindala-etal2025}
{Nitindala} A.~P.,  {Veledina} A.,   {Poutanen} J.,  2025, \mn@doi [\aap]
  {10.1051/0004-6361/202453188}, \href
  {https://ui.adsabs.harvard.edu/abs/2025A&A...694A.230N} {694, A230}

\bibitem[\protect\citeauthoryear{{Orosz} et~al.,}{{Orosz}
  et~al.}{2009}]{Orosz-etal2009}
{Orosz} J.~A.,  et~al., 2009, \mn@doi [\apj] {10.1088/0004-637X/697/1/573},
  \href {https://ui.adsabs.harvard.edu/abs/2009ApJ...697..573O} {697, 573}

\bibitem[\protect\citeauthoryear{{Orosz}, {Steiner}, {McClintock}, {Buxton},
  {Bailyn}, {Steeghs}, {Guberman}  \& {Torres}}{{Orosz}
  et~al.}{2014}]{Orosz-etal2014}
{Orosz} J.~A.,  {Steiner} J.~F.,  {McClintock} J.~E.,  {Buxton} M.~M.,
  {Bailyn} C.~D.,  {Steeghs} D.,  {Guberman} A.,   {Torres} M. A.~P.,  2014,
  \mn@doi [\apj] {10.1088/0004-637X/794/2/154}, \href
  {https://ui.adsabs.harvard.edu/abs/2014ApJ...794..154O} {794, 154}

\bibitem[\protect\citeauthoryear{Pahari et~al.,}{Pahari
  et~al.}{2018}]{Pahari-etal2018}
Pahari M.,  et~al., 2018, \mn@doi [The Astrophysical Journal]
  {10.3847/1538-4357/aae53b}, 867, 86

\bibitem[\protect\citeauthoryear{Parker et~al.,}{Parker
  et~al.}{2016}]{Parker-etal2016}
Parker M.~L.,  et~al., 2016, \mn@doi [The Astrophysical Journal Letters]
  {10.3847/2041-8205/821/1/L6}, 821, L6

\bibitem[\protect\citeauthoryear{{Parmar}, {Angelini}  \& {White}}{{Parmar}
  et~al.}{1995}]{Parmar-etal1995}
{Parmar} A.~N.,  {Angelini} L.,   {White} N.~E.,  1995, \mn@doi [\apjl]
  {10.1086/309730}, \href
  {https://ui.adsabs.harvard.edu/abs/1995ApJ...452L.129P} {452, L129}

\bibitem[\protect\citeauthoryear{Peng et~al.,}{Peng
  et~al.}{2024a}]{Peng-etal2024a}
Peng J.-Q.,  et~al., 2024a, \mn@doi [The Astrophysical Journal Letters]
  {10.3847/2041-8213/ad17ca}, 960, L17

\bibitem[\protect\citeauthoryear{{Peng} et~al.,}{{Peng}
  et~al.}{2024b}]{Peng-etal2024b}
{Peng} J.-Q.,  et~al., 2024b, \mn@doi [\apjl] {10.3847/2041-8213/ad74ec}, \href
  {https://ui.adsabs.harvard.edu/abs/2024ApJ...973L...7P} {973, L7}

\bibitem[\protect\citeauthoryear{{Peters}, {Polisensky}, {Clarke},
  {Giacintucci}  \& {Kassim}}{{Peters} et~al.}{2023}]{Peters-etal2023}
{Peters} W.~M.,  {Polisensky} E.,  {Clarke} T.~E.,  {Giacintucci} S.,
  {Kassim} N.~E.,  2023, The Astronomer's Telegram, \href
  {https://ui.adsabs.harvard.edu/abs/2023ATel16279....1P} {16279, 1}

\bibitem[\protect\citeauthoryear{{Pietrzy{\'n}ski} et~al.,}{{Pietrzy{\'n}ski}
  et~al.}{2013}]{Pietrzynski-etal2013}
{Pietrzy{\'n}ski} G.,  et~al., 2013, \mn@doi [\nat] {10.1038/nature11878},
  \href {https://ui.adsabs.harvard.edu/abs/2013Natur.495...76P} {495, 76}

\bibitem[\protect\citeauthoryear{{Podgorn{\'y}} et~al.,}{{Podgorn{\'y}}
  et~al.}{2023}]{Podgorny-etal2023}
{Podgorn{\'y}} J.,  et~al., 2023, \mn@doi [\mnras] {10.1093/mnras/stad3103},
  \href {https://ui.adsabs.harvard.edu/abs/2023MNRAS.526.5964P} {526, 5964}

\bibitem[\protect\citeauthoryear{{Podgorn{\'y}} et~al.,}{{Podgorn{\'y}}
  et~al.}{2024}]{Podgorny-etal2024}
{Podgorn{\'y}} J.,  et~al., 2024, \mn@doi [\aap] {10.1051/0004-6361/202450566},
  \href {https://ui.adsabs.harvard.edu/abs/2024A&A...686L..12P} {686, L12}

\bibitem[\protect\citeauthoryear{{Poutanen}, {Krolik}  \& {Ryde}}{{Poutanen}
  et~al.}{1997}]{Poutanen-etal1997}
{Poutanen} J.,  {Krolik} J.~H.,   {Ryde} F.,  1997, \mn@doi [\mnras]
  {10.1093/mnras/292.1.L21}, \href
  {https://ui.adsabs.harvard.edu/abs/1997MNRAS.292L..21P} {292, L21}

\bibitem[\protect\citeauthoryear{{Poutanen}, {Veledina}  \&
  {Beloborodov}}{{Poutanen} et~al.}{2023}]{Poutanen-etal2023}
{Poutanen} J.,  {Veledina} A.,   {Beloborodov} A.~M.,  2023, \mn@doi [\apjl]
  {10.3847/2041-8213/acd33e}, \href
  {https://ui.adsabs.harvard.edu/abs/2023ApJ...949L..10P} {949, L10}

\bibitem[\protect\citeauthoryear{{Prabhakar}, {Mandal}, {Bhuvana}  \&
  {Nandi}}{{Prabhakar} et~al.}{2023}]{Prabhakar-etal2023}
{Prabhakar} G.,  {Mandal} S.,  {Bhuvana} G.~R.,   {Nandi} A.,  2023, \mn@doi
  [\mnras] {10.1093/mnras/stad080}, \href
  {https://ui.adsabs.harvard.edu/abs/2023MNRAS.520.4889P} {520, 4889}

\bibitem[\protect\citeauthoryear{{Radhika} \& {Nandi}}{{Radhika} \&
  {Nandi}}{2014}]{Radhika-etal2014}
{Radhika} D.,  {Nandi} A.,  2014, \mn@doi [Advances in Space Research]
  {10.1016/j.asr.2014.06.039}, \href
  {https://ui.adsabs.harvard.edu/abs/2014AdSpR..54.1678R} {54, 1678}

\bibitem[\protect\citeauthoryear{{Radhika}, {Nandi}, {Agrawal}  \&
  {Seetha}}{{Radhika} et~al.}{2016}]{Radhika-etal2016}
{Radhika} D.,  {Nandi} A.,  {Agrawal} V.~K.,   {Seetha} S.,  2016, \mn@doi
  [\mnras] {10.1093/mnras/stw1239}, \href
  {https://ui.adsabs.harvard.edu/abs/2016MNRAS.460.4403R} {460, 4403}

\bibitem[\protect\citeauthoryear{{Radhika}, {Sreehari}, {Nandi}, {Iyer}  \&
  {Mandal}}{{Radhika} et~al.}{2018}]{Radhika-etal2018}
{Radhika} D.,  {Sreehari} H.,  {Nandi} A.,  {Iyer} N.,   {Mandal} S.,  2018,
  \mn@doi [\apss] {10.1007/s10509-018-3411-1}, \href
  {https://ui.adsabs.harvard.edu/abs/2018Ap&SS.363..189D} {363, 189}

\bibitem[\protect\citeauthoryear{{Ratheesh} et~al.,}{{Ratheesh}
  et~al.}{2024}]{Ratheesh-etal2024}
{Ratheesh} A.,  et~al., 2024, \mn@doi [\apj] {10.3847/1538-4357/ad226e}, \href
  {https://ui.adsabs.harvard.edu/abs/2024ApJ...964...77R} {964, 77}

\bibitem[\protect\citeauthoryear{{Rawat}, {Garg}  \& {M{\'e}ndez}}{{Rawat}
  et~al.}{2023a}]{Rawat-etal2023b}
{Rawat} D.,  {Garg} A.,   {M{\'e}ndez} M.,  2023a, \mn@doi [\mnras]
  {10.1093/mnras/stad2327}, \href
  {https://ui.adsabs.harvard.edu/abs/2023MNRAS.525..661R} {525, 661}

\bibitem[\protect\citeauthoryear{{Rawat}, {Garg}  \& {M{\'e}ndez}}{{Rawat}
  et~al.}{2023b}]{Rawat-etal2023a}
{Rawat} D.,  {Garg} A.,   {M{\'e}ndez} M.,  2023b, \mn@doi [\apjl]
  {10.3847/2041-8213/acd77b}, \href
  {https://ui.adsabs.harvard.edu/abs/2023ApJ...949L..43R} {949, L43}

\bibitem[\protect\citeauthoryear{{Reid} \& {Miller-Jones}}{{Reid} \&
  {Miller-Jones}}{2023}]{Reid-etal2023}
{Reid} M.~J.,  {Miller-Jones} J.~C.~A.,  2023, \mn@doi [\apj]
  {10.3847/1538-4357/acfe0c}, \href
  {https://ui.adsabs.harvard.edu/abs/2023ApJ...959...85R} {959, 85}

\bibitem[\protect\citeauthoryear{{Remillard} \& {McClintock}}{{Remillard} \&
  {McClintock}}{2006}]{Remillard-etal2006}
{Remillard} R.~A.,  {McClintock} J.~E.,  2006, \mn@doi [\araa]
  {10.1146/annurev.astro.44.051905.092532}, \href
  {https://ui.adsabs.harvard.edu/abs/2006ARA&A..44...49R} {44, 49}

\bibitem[\protect\citeauthoryear{{Remillard} et~al.,}{{Remillard}
  et~al.}{2022}]{Remillard-etal2022}
{Remillard} R.~A.,  et~al., 2022, \mn@doi [\aj] {10.3847/1538-3881/ac4ae6},
  \href {https://ui.adsabs.harvard.edu/abs/2022AJ....163..130R} {163, 130}

\bibitem[\protect\citeauthoryear{{Rodriguez Cavero} et~al.,}{{Rodriguez Cavero}
  et~al.}{2023}]{Rodriguez-etal2023}
{Rodriguez Cavero} N.,  et~al., 2023, \mn@doi [\apjl]
  {10.3847/2041-8213/acfd2c}, \href
  {https://ui.adsabs.harvard.edu/abs/2023ApJ...958L...8R} {958, L8}

\bibitem[\protect\citeauthoryear{{Rodriguez}, {Corbel}, {Caballero}, {Tomsick},
  {Tzioumis}, {Paizis}, {Cadolle Bel}  \& {Kuulkers}}{{Rodriguez}
  et~al.}{2011}]{Rodriguez-etal2011}
{Rodriguez} J.,  {Corbel} S.,  {Caballero} I.,  {Tomsick} J.~A.,  {Tzioumis}
  T.,  {Paizis} A.,  {Cadolle Bel} M.,   {Kuulkers} E.,  2011, \mn@doi [\aap]
  {10.1051/0004-6361/201117511}, \href
  {https://ui.adsabs.harvard.edu/abs/2011A&A...533L...4R} {533, L4}

\bibitem[\protect\citeauthoryear{Rodriguez et~al.,}{Rodriguez
  et~al.}{2015}]{Rodriguez-etal2015}
Rodriguez J.,  et~al., 2015, \mn@doi [The Astrophysical Journal]
  {10.1088/0004-637X/807/1/17}, 807, 17

\bibitem[\protect\citeauthoryear{{Schnittman} \& {Krolik}}{{Schnittman} \&
  {Krolik}}{2009}]{Schnittman-etal2009}
{Schnittman} J.~D.,  {Krolik} J.~H.,  2009, \mn@doi [\apj]
  {10.1088/0004-637X/701/2/1175}, \href
  {https://ui.adsabs.harvard.edu/abs/2009ApJ...701.1175S} {701, 1175}

\bibitem[\protect\citeauthoryear{{Seifina}, {Titarchuk}  \&
  {Shaposhnikov}}{{Seifina} et~al.}{2014}]{Seifina-etal2014}
{Seifina} E.,  {Titarchuk} L.,   {Shaposhnikov} N.,  2014, \mn@doi [\apj]
  {10.1088/0004-637X/789/1/57}, \href
  {https://ui.adsabs.harvard.edu/abs/2014ApJ...789...57S} {789, 57}

\bibitem[\protect\citeauthoryear{{Shakura} \& {Sunyaev}}{{Shakura} \&
  {Sunyaev}}{1973}]{Shakura-etal1973}
{Shakura} N.~I.,  {Sunyaev} R.~A.,  1973, \aap, \href
  {https://ui.adsabs.harvard.edu/abs/1973A&A....24..337S} {24, 337}

\bibitem[\protect\citeauthoryear{{Shui} et~al.,}{{Shui}
  et~al.}{2024}]{Shui-etal2024}
{Shui} Q.-C.,  et~al., 2024, \mn@doi [\apj] {10.3847/1538-4357/ad676a}, \href
  {https://ui.adsabs.harvard.edu/abs/2024ApJ...973...59S} {973, 59}

\bibitem[\protect\citeauthoryear{{Smale} \& {Boyd}}{{Smale} \&
  {Boyd}}{2012}]{Smale-etal2012}
{Smale} A.~P.,  {Boyd} P.~T.,  2012, \mn@doi [\apj]
  {10.1088/0004-637X/756/2/146}, \href
  {https://ui.adsabs.harvard.edu/abs/2012ApJ...756..146S} {756, 146}

\bibitem[\protect\citeauthoryear{{Soleri}, {Belloni}  \& {Casella}}{{Soleri}
  et~al.}{2008}]{Soleri-etal2008}
{Soleri} P.,  {Belloni} T.,   {Casella} P.,  2008, \mn@doi [\mnras]
  {10.1111/j.1365-2966.2007.12596.x}, \href
  {https://ui.adsabs.harvard.edu/abs/2008MNRAS.383.1089S} {383, 1089}

\bibitem[\protect\citeauthoryear{{Sreehari}, {Iyer}, {Radhika}, {Nandi}  \&
  {Mandal}}{{Sreehari} et~al.}{2019a}]{Sreehari-etal2019a}
{Sreehari} H.,  {Iyer} N.,  {Radhika} D.,  {Nandi} A.,   {Mandal} S.,  2019a,
  \mn@doi [Advances in Space Research] {10.1016/j.asr.2018.10.042}, \href
  {https://ui.adsabs.harvard.edu/abs/2019AdSpR..63.1374S} {63, 1374}

\bibitem[\protect\citeauthoryear{{Sreehari}, {Ravishankar}, {Iyer}, {Agrawal},
  {Katoch}, {Mandal}  \& {Nand i}}{{Sreehari}
  et~al.}{2019b}]{Sreehari-etal2019}
{Sreehari} H.,  {Ravishankar} B.~T.,  {Iyer} N.,  {Agrawal} V.~K.,  {Katoch}
  T.~B.,  {Mandal} S.,   {Nand i} A.,  2019b, \mn@doi [\mnras]
  {10.1093/mnras/stz1327}, \href
  {https://ui.adsabs.harvard.edu/abs/2019MNRAS.487..928S} {487, 928}

\bibitem[\protect\citeauthoryear{{Sreehari}, {Nandi}, {Das}, {Agrawal},
  {Mandal}, {Ramadevi}  \& {Katoch}}{{Sreehari}
  et~al.}{2020}]{Sreehari-etal2020}
{Sreehari} H.,  {Nandi} A.,  {Das} S.,  {Agrawal} V.~K.,  {Mandal} S.,
  {Ramadevi} M.~C.,   {Katoch} T.,  2020, \mn@doi [\mnras]
  {10.1093/mnras/staa3135}, \href
  {https://ui.adsabs.harvard.edu/abs/2020MNRAS.499.5891S} {499, 5891}

\bibitem[\protect\citeauthoryear{{Steiner}, {McClintock}, {Remillard}, {Gou},
  {Yamada}  \& {Narayan}}{{Steiner} et~al.}{2010}]{Steiner-etal2010}
{Steiner} J.~F.,  {McClintock} J.~E.,  {Remillard} R.~A.,  {Gou} L.,  {Yamada}
  S.,   {Narayan} R.,  2010, \mn@doi [\apjl] {10.1088/2041-8205/718/2/L117},
  \href {https://ui.adsabs.harvard.edu/abs/2010ApJ...718L.117S} {718, L117}

\bibitem[\protect\citeauthoryear{{Steiner} et~al.,}{{Steiner}
  et~al.}{2024}]{Steiner-etal2024}
{Steiner} J.~F.,  et~al., 2024, \mn@doi [\apjl] {10.3847/2041-8213/ad58e4},
  \href {https://ui.adsabs.harvard.edu/abs/2024ApJ...969L..30S} {969, L30}

\bibitem[\protect\citeauthoryear{{Stella} \& {Vietri}}{{Stella} \&
  {Vietri}}{1998}]{Stella-etal1998}
{Stella} L.,  {Vietri} M.,  1998, \mn@doi [\apjl] {10.1086/311075}, \href
  {https://ui.adsabs.harvard.edu/abs/1998ApJ...492L..59S} {492, L59}

\bibitem[\protect\citeauthoryear{Stern, Poutanen, Svensson, Sikora  \&
  Begelman}{Stern et~al.}{1995}]{Stern-etal1995}
Stern B.~E.,  Poutanen J.,  Svensson R.,  Sikora M.,   Begelman M.~C.,  1995,
  \mn@doi [The Astrophysical Journal] {10.1086/309617}, 449, L13

\bibitem[\protect\citeauthoryear{{Strohmayer}}{{Strohmayer}}{2017}]{Strohmayer-etal2017}
{Strohmayer} T.~E.,  2017, \mn@doi [\apj] {10.3847/1538-4357/aa643d}, \href
  {https://ui.adsabs.harvard.edu/abs/2017ApJ...838...72S} {838, 72}

\bibitem[\protect\citeauthoryear{{Svoboda} et~al.,}{{Svoboda}
  et~al.}{2024a}]{Svoboda-etal2024a}
{Svoboda} J.,  et~al., 2024a, \mn@doi [\apj] {10.3847/1538-4357/ad0842}, \href
  {https://ui.adsabs.harvard.edu/abs/2024ApJ...960....3S} {960, 3}

\bibitem[\protect\citeauthoryear{{Svoboda} et~al.,}{{Svoboda}
  et~al.}{2024b}]{Svoboda-etal2024b}
{Svoboda} J.,  et~al., 2024b, \mn@doi [\apjl] {10.3847/2041-8213/ad402e}, \href
  {https://ui.adsabs.harvard.edu/abs/2024ApJ...966L..35S} {966, L35}

\bibitem[\protect\citeauthoryear{{Tigelaar}, {Fender}, {Tilanus}, {Gallo}  \&
  {Pooley}}{{Tigelaar} et~al.}{2004}]{Tigelaar-etal2004}
{Tigelaar} S.~P.,  {Fender} R.~P.,  {Tilanus} R.~P.~J.,  {Gallo} E.,   {Pooley}
  G.~G.,  2004, \mn@doi [\mnras] {10.1111/j.1365-2966.2004.07992.x}, \href
  {https://ui.adsabs.harvard.edu/abs/2004MNRAS.352.1015T} {352, 1015}

\bibitem[\protect\citeauthoryear{{Titarchuk}}{{Titarchuk}}{1994}]{Titarchuk-etal1994}
{Titarchuk} L.,  1994, \mn@doi [\apj] {10.1086/174760}, \href
  {https://ui.adsabs.harvard.edu/abs/1994ApJ...434..570T} {434, 570}

\bibitem[\protect\citeauthoryear{{Tomaru}, {Done}  \& {Odaka}}{{Tomaru}
  et~al.}{2024}]{Tomaru-etal2024}
{Tomaru} R.,  {Done} C.,   {Odaka} H.,  2024, \mn@doi [\mnras]
  {10.1093/mnras/stad3649}, \href
  {https://ui.adsabs.harvard.edu/abs/2024MNRAS.527.7047T} {527, 7047}

\bibitem[\protect\citeauthoryear{{Tomsick}, {Lapshov}  \& {Kaaret}}{{Tomsick}
  et~al.}{1998}]{Tomsick-etal1998}
{Tomsick} J.~A.,  {Lapshov} I.,   {Kaaret} P.,  1998, \mn@doi [\apj]
  {10.1086/305240}, \href
  {https://ui.adsabs.harvard.edu/abs/1998ApJ...494..747T} {494, 747}

\bibitem[\protect\citeauthoryear{Tomsick et~al.,}{Tomsick
  et~al.}{2013}]{Tomsick-etal2013}
Tomsick J.~A.,  et~al., 2013, \mn@doi [The Astrophysical Journal]
  {10.1088/0004-637X/780/1/78}, 780, 78

\bibitem[\protect\citeauthoryear{{Torpin}, {Boyd}, {Smale}  \&
  {Valencic}}{{Torpin} et~al.}{2017}]{Torpin-etal2017}
{Torpin} T.~J.,  {Boyd} P.~T.,  {Smale} A.~P.,   {Valencic} L.~A.,  2017,
  \mn@doi [\apj] {10.3847/1538-4357/aa8f96}, \href
  {https://ui.adsabs.harvard.edu/abs/2017ApJ...849...32T} {849, 32}

\bibitem[\protect\citeauthoryear{{Veledina} et~al.,}{{Veledina}
  et~al.}{2023}]{Veledina-etal2023}
{Veledina} A.,  et~al., 2023, \mn@doi [\apjl] {10.3847/2041-8213/ad0781}, \href
  {https://ui.adsabs.harvard.edu/abs/2023ApJ...958L..16V} {958, L16}

\bibitem[\protect\citeauthoryear{{Veledina} et~al.,}{{Veledina}
  et~al.}{2024a}]{Veledina-etal2024a}
{Veledina} A.,  et~al., 2024a, \mn@doi [Nature Astronomy]
  {10.1038/s41550-024-02294-9}, \href
  {https://ui.adsabs.harvard.edu/abs/2024NatAs...8.1031V} {8, 1031}

\bibitem[\protect\citeauthoryear{{Veledina} et~al.,}{{Veledina}
  et~al.}{2024b}]{Veledina-etal2024b}
{Veledina} A.,  et~al., 2024b, \mn@doi [\aap] {10.1051/0004-6361/202451356},
  \href {https://ui.adsabs.harvard.edu/abs/2024A&A...688L..27V} {688, L27}

\bibitem[\protect\citeauthoryear{{Wang}, {M{\'e}ndez}, {Altamirano}, {Court},
  {Beri}  \& {Cheng}}{{Wang} et~al.}{2018}]{Wang-etal2018}
{Wang} Y.,  {M{\'e}ndez} M.,  {Altamirano} D.,  {Court} J.,  {Beri} A.,
  {Cheng} Z.,  2018, \mn@doi [\mnras] {10.1093/mnras/sty1372}, \href
  {https://ui.adsabs.harvard.edu/abs/2018MNRAS.478.4837W} {478, 4837}

\bibitem[\protect\citeauthoryear{{Wang} et~al.,}{{Wang}
  et~al.}{2024}]{Wang-etal2024}
{Wang} J.,  et~al., 2024, \mn@doi [\apj] {10.3847/1538-4357/ad1595}, \href
  {https://ui.adsabs.harvard.edu/abs/2024ApJ...963...14W} {963, 14}

\bibitem[\protect\citeauthoryear{{Webster} \& {Murdin}}{{Webster} \&
  {Murdin}}{1972}]{Webster-etal1972}
{Webster} B.~L.,  {Murdin} P.,  1972, \mn@doi [\nat] {10.1038/235037a0}, \href
  {https://ui.adsabs.harvard.edu/abs/1972Natur.235...37W} {235, 37}

\bibitem[\protect\citeauthoryear{{Weisskopf} et~al.,}{{Weisskopf}
  et~al.}{2022}]{Weisskopf-etal2022}
{Weisskopf} M.~C.,  et~al., 2022, \mn@doi [Journal of Astronomical Telescopes,
  Instruments, and Systems] {10.1117/1.JATIS.8.2.026002}, \href
  {https://ui.adsabs.harvard.edu/abs/2022JATIS...8b6002W} {8, 026002}

\bibitem[\protect\citeauthoryear{{Williams-Baldwin}, {Motta}, {Rhodes},
  {Carotenuto}, {Fender}  \& {Beswick}}{{Williams-Baldwin}
  et~al.}{2023}]{Williams-Baldwin-etal2023}
{Williams-Baldwin} D.,  {Motta} S.,  {Rhodes} L.,  {Carotenuto} F.,  {Fender}
  R.,   {Beswick} R.,  2023, The Astronomer's Telegram, \href
  {https://ui.adsabs.harvard.edu/abs/2023ATel16231....1W} {16231, 1}

\bibitem[\protect\citeauthoryear{{Wilms}, {Allen}  \& {McCray}}{{Wilms}
  et~al.}{2000}]{Wilms-etal2000}
{Wilms} J.,  {Allen} A.,   {McCray} R.,  2000, \mn@doi [\apj] {10.1086/317016},
  \href {https://ui.adsabs.harvard.edu/abs/2000ApJ...542..914W} {542, 914}

\bibitem[\protect\citeauthoryear{{Wood} et~al.,}{{Wood}
  et~al.}{2024}]{Wood-etal2024}
{Wood} C.~M.,  et~al., 2024, \mn@doi [\apjl] {10.3847/2041-8213/ad6572}, \href
  {https://ui.adsabs.harvard.edu/abs/2024ApJ...971L...9W} {971, L9}

\bibitem[\protect\citeauthoryear{Xu et~al.,}{Xu et~al.}{2017}]{Xu-etal2017}
Xu Y.,  et~al., 2017, \mn@doi [The Astrophysical Journal]
  {10.3847/1538-4357/aa9ab4}, 851, 103

\bibitem[\protect\citeauthoryear{{Zdziarski}, {Johnson}  \&
  {Magdziarz}}{{Zdziarski} et~al.}{1996}]{Zdziarski-etal1996}
{Zdziarski} A.~A.,  {Johnson} W.~N.,   {Magdziarz} P.,  1996, \mn@doi [\mnras]
  {10.1093/mnras/283.1.193}, \href
  {https://ui.adsabs.harvard.edu/abs/1996MNRAS.283..193Z} {283, 193}

\bibitem[\protect\citeauthoryear{{Zdziarski}, {Szanecki}, {Poutanen},
  {Gierli{\'n}ski}  \& {Biernacki}}{{Zdziarski}
  et~al.}{2020}]{Zdziarski-etal2020}
{Zdziarski} A.~A.,  {Szanecki} M.,  {Poutanen} J.,  {Gierli{\'n}ski} M.,
  {Biernacki} P.,  2020, \mn@doi [\mnras] {10.1093/mnras/staa159}, \href
  {https://ui.adsabs.harvard.edu/abs/2020MNRAS.492.5234Z} {492, 5234}

\bibitem[\protect\citeauthoryear{{Zdziarski}, {Banerjee}, {Szanecki}, {Misra}
  \& {Dewangan}}{{Zdziarski} et~al.}{2024}]{Zdziarski-etal2024}
{Zdziarski} A.~A.,  {Banerjee} S.,  {Szanecki} M.,  {Misra} R.,   {Dewangan}
  G.,  2024, \mn@doi [arXiv e-prints] {10.48550/arXiv.2412.15705}, \href
  {https://ui.adsabs.harvard.edu/abs/2024arXiv241215705Z} {p. arXiv:2412.15705}

\bibitem[\protect\citeauthoryear{{Zhang} et~al.,}{{Zhang}
  et~al.}{2023}]{Zhang-etal2023}
{Zhang} Y.,  et~al., 2023, \mn@doi [\mnras] {10.1093/mnras/stad460}, \href
  {https://ui.adsabs.harvard.edu/abs/2023MNRAS.520.5144Z} {520, 5144}

\bibitem[\protect\citeauthoryear{Zhao et~al.,}{Zhao
  et~al.}{2021}]{Zhao-etal2021}
Zhao X.,  et~al., 2021, \mn@doi [The Astrophysical Journal]
  {10.3847/1538-4357/abbcd6}, 908, 117

\bibitem[\protect\citeauthoryear{{van Kerkwijk} et~al.,}{{van Kerkwijk}
  et~al.}{1992}]{van-Kerkwijk-etal1992}
{van Kerkwijk} M.~H.,  et~al., 1992, \mn@doi [\nat] {10.1038/355703a0}, \href
  {https://ui.adsabs.harvard.edu/abs/1992Natur.355..703V} {355, 703}

\makeatother
\end{thebibliography}
\input{ms.bbl}

\end{document}